\documentclass[aps,twocolumn,superscriptaddress,floatfix,longbibliography,showpacs,showkeys]{revtex4-2}

\usepackage[utf8]{inputenc}
\usepackage[T1]{fontenc}
\usepackage{lmodern}
\usepackage{newtxtext}

\usepackage{amsmath}
\usepackage{amssymb}
\usepackage{amsthm}
\usepackage{amscd}
\usepackage{mathtools}
\usepackage{empheq}
\usepackage{mathrsfs}
\usepackage{dsfont}
\usepackage{scalerel}

\usepackage{graphicx}
\usepackage{grffile}
\usepackage{xcolor}
\usepackage{subcaption}
\usepackage{sidecap}

\usepackage{tikz}
\usetikzlibrary{arrows.meta, positioning, shapes.geometric, decorations.pathmorphing}

\usepackage{etoolbox}
\usepackage[mathlines]{lineno}
\usepackage[ruled]{algorithm2e}

\usepackage[colorlinks=true, pdfstartview=FitV, linkcolor=blue, citecolor=blue, urlcolor=blue]{hyperref}

\makeatletter
\renewcommand{\p@subsubsection}{\thesection\,}

\renewcommand{\subsubsection}{\@startsection{subsubsection}{3}{\z@}
  {2ex \@plus 1ex \@minus .1ex}
  {1ex \@plus .1ex}
  {\normalfont\small\bfseries}}
\makeatother

\makeatletter
\def\appendix@toc@setup{
  \let\oldl@section\l@section
  \def\l@section##1##2{
    \begingroup
    \def\numberline####1{\makebox[6.5em][l]{Appendix ####1.\quad}}
    \oldl@section{##1}{##2}
    \endgroup
  }
}
\apptocmd{\appendix}{\addtocontents{toc}{\protect\appendix@toc@setup}}{}{}
\makeatother

\usepackage{amscd}
\usepackage{amsmath}
\usepackage{amssymb}
\usepackage{graphicx}
\usepackage{xcolor}
\definecolor{rred}{rgb}{0.7,0,0.1}
\definecolor{greenrb}{rgb}{0.2,0.6,0.2}
\definecolor{ccyan}{rgb}{0,.5,1}

\definecolor{revisionpurple}{RGB}{128,0,128}

\usepackage[ruled]{algorithm2e}
\usepackage{needspace}
\usepackage{newtxtext}
\usepackage{booktabs}
\usepackage{bm}

\usepackage[colorlinks=true, pdfstartview=FitV, linkcolor=blue,
            citecolor=blue, urlcolor=blue]{hyperref} 
            
\usepackage{subcaption}            
\usepackage{sidecap}
\usepackage{grffile}

\def\x{\mathbf{x}}
\def\y{\mathbf{y}}
\def\kvec{\mathbf{k}}

\def\d{\mathrm{d}}
\def\Torus{\mathbb{T}^2}

\def\bea{\begin{equation} \begin{aligned}}
\def\eea{\end{aligned} \end{equation}}
\def\beas{\begin{equation*} \begin{aligned}}
\def\eeas{\end{aligned} \end{equation*}}
\def\bes{\begin{equation*}}
\def\ees{\end{equation*}}

\def\d{\, \mathrm{d}}

\def\be{\begin{equation}}
\def\ee{\end{equation}}
\def\adots{
  \mathinner{\mkern1mu\raise1pt\hbox{.}\mkern2mu\raise4pt\hbox{.}
  \mkern2mu\raise7pt\vbox{\kern7pt\hbox{.}}\mkern1mu}}

\newtheorem{rem}{Remark}[section]
\def\br{\begin{rem}}
\def\er{\end{rem}}
\def\bi{\begin{itemize}}
\def\ei{\end{itemize}}
\def\ben{\begin{enumerate}}
\def\een{\end{enumerate}}

\begin{document}
\bibliographystyle{ieeetr}

\title[Emergent cloud patterns from stochastic jump models]{Emergent cloud patterns from stochastic jump models of parcel microphysics}

\author{Micka{\"e}l D. Chekroun}
\affiliation{Department of Atmospheric \& Oceanic Sciences, University of California, Los Angeles, CA 90095-1565, USA}
\affiliation{Department of Earth and Planetary Sciences, Weizmann Institute of Science, Rehovot 76100, Israel} 
\email{mchekroun@atmos.ucla.edu}

\author{Ilan Koren}
\affiliation{Department of Earth and Planetary Sciences, Weizmann Institute of Science, Rehovot 76100, Israel} 

\author{Honghu Liu}
\affiliation{Department of Mathematics, Virginia Tech, Blacksburg, VA 24061, USA} 

\date{\today}

\begin{abstract}
Warm clouds organize into bands, cells, and intricate wave networks, yet the dynamical mechanisms connecting these morphologies remain difficult to isolate. To bridge unresolved parcel microphysics and cloud organization, we formulate a spatial jump-process model of cloud-water and rain-water anomalies in an activated regime and analyze its mean-field closure. Eliminating fast supersaturation adjustment from droplet growth and vapor transport supplies a screened environmental kernel, which we couple to a cubic oscillatory law near Hopf onset. Averaging over Poisson parcel episodes and closing third-order moments yields a pseudoparabolic Ginzburg--Landau equation (PGLE), whose mixed space--time derivative encodes dynamic inertia, slowing fine-scale relaxation. Unlike classical diffusive amplitude equations, the PGLE decomposes into a regularized nonlinear Schr\"odinger Hamiltonian sector and a pseudoparabolic Allen--Cahn gradient sector governed by a conserved Noether charge, alongside an open-system liquid-water budget distinguishing conversion from vapor exchange, fallout, and triad-driven domain-mean oscillations. Exact stability analyses establish how dynamic inertia and amplitude-dependent frequency shifts (twist) control modulational loss of synchrony and resonant cross-carrier sideband amplification above a critical twist threshold. Numerical integrations demonstrate that increasing twist transforms coherent undulatus-like bands into defect-rich asperitas-like waves via defect proliferation, whereas subcritical mode competition consolidates cellular interference into parallel rolls. Even without twist, unstructured random anomalies evolve into a persistent cellular exclusion scale and outward-radiating actinoform-like spiral arms, while one-dimensional wave breakup reorganizes into a two-frequency combination-tone spectrum. The jump formulation identifies the omitted fluctuation moments for future calibration against large-eddy simulations.
\end{abstract}



\maketitle

{\small
\makeatletter
\def\appendix@toc@setup{
  \def\l@f@section{\bfseries
    \let\appendix@numberline\numberline
    \def\numberline####1{
      \@if@empty{####1}{\appendix@numberline{}}
        {\appendix@numberline{Appendix ####1}}}
  }
}
\makeatother
\tableofcontents
}
\section{Introduction}

 Warm clouds organize into bands, cellular structures, and clusters whose geometry changes as the cloud field evolves. The recurring trade-wind patterns known as Sugar, Gravel, Fish, and Flowers illustrate this diversity~\cite{stevens2020patterns}; their occurrence varies with environmental conditions and is associated with differences in cloud cover~\cite{bony2020patterns}. Understanding these arrangements requires more than identifying their shapes. It also requires explaining how organized structures emerge and how one arrangement gives way to another. Cloud-resolving studies show that precipitation, cold-pool interactions, and wave-mediated disturbances can reorganize cloud fields~\cite{feingold_2010,dagan2018organization}. The spatial distribution of precipitation influences the development of open cellular convection~\cite{yamaguchi2015precipitation}, while analyses of transitions between stratocumulus states show how precipitation and entrainment jointly shape the evolving water budget~\cite{hoffmann2023roles}. These results motivate a complementary theoretical question: which collective mechanisms are sufficient to connect different patterns within a tractable evolution law?

 The question spans scales. Condensation and evaporation act at droplet surfaces, whereas coherent cloud structures extend across many parcels. Droplet growth consumes vapor and releases latent heat, modifying the supersaturation that drives further growth; transport and mixing redistribute this thermodynamic response~\cite{khain_pinsky_2018}. Consequently, neighboring droplet populations interact through an environment that they also modify. A reduced description must retain some trace of this shared adjustment if it is to connect microphysical activity with collective cloud evolution.

{Optically faint twilight clouds occupy the transition between haze and fully developed clouds, with consequences for the radiative properties of nominally clear sky \cite{koren2007twilight,hirsch2014transition,eytan2020}. A recent droplet--supersaturation theory of cloud formation provides such a connection \cite{chekroun_koren_hovering2026}. Starting from droplet growth and supersaturation transport, it obtains coupled population fields in which the supersaturation is spatially redistributed while the droplet-population variable responds locally. Eliminating the environmental field transfers its spatial response into the population equation: a mixed space--time derivative emerges. The resulting equation is pseudoparabolic, and the mixed derivative expresses \emph{dynamic inertia}---the resistance to changing a droplet population together with its surrounding supersaturation envelope. The theory can describe cloud patches emerging from haze; here its relevance is the physical origin of an evolution law whose rate of change is itself spatially coupled~\cite{chekroun_koren_hovering2026}.}

Such mixed derivatives have an established place in continuum physics~\cite{showalter1970pseudoparabolic,carroll1977singular,Showalter1994}. Exchange between fractures and porous blocks introduces internal adjustment into flow through fissured rocks~\cite{barenblatt1960basic}; related structures arise in two-temperature heat conduction~\cite{Chen_Gurtin1968theory} and in constitutive descriptions of viscoelastic response~\cite{Coleman_Noll1974,huilgol1968second,ting1963certain}. Constitutive coupling can thereby change the character of macroscopic evolution~\cite{joseph1985hyperbolicity}, while radiative-transfer theory provides further examples of nonlocal environmental response~\cite{milne1926diffusion,sobolev1963treatise}. For clouds, the specific coupling is between droplet populations and the transported supersaturation field. Its elimination supplies a screened spatial response: disturbances influence neighboring regions over a characteristic adjustment length~\cite{chekroun_koren_hovering2026}.

To describe organization in an activated, potentially precipitating cloud field, spatial adjustment must be coupled to the evolution of cloud water and rain water. Precipitation-generated oscillations in open cells~\cite{feingold_2010} and bin-microphysics large-eddy simulations (LES) resolving closed recharge--discharge cycles in the cloud-mass versus rain-mass phase plane~\cite{dagan2018organization} provide a direct physical basis for treating that evolution as oscillatory. Conceptual cloud--rain models likewise exhibit oscillations through delayed feedback, with Hopf bifurcations organizing their onset~\cite{koren2011aerosol,feingold_2013,chekroun2020efficient}. Near such an onset, a complex amplitude combines the two oscillating components, and a cubic Stuart--Landau law describes amplitude saturation and the dependence of timing on amplitude~\cite{guckenheimer2013nonlinear,Tantet_al_Hopf,Chekroun_al22SciAdv}. These properties suggest an amplitude-level description of cloud--rain organization in which the screened environmental response couples neighboring oscillations. This reduction introduces cloud--rain amplitude coordinates for the oscillatory regime, carrying the screened environmental response from the droplet-population theory into the new variables.

The nonspatial Koren--Tziperman--Feingold (KTF) cloud-rain delay model provides a concrete precedent~\cite{koren2017exploring}; see also \cite{koren2011aerosol}. Its Hopf reduction guides stochastic perturbations representing intermittent rain and turbulent effects; through shear-induced stretching and folding, these perturbations turn periodic cycles into irregular oscillations whose dominant period and spectral structure compare closely with WRF simulations of open cells~\cite{Chekroun_al22SciAdv}. This gives the Stuart--Landau law a cloud-specific motivation. The oscillation frequency depends on the amplitude of the cloud-water and rain-water excursions. A disturbance that changes this amplitude therefore changes the rate at which the system progresses through its cycle, advancing or delaying subsequent maxima. This amplitude dependence of frequency is called \emph{twist}, and it makes the timing of the cycle sensitive to amplitude disturbances \cite{guckenheimer1975isochrons,Ashwin2016,izhikevich2007dynamical,winfree1980geometry,Chekroun_al22SciAdv}. The present model asks how that response organizes a field when neighboring oscillations interact through a screened environment.

{Unresolved parcel activity supplies a second ingredient. Episodes of entrainment and evaporation, or of rain formation through collision--coalescence and subsequent removal, change water content over times shorter than those resolved by a cloud-field description~\cite{khain_pinsky_2018}. Stochastic convection models already provide several ways to retain unresolved variability. Lattice multicloud models describe transitions between discrete cloud types and their evolving area fractions~\cite{khouider2010multicloud,Frenkel2012,Frenkel2013}. Conditional Markov chains infer cloud-state transitions from resolved simulations or observations, including formulations with local spatial coupling~\cite{Dorrestijn2013,Dorrestijn2015}. Other parameterizations introduce fluctuations into convective forcing and closure relations, or describe noise-driven transitions to strong convection~\cite{LinNeelin2000,LinNeelin2003,stechmann2011stochastic,majda2008stochastic}. These approaches establish the importance of stochastic variability for convection; they also make the choice of the variable changed by an event central to the model's physical meaning.}

Here, each jump represents the cumulative change in continuous cloud-water and rain-water anomalies during an unresolved parcel episode, with a systematic increment governed by the Hopf oscillatory law~\cite{chekroun2020efficient,Chekroun_al22SciAdv} and a lateral influence distributed across neighboring elements by the screened environmental kernel. Specifying an explicit Poisson event law with random water-path departures separates the transport-based origin of spatial coupling from the statistical closure of unresolved microphysics, which in future work can be calibrated directly against large-eddy simulation (LES) diagnostics of entrainment, precipitation, and water-budget evolution~\cite{dawe2013entrainment,hoffmann2023roles} and analyzed with its fluctuation variance~\cite{chekroun_koren_liu_variance2026}. Taking the ensemble mean and closing the cubic fluctuation moments yields the pseudoparabolic Ginzburg--Landau equation (PGLE) as the leading continuum law for cloud--rain variability (Sec.~\ref{Sec_jump_construction}). Unlike lattice cloud models whose mean-field limit collapses spatial morphology into bulk cloud-fraction equations~\cite{khouider2019new}, the PGLE retains nonlocal spatial coupling and dynamic inertia at the mean-field level, while identifying the exact moment corrections omitted by the deterministic closure.

To establish what this mean-field dynamics sustains and why it differs from classical diffusive amplitude equations, we develop its dispersion, budget, and stability structure before turning to numerical experiments. Contrasting the PGLE with the standard complex Ginzburg--Landau equation and linear stochastic-diffusion cloud models~\cite{staliunas2001spectra,stechmann2016cloud} shows that dynamic inertia reverses short-wavelength kinematics, slowing the relaxation of fine structure rather than smoothing it diffusively (Sec.~\ref{Sec_dispersion_nonlinear}). We formulate the open-system liquid-water mass budget---showing how conversion-only events conserve liquid water while resonant wave triads drive bounded domain-mean water oscillations---and uncover a metric-coupled Hamiltonian and gradient decomposition into a Benjamin--Bona--Mahony-regularized nonlinear Schr\"odinger sector~\cite{benjamin1972model} and a pseudoparabolic Allen--Cahn sector, whose conserved $U(1)$ Noether charge $\mathcal N_\epsilon$ separates conservative phase-shearing from global amplitude saturation (Sec.~\ref{Sec_budgets}). Exact linear and weakly nonlinear stability analyses then establish how dynamic inertia and amplitude-dependent timing shifts destabilize single traveling waves and homogeneous oscillations through non-harmonic sidebands (Sec.~\ref{Sec_Instab_Neimark}), and yield a closed gain--detuning criterion for cross-carrier amplification governed by the twist ratio $|\beta/\alpha|>\sqrt{5}/2$ (Sec.~\ref{Sec_Nonlin_Wave_Interact}).

Numerical integrations of the deterministic PGLE demonstrate how these mechanisms unify seemingly disparate cloud morphologies (Sec.~\ref{Sec_modeling_results}). In one dimension, an unstable traveling wave transitions into a modulated state whose saturated spectrum organizes into a sharp combination-tone comb distinct from the initiating linear frequencies (Sec.~\ref{Sec_sidebands}). In two dimensions, strengthening the amplitude-dependent timing twist at fixed growth and saturation coefficients redistributes the Noether charge into intense gradient filaments and transforms coherent \emph{undulatus}-like bands---punctuated by isolated $\pm 1$ phase defects analogous to natural band breaks---into intricate \emph{asperitas}-like wave networks through rapid defect proliferation (Secs.~\ref{Sec_banded_solutions}--\ref{Sec_defect_analysis}), offering an oscillatory wave-interaction perspective alongside shear-settling mechanisms~\cite{harrison2017asperitas,ravichandran2022asperitas}. 

Conversely, below the cross-carrier amplification threshold, cubic mode competition consolidates a cellular interference pattern into parallel convective rolls as spatial amplitude variance vanishes (Sec.~\ref{Sec_spontaneous_stabilization}). Most strikingly, even without phase twist or prescribed carriers, small spatially uncorrelated anomalies organize into a persistent cellular exclusion scale \cite{koren2013adaptive}---exhibiting a sustained negative amplitude-correlation lobe absent in a linear control---and develop outward-radiating spiral cloud-water arms around phase singularities reminiscent of marine actinoform clouds~\cite{garay2004actinoform}, without invoking a built-in preferred wavenumber as in Swift--Hohenberg descriptions~\cite{monroy2021description} (Sec.~\ref{Sec_random_organization}). Together, these results establish the organizing capacity of the deterministic mean-field dynamics and provide a quantitative baseline for future studies restoring LES-calibrated parcel fluctuations (Sec.~\ref{Sec_concluding_perspective}).

\section{Physical Foundations of the Pseudoparabolic Model}
\label{Sec_PGLE_physics}

\subsection{From droplet growth to spatial population fields}

The spatial coupling used below originates in the environment shared by neighboring droplet populations. Droplet growth consumes water vapor and releases latent heat; air motion, mixing, and cooling replenish or redistribute the supersaturation that drives that growth. Before specifying an event model, we recall how this environmental response enters a population-field description. Let $s=e/e_s(T)-1$ denote supersaturation, where $e$ is vapor pressure and $e_s(T)$ its saturation value at temperature $T$. Combining the vapor and temperature balances gives the transport equation~\cite[Sec.~5.2.7]{khain_pinsky_2018}
\be\label{Eq_Transport_Siewert_al}
\partial_t s+\mathbf v\cdot\nabla s=\kappa_\chi \Delta s+A_1w-A_2\dot q_l-s\frac{1}{\bar\rho_a}\partial_{x_i}
(\bar\rho_a v_i).
\ee
Here $s$ is dimensionless supersaturation, $\mathbf v$ the air velocity, $w$ its vertical component, and $\bar\rho_a$ the reference air density. We use $\mathbf v$ and $\bar\rho_a$ to distinguish the transport variables from the cloud-field amplitude introduced below. {The coefficient $\kappa_\chi$, with units of area per time, is an effective molecular diffusivity for the supersaturation balance, combining vapor and thermal transport.} The term $A_1w$ is the production of supersaturation by adiabatic cooling, while $A_2\dot q_l$ is its depletion by condensation, with $q_l$ the liquid-water mixing ratio and $\dot q_l$ its phase-change tendency. The final term is the density-weighted expansion or dilution contribution, with summation over the coordinate index $i$~\cite{clark1973warmcumulus,khain_pinsky_2018}. {Under an anelastic resolved-flow constraint, $\nabla\cdot(\bar\rho_a\mathbf v)=0$, the final term vanishes; unresolved environmental exchange must still be modeled in the coarse-grained balance.} For a mass-ratio definition of $q_l$, $[A_1]=\mathsf L^{-1}$ and $A_2$ is dimensionless, so every term has units of inverse time.

The supersaturation balance is coupled to diffusional droplet growth through
\be\label{Eq_droplet_growth}
\frac{\d r^2}{\d t}=2D\,[s-f(r^2)],
\ee
where $r$ is droplet radius, $D$ is the condensational growth coefficient with units of area per time, and $f(r^2)$ is the dimensionless K\"ohler equilibrium supersaturation, accounting for curvature and solute effects~\cite{kohler_1936,gutierrez2024gibbs,santos2026brownian}. The sign of $s-f(r^2)$ determines growth or evaporation. Equations~\eqref{Eq_Transport_Siewert_al} and \eqref{Eq_droplet_growth} thus couple a local microphysical response to a transported environment. Their coarse-grained reduction is developed in Ref.~\cite{chekroun_koren_hovering2026}.

The local population averages define the macroscopic fields
\be\label{Eq_population_fields}
\phi(\x,t)\equiv\langle r^2\rangle_{\x,t},
\qquad \chi(\x,t)=\langle s\rangle_{\x,t}.
\ee
Here the brackets denote the droplet-population average in a coarse spatial neighborhood of $\x$ at time $t$; $\chi$ is the corresponding droplet-sampled supersaturation. These population fields resolve spatial variations across haze and cloud regions throughout the domain. Under its local-bath approximation, preferential sampling of unresolved supersaturation is neglected and $\chi$ also represents the spatially filtered environmental field. Unresolved scalar transport contributes an eddy diffusivity, while fluctuations of environmental exchange supply a multiplicative forcing. In coordinates following the resolved flow, the resulting system is \cite{chekroun_koren_hovering2026}:
\bea\label{Eq_hover_parcel}
\partial_t\phi&=-\mu\phi+2D[\chi-f(\phi)],\\
\partial_t\chi&=D_\chi\Delta\chi+\gamma-\lambda_\chi\chi\\
&\quad-\alpha_N\sqrt{\phi}[\chi-f(\phi)]+\sigma \chi\circ\eta(\x,t).
\eea
{Here $\phi$ is the local mean-square droplet radius and $\chi$ the mean supersaturation defined in Eq.~\eqref{Eq_population_fields}. The source $\gamma$ includes the resolved updraft contribution $A_1\overline w$, where $\overline w$ is the coarse-grained vertical velocity; a prescribed environmental supersaturation can add an exchange source. The coefficient $\lambda_\chi>0$ is the mean mixing or environmental-exchange rate: the term $-\lambda_\chi\chi$ relaxes supersaturation toward the reference environment represented in $\gamma$. Together, $\gamma$ and $\lambda_\chi$ set the environmental supply and relaxation that compete with condensation.
The noise term in Eq.~\eqref{Eq_hover_parcel} is understood in the Stratonovich sense represented by the symbol $\circ$.  The factor $\sigma \eta(\x,t)$ represents fluctuations of the environmental-exchange rate where 
$\eta$ is a centered Gaussian noise, white in time and correlated in space. Conversion to It\^o form adds the drift proportional to $\tfrac{1}{2}\sigma^2 \chi$ to the bath equation \cite{chekroun_koren_hovering2026}; the effective relaxation rate $\Gamma_0^*$ used below includes this correction.

The projected (thermodynamic) equilibrium curve is
\be\label{Eq_hover_kohler}
f(\phi)=a_K\phi^{-1/2}-b_K\phi^{-3/2}
\ee
with curvature and solute coefficients $a_K,b_K>0$~\cite{kohler_1936,khain_pinsky_2018,gutierrez2024gibbs,santos2026brownian}. This projection closes the droplet-spectrum average at the mean-square radius. The correction depends on the width and higher moments of the size distribution, which control the average nonlinear response.

As shown in \cite{chekroun_koren_hovering2026}, the coefficient $D$ describes condensation and evaporation at droplet surfaces, while $D_\chi=\kappa_\chi+K_\chi$ combines molecular transport with an effective turbulent diffusivity. {We reserve $\lambda_\chi$ for environmental mixing, distinguishing it from the cloud--rain amplitude growth rate introduced below.} The condensation sink has strength $\alpha_N$, incorporating the prescribed droplet number concentration $N$ and the closure of unresolved size--supersaturation correlations. The effective loss $-\mu\phi$ describes residence-time and parcel-exchange effects.

These parameter interpretations also fix their units. With length and time dimensions $\mathsf L$ and $\mathsf T$, $[\phi]=\mathsf L^2$, $[D]=[D_\chi]=\mathsf L^2\mathsf T^{-1}$, $[\mu]=[\lambda_\chi]=[\gamma]=\mathsf T^{-1}$, and $[\alpha_N]=\mathsf L^{-1}\mathsf T^{-1}$. Also $[a_K]=\mathsf L$ and $[b_K]=\mathsf L^3$. If $\eta$ denotes temporally white noise with a dimensionless spatial covariance normalized at zero separation, $[\eta]=\mathsf T^{-1/2}$ and $[\sigma]=\mathsf T^{-1/2}$; a different spatial noise normalization changes the units of $\sigma$ accordingly.

Equation~\eqref{Eq_hover_parcel} is a \emph{degenerate reaction--diffusion system}: supersaturation diffuses, whereas $\phi$ has no direct diffusion term. Spatial interaction is nevertheless present throughout the population field. Changes in one parcel affect the supersaturation environment shared with neighboring parcels, and this environment feeds back on their droplet growth.

\subsection{Emergence of dynamic inertia}
\label{Sec_screened_origin}

To express this interaction through the droplet field alone, the cloud-formation theory of \cite{chekroun_koren_hovering2026} exploits the system's slow--fast structure: supersaturation relaxes exponentially toward an attracting slow manifold, along which droplet growth governs the subsequent evolution. Eliminating the fast variable must therefore retain the response of the environment that the droplets themselves modify. A growing population draws on the surrounding supersaturation and induces a halo of thermodynamic adjustment. The halo communicates that change to neighboring populations, and its finite relaxation time delays the environmental response to further growth.

 This is the physical content of the concept of \emph{dragged supersaturation envelope} discussed in \cite{chekroun_koren_hovering2026}. Its connection to polaron theory follows from the same type of feedback: a particle polarizes its surrounding medium, whose adjustment modifies the particle's effective dynamics~\cite{gurari1953,feynman1955slow}. The cloud-formation theory of \cite{chekroun_koren_hovering2026} extends this mechanism to the fields of Eq.~\eqref{Eq_hover_parcel} through a Gurari--Feynman variational reduction within the Martin--Siggia--Rose response-field formalism~\cite{MartinSiggiaRose,chekroun_koren_hovering2026}. The droplet population is dressed by the supersaturation halo it induces. Eliminating that halo yields screened spatial interactions and, through its delayed adjustment, an emergent dynamic inertia analogous to the polaron's effective mass. Turbulent fluctuations of the bath also leave temporally correlated forcing in the reduced description; see \cite{chekroun_koren_hovering2026}.

 The origin of the spatial and temporal response can be displayed directly at the deterministic level. Grouping the bath terms in Eq.~\eqref{Eq_hover_parcel} gives
\be\label{Eq_bath_grouped}
\partial_t\chi=D_\chi\Delta\chi-\Gamma(\phi)\chi+J(\phi),
\ee
where $\Gamma(\phi)=\lambda_\chi+\alpha_N\sqrt\phi$ and $J(\phi)=\gamma+\alpha_N\sqrt\phi f(\phi)$. {Turbulent exchange fluctuations modify the effective bath response in the cloud-formation theory of \cite{chekroun_koren_hovering2026}. Write $\Gamma^*(\phi)$ for the effective relaxation rate after the Stratonovich-to-It\^o drift correction, and $\Gamma_0^*>0$ for its value at a representative background. The star distinguishes this corrected rate from the bare $\Gamma(\phi)$ in Eq.~\eqref{Eq_bath_grouped}; without exchange-noise dressing they coincide. Freezing the corrected rate while retaining the evolving source gives the deterministic response of the effective bath
\be\label{Eq_bath_dynamic}
\widetilde\chi(\kvec,\nu)
=\frac{\widetilde{J(\phi)}(\kvec,\nu)}
{\Gamma_0^*+D_\chi|\kvec|^2-i\nu},
\ee
where tildes denote space--time Fourier transforms, $\kvec$ is wavenumber, and $\nu$ is angular frequency. The denominator displays both spatial screening and temporal adjustment: a parcel perturbation acts through an environment that redistributes supersaturation and relaxes over a finite time.

To express the corresponding droplet-field equation, define the bath relaxation time, screening coefficient, and screening length by
\be\label{Eq_bath_scales}
\tau_0^*=(\Gamma_0^*)^{-1},\qquad
\epsilon=D_\chi\tau_0^*,\qquad L_{\rm scrn}=\sqrt\epsilon.
\ee
Substituting the bath response into the droplet equation reduces the two-field system, Eq.~\eqref{Eq_hover_parcel}, to the following deterministic equation at the same frozen-coefficient level:
\bea\label{Eq_hover_dynamic}
\bigl(I-\epsilon\Delta+\tau_0^*\partial_t\bigr)
\bigl[\partial_t\phi+\mu\phi+2Df(\phi)\bigr]\\
=2D\tau_0^*J(\phi).
\eea
For resolved frequencies $|\nu|\tau_0^*\ll1$, the leading low-frequency approximation retains the spatial response and gives
\be\label{Eq_hover_reduced}
(I-\epsilon\Delta)
[\partial_t\phi+\mu\phi+2Df(\phi)]
\simeq2D\tau_0^*J(\phi).
\ee
The slow derivative $\partial_t\phi$ remains inside this operator even though the explicit higher-frequency bath correction has been dropped. Consequently the reduced equation contains
\be\label{Eq_dynamic_inertia}
\partial_t\phi-\epsilon\Delta\partial_t\phi.
\ee
This is the emergence of \emph{dynamic inertia} in the cloud description. Spatial coupling, initially carried entirely by supersaturation transport, now acts on the rate of droplet-field evolution. The population can no longer adjust as a collection of isolated local variables: its evolution must accommodate the thermodynamic response extending into neighboring parcels.

This result supplies the physical significance of the pseudoparabolic operator. The coefficient $\epsilon$, with dimensions of length squared, measures the spatial extent over which environmental adjustment enters the slow dynamics. {For a varying population field, the cloud-formation theory of \cite{chekroun_koren_hovering2026}  uses $\epsilon(\phi)=D_\chi/\Gamma^*(\phi)$ and discusses a conservative extension $I-\nabla\cdot[\epsilon(\phi)\nabla]$. This extension is a closure choice: it adds a gradient term relative to the exact instantaneous operator $I-\epsilon(\phi)\Delta$. We retain a constant effective $\epsilon$ here, for which the two forms coincide.} The resulting operator has the Barenblatt--Zheltov--Kochina pseudoparabolic structure familiar from fissured-rock flow~\cite{barenblatt1960basic}. {The mixed derivative also appears in two-temperature heat conduction~\cite{Chen_Gurtin1968theory} and in pseudoparabolic transport theory~\cite{showalter1970pseudoparabolic,carroll1977singular,Showalter1994}. Its interpretation recalls a viscoelastic resistance to reorganization:} the effective rate of change reflects the adjustment of an accompanying medium~\cite{Coleman_Noll1974,huilgol1968second,ting1963certain}. In a cloud, this resistance has a thermodynamic origin. The evolving droplet population draws its environment out of balance, and the spatially extended supersaturation response enters the law governing its subsequent evolution.

For cloud modeling, the gain is substantial: the spatial response and the thermodynamic feedback are retained together when the rapidly evolving environment is eliminated. Localized changes in droplet growth acquire a collective spatial influence, and their effective rate of adjustment depends on spatial scale. The mixed derivative provides a compact way to represent that influence while retaining a first-order evolution for the resolved field. Equation~\eqref{Eq_hover_dynamic} identifies the additional temporal corrections required when the separation from the bath relaxation time is no longer adequate. The PGLE developed below retains the leading screened response and expresses this dynamic inertia in a minimal model of cloud-field variability.

\subsection{The Green function of the screened response}

The reduction of the two-field system (Eq.~\eqref{Eq_hover_parcel}) to  its one-field closure, Eq.~\eqref{Eq_hover_dynamic}, also has a nonlocal representation. At low frequency the bath response is
\bea\label{Eq_screened_response}
\chi&\simeq(\Gamma_0^*-D_\chi\Delta)^{-1}J(\phi)\\
&=\tau_0^*(I-\epsilon\Delta)^{-1}J(\phi).
\eea
The inverse operator distributes the influence of a local source over the surrounding supersaturation envelope. On the horizontal plane its normalized Green function is
\bea\label{Eq_Green}
(I-\epsilon\Delta)G_\epsilon&=\delta,\qquad
\int_{\mathbb R^2}G_\epsilon(\x)\d\x=1,\\
G_\epsilon(\x)&=\frac{1}{2\pi\epsilon}
K_0\!\left(\frac{|\x|}{\sqrt\epsilon}\right),
\eea
where $K_0$ is the modified Bessel function of the second kind of order zero. The kernel has units $\mathsf L^{-2}$. It possesses an integrable logarithmic singularity at the source, and an exponentially decaying tail on the scale $L_{\rm scrn}$. On a periodic domain we use its periodization. In either representation its Fourier multiplier is
\be\label{Eq_Green_symbol}
\widehat G_\epsilon(\kvec)=\frac{1}{1+\epsilon|\kvec|^2}.
\ee
Thus the Green function and the mixed derivative describe the same spatial response, in integral and differential form. This derivation motivates the form of the kernel used below. Using this response for cloud-water and rain-water variability is a further closure, with an effective screening length at the chosen resolution.  Thus, the derivation described above from the transport equation Eq.~\eqref{Eq_Transport_Siewert_al} supplies the operator structure to describe the evolution of cloudlets. The cloud--rain amplitude law and event statistics complete the model at the resolved water-path scale.

An elliptic operator acting on a rate of change also arises in convective momentum dynamics, through a different mechanism \cite{kuo2022conditions,kuo2025anelastic}. Eliminating perturbation pressure from the anelastic equations gives $\Delta_3\partial_t m=\Delta_h\mathcal F+\mathcal M$, where $m$ is the vertical mass flux, $\Delta_3$ and $\Delta_h$ are the three-dimensional and horizontal Laplacians, $\mathcal F$ combines buoyancy with vertical momentum-flux convergence, and $\mathcal M$ collects horizontal momentum contributions~\cite{kuo2022conditions,kuo2025anelastic}. For one horizontal wavenumber $k_h\neq0$ this becomes 
\bes
(I-k_h^{-2}\partial_z^2)\partial_t m_{k_h}=\mathcal F_{k_h}-k_h^{-2}\mathcal M_{k_h},
\ees
 a Helmholtz operator in the vertical coordinate. Its length $1/k_h$ is set by the horizontal scale of the disturbance, and the pressure adjustment is imposed instantaneously by mass continuity. Here the operator acts along the horizontal, its length follows from supersaturation transport and relaxation, and it originates in droplet growth. The horizontal scale dependence also differs: for fixed forcing amplitude and vertical profile, narrower disturbances accelerate more strongly in the pressure response, whereas the multiplier in Eq.~\eqref{Eq_Green_symbol} slows finer structure. Both results nevertheless show that local forcing need not equal local evolution when the environment of a cloud element adjusts spatially.

\subsection{Cloud--rain observables and their oscillatory feedback}
\label{Sec_PGLE_variables}

Whereas the population fields $\phi$ and $\chi$ in Eq.~\eqref{Eq_hover_parcel} describe diffusional droplet growth, activation, and the emergence of cloud patches from haze, organized convection in an activated cloud layer involves recurring cycles of cloud thickening, rain production, and post-precipitation recovery. Observations and bin-microphysics LES of shallow convective fields document these phase-lagged recharge--discharge loops between cloud mass and rain mass~\cite{feingold_2010,dagan2018organization}, and conceptual cloud--rain delay models show that delayed depletion and replenishment organize the onset of oscillations through a Hopf bifurcation~\cite{koren2011aerosol,feingold_2013,chekroun2020efficient,Chekroun_al22SciAdv}.
To represent variability in this activated, potentially precipitating regime, let $C(\x,t)$ and $P(\x,t)$ denote the vertically integrated cloud-water and rain-water paths (with units of mass per unit area, $\mathrm{kg\,m^{-2}}$), and define dimensionless signed departures about positive reference backgrounds $(C_0,P_0)$ by
\be\label{Eq_water_anomalies}
\rho=\frac{C-C_0}{C_*},\qquad
q=\frac{P-P_0}{P_*},\qquad u=\rho+iq,
\ee
where $C_*,P_*>0$ carry the units of the respective water paths. In this representation, $u=0$ corresponds to the activated background state $(C_0,P_0)$ rather than clear sky; the physical interpretation of $(\rho,q)$ in relation to droplet-spectrum moments, normal-form coordinates, and column water budgets is developed in Secs.~\ref{Sec_PGLE_interpretation} and \ref{Sec_liquid_water_budget}.

Near a supercritical Hopf onset, a complex-conjugate pair of modes captures the dominant cloud--rain cycle, and normal-form reduction of delayed cloud--rain models motivates the cubic Stuart--Landau tendency~\cite{guckenheimer2013nonlinear,chekroun2020efficient,Chekroun_al22SciAdv}
\be\label{Eq_effective_F}
F(u)=(\lambda+i\omega)u-\kappa|u|^2u,
\qquad\kappa=\alpha+i\beta,\quad\alpha>0,
\ee
where the linear growth rate $\lambda$, linear angular frequency $\omega$, nonlinear saturation coefficient $\alpha$, and amplitude-dependent frequency-shift (twist) coefficient $\beta$ all have units of inverse time for dimensionless $u$. Together, the screened environmental kernel $G_\epsilon$ of Eq.~\eqref{Eq_Green} and the oscillatory tendency $F(u)$ of Eq.~\eqref{Eq_effective_F} supply the spatial response and local feedback that we now embed into a stochastic parcel-event description.

\section{Parcel Events, Mean-Field Closure, and Cloud-Field Interpretation}
\label{Sec_jump_construction}

At the mesoscale resolution of the model, a cloud element continually experiences subgrid bursts of condensation, entrainment-driven evaporation, collision--coalescence, and precipitation fallout. Each parcel event represents the cumulative outcome of such an episode over a time interval long enough to encompass the microphysical burst and its rapid thermodynamic adjustment, yet short compared with the resolved cloud--rain oscillation cycle. Within this separation of timescales, condensing the episode into a discrete jump $(\Delta C,\Delta P)$ retains when and where water-path changes occur, how strongly they fluctuate, and how cloud water and rain water co-vary, preserving the intermittent precipitation and turbulent forcing omitted by a smooth deterministic tendency~\cite{palmer2001nonlinear,berner2017stochastic,Chekroun_al22SciAdv}. Moreover, retaining both components in $(\Delta C,\Delta P)$ captures not only mass-conserving cloud-to-rain conversion ($\Delta C+\Delta P=0$ via autoconversion and accretion), but also open-system exchanges with water vapor ($\Delta C\gtrless 0$) and rain removal by evaporation or surface fallout ($\Delta P<0$)~\cite{kessler1969distribution,khain_pinsky_2018}.

Because each episode also perturbs the supersaturation environment shared with neighboring cloud elements, we distribute its influence laterally using the common screened kernel $G_\epsilon$ of Sec.~\ref{Sec_PGLE_physics} across the screening length $L_{\rm scrn}=\sqrt{\epsilon}$. Adopting this common kernel closes the thermodynamic response of the shared environment rather than implying identical physical transport of cloud water and rain, with $L_{\rm scrn}$ setting the lateral reach of that environmental adjustment around a localized parcel episode. Together, the event occurrence statistics, joint water-path increments, and screened spatial response provide a direct interface for incorporating unresolved cloud physics, which in a calibrated model could be inferred from large-eddy simulation (LES) diagnostics of entrainment, detrainment, precipitation, and water-budget evolution~\cite{dawe2013entrainment,yamaguchi2015precipitation,hoffmann2023roles}. Leaving that physical calibration to future work, we specify below a tractable Poisson event law that allows exact ensemble-mean calculations while isolating the fluctuations omitted by the deterministic PGLE.

\subsection{A spatial jump process for unresolved cloud physics}
\label{Sec_spatial_jump_process}
To connect these parcel episodes to an evolving cloud field, let $U_t(\x)$ denote a single stochastic realization of the dimensionless cloud--rain anomaly field in Eq.~\eqref{Eq_water_anomalies},
\be\label{Eq_random_water_mapping}
U_t(\x)=\frac{C_t(\x)-C_0}{C_*}
+i\frac{P_t(\x)-P_0}{P_*},
\ee
or its corresponding complex normal-form mode when a coordinate transformation from a parent model is retained. Under identical large-scale environmental conditions, different space--time sequences of mixing, condensation, and precipitation episodes generate different realization histories $U_t(\x)$, whose ensemble mean $u(\x,t)=\mathbb E[U_t(\x)]$ forms the basis for the deterministic closure in Sec.~\ref{Sec_mean_field_linear}.

Much of the unresolved cloud physics enters through the probability law governing event occurrence and outcomes, represented below by the marked counting measure $N$. This law specifies how frequently episodes occur, where they occur, and how their water-path changes are distributed; its physical interpretation is completed by the state-dependent increment and the spatial response assigned to each event. A physically calibrated law could be estimated, for example, from large-eddy simulations (LES). Cloud tracking in LES has been used to determine distributions of entrainment and detrainment rates~\cite{dawe2013entrainment}, while LES studies of stratocumulus quantify the roles of precipitation and entrainment in water-budget evolution and the influence of the spatial distribution of rain~\cite{hoffmann2023roles,yamaguchi2015precipitation}. At a chosen parcel area and averaging interval, such simulations could be used to identify episodes, estimate their occurrence rates and joint $(\Delta C,\Delta P)$ distributions after accounting for resolved transport, and determine how these statistics and the episode footprints depend on the local cloud and environmental state. This would also test whether independent event counts are an adequate approximation.

We leave this physical calibration to a subsequent study, as discussed in Sec.~\ref{Sec_concluding_perspective}, and specify here a simple, complete event law that permits explicit mean calculations. Our principal focus is its mean-field approximation, the deterministic PGLE, through which we analyze how cloud--rain variability is sustained and reorganized at this first level of description.

To specify this stochastic model in the simplest form that permits exact mean calculations, we idealize the cloud layer $\Torus$ (of area $|\Torus|$) as hosting discrete parcel episodes---such as bursts of entrainment, evaporation, collision--coalescence, or rain fallout---each characterized by a reference horizontal parcel area $a_p$ (units $\mathsf L^2$) and a mean occurrence frequency $r_p$ per parcel element (units $\mathsf T^{-1}$). At this leading level of approximation, we assume that the episode rate per unit area $r_p/a_p$ is spatially uniform and that episode counts in disjoint space--time regions are independent. These two assumptions define a space--time Poisson process for the episode arrival times and locations: any horizontal subregion $E\subset\Torus$ experiences on average $(r_p|E|/a_p)$ episodes per unit time, and the probability that a single episode occurs inside $E$ during an infinitesimal interval $\d t$ is $(r_p|E|/a_p)\,\d t+o(\d t)$. Dependence on the evolving cloud--rain state enters not through the arrival rate, but through the water-path change produced by each episode~\cite{Applebaum2009,peszat2007stochastic}.

Even when two episodes begin from the exact same resolved cloud--rain state $U_{t-}(\y)$ immediately prior to the event, they generally produce different net water-path changes because of unresolved differences in entrained environmental humidity, episode duration, or droplet size distribution. To represent this subgrid scatter, we attach to each episode a dimensionless complex random variable $\zeta=\zeta_r+i\zeta_i$---termed a \emph{mark} in jump-process theory~\cite{Applebaum2009}, and serving here as a pair of random auxiliary variables that account for unresolved microphysical and thermodynamic fluctuations, such as variable vapor exchange, entrained humidity, or conversion efficiency, that add or remove cloud water ($\Delta C$) and rain water ($\Delta P$) during the episode.
These auxiliary  variables are drawn from a joint probability distribution $\Pi$ such that $\Pi(B_\zeta)$ is the probability that $\zeta$ falls in a subset $B_\zeta$ of the complex plane $\mathbb C$. 

Allowing the real and imaginary components of $\zeta$ to be correlated captures microphysical couplings such as enhanced rain formation accompanying a loss of cloud water ($\zeta_r<0$, $\zeta_i>0$). We take the marks to be centered ($\mathbb E\zeta=0$), mutually independent across episodes, and independent of the pre-event field, with finite moments through third order and an overall dimensionless amplitude factor $b$. Centering cleanly separates unresolved scatter from the systematic tendency, while physical constraints such as liquid-water conservation during conversion-only episodes apply to the total increment (Sec.~\ref{Sec_liquid_water_budget}).

Rather than indexing each individual episode by its arrival time, center location, and random outcome separately, it is compact to record the entire sequence of events through the \emph{marked counting measure} $N(\d t,\d\y,\d\zeta)$: integrating $N$ over a time interval, a horizontal patch, and a set of outcomes simply counts how many episodes occurred in that region with those water-path departures. Under the Poisson and independence assumptions above, the expected number of episodes in an infinitesimal space--time--outcome element is given by the intensity
\be\label{Eq_event_intensity}
\frac{r_p}{a_p}\,\d t\,\d\y\,\Pi(\d\zeta),
\ee
which multiplies the episode rate per unit area by the duration $\d t$, patch area $\d\y$, and outcome probability $\Pi(\d\zeta)$~\cite{Applebaum2009,Applebaum_2019}.

Each episode also delivers a systematic, state-dependent change $F_B(U_{t-}(\y))/r_p$ reflecting the local cloud--rain feedback near Hopf onset. Here $F_B$ is a smooth bounded extension of the cubic Stuart--Landau tendency $F$ of Eq.~\eqref{Eq_effective_F} that coincides with $F$ inside the normal-form range $|U|\le B$ while preventing unphysical blow-up during rare large excursions (the PGLE closure holds when excursions beyond $|U|=B$ contribute negligibly to the mean). Dividing $F_B$ by the frequency $r_p$ ensures that averaging events arriving at rate $r_p$ recovers the systematic tendency $F_B$ regardless of how intermittently episodes occur. Combining this systematic increment with the random departure $b\zeta$ and distributing the resulting perturbation laterally across the surrounding supersaturation envelope via the periodized Green function $G_\epsilon$ of Eq.~\eqref{Eq_Green} shows that a single episode centered at $\y$ shifts the entire field by
\be\label{Eq_Jump_function}
\delta U_t (\x)=a_pG_\epsilon(\x-\y)
\left[\frac{F_B(U_{t-}(\y))}{r_p}+b\zeta\right].
\ee
Because $\int_{\Torus}G_\epsilon\,\d\x=1$, the dimensionless spatial weight $a_p G_\epsilon(\x-\y)$ spreads a point-localized parcel source at $\y$ over the screening scale $L_{\rm scrn}=\sqrt{\epsilon}$ while preserving the spatially integrated water-path change $a_p[F_B(U_{t-}(\y))/r_p+b\zeta]$.

Accumulating the impulses $\delta U_t(\x)$ from all episodes arriving during $\d t$ across $\Torus$ yields the stochastic jump evolution equation
\bea\label{Eq_SIDE}
\d U_t(\x)&=\int_{\Torus\times\mathbb C}a_pG_\epsilon(\x-\y)\\
&\quad\times\left[\frac{F_B(U_{t-}(\y))}{r_p}+b\zeta\right]
N(\d t,\d\y,\d\zeta),
\eea
where the integral against $N(\d t,\d\y,\d\zeta)$ is simply shorthand for summing the discrete event jumps Eq.~\eqref{Eq_Jump_function} as they occur in time and space~\cite{Applebaum2009,peszat2007stochastic}. A key physical feature of Eq.~\eqref{Eq_SIDE} is that scaling the systematic increment by $1/r_p$ holds the conditional mean tendency at a fixed state invariant as the event frequency varies, whereas the parcel parameters $(r_p, a_p, b, \Pi)$ govern how intermittently that tendency is realized: sparser, larger-area, or more variable episodes inject stronger fluctuations that can feed back onto the nonlinear mean. Introducing the jump process before taking its ensemble average therefore makes explicit both the physical origin of the mean-field PGLE and the fluctuation terms omitted by the deterministic closure, which we derive next in Sec.~\ref{Sec_mean_field_linear} and revisit in Sec.~\ref{Sec_concluding_perspective}.

\subsection{Exact mean equation and the nonlinear closure}
\label{Sec_mean_field_linear}

To extract the systematic evolution shared by the stochastic event histories for a prescribed initial state and large-scale environment, we define the ensemble-mean field $u(\x,t)=\mathbb E[U_t(\x)]$ by averaging over event times, locations, and marks at fixed $(\x,t)$; an ensemble average across realizations, distinct from the local within-population droplet averages defining $\phi$ and $\chi$ in Eq.~\eqref{Eq_population_fields}. Taking the expectation of Eq.~\eqref{Eq_SIDE} via the compensator identity for Poisson integrals~\cite{Applebaum2009,peszat2007stochastic} replaces the counting measure $N(\d t,\d\y,\d\zeta)$ by its intensity in Eq.~\eqref{Eq_event_intensity}; the factors $r_p/a_p$ and $a_p/r_p$ in Eqs.~\eqref{Eq_event_intensity} and \eqref{Eq_Jump_function} then cancel, the centered marks ($\mathbb E\zeta=0$) contribute zero direct drift, and the spatial integral reduces to the exact first-moment convolution
\be\label{Eq_mean_exact_cutoff}
\partial_tu=G_\epsilon*\mathbb E[F_B(U)],
\ee
which is well defined on the finite domain $\Torus$ for bounded $F_B$ and finite-moment initial data. Separating the cubic Stuart--Landau law $F(U)=(\lambda+i\omega)U-\kappa|U|^2U$ of Eq.~\eqref{Eq_effective_F} from its high-amplitude cutoff via the remainder
\be\label{Eq_extension_remainder}
\mathcal R_B=\mathbb E[F_B(U)-F(U)],
\ee
allows for the expected tendency in Eq.~\eqref{Eq_mean_exact_cutoff} to be expanded as
\bea\label{Eq_mean_rate_expanded}
\mathbb E[F_B(U)]
&=(\lambda+i\omega)\mathbb E[U]
-\kappa\mathbb E[|U|^2U]+\mathcal R_B\\
&=(\lambda+i\omega)u
-\kappa\mathbb E[|U|^2U]+\mathcal R_B.
\eea
By linearity of expectation, the growth and oscillation terms pass unchanged from the parcel increment into the mean equation, confining the closure problem entirely to the cubic moment. Applying the inverse Helmholtz operator $I-\epsilon\Delta$ to Eq.~\eqref{Eq_mean_exact_cutoff} and using $(I-\epsilon\Delta)G_\epsilon=\delta$ from Eq.~\eqref{Eq_Green} removes the spatial convolution and yields the exact mean balance
\be\label{Eq_PGLE_mean_field}
(I-\epsilon\Delta)\partial_tu
=(\lambda+i\omega)u-\kappa\mathbb E[|U|^2U]+\mathcal R_B.
\ee
The pseudoparabolic operator $(I-\epsilon\Delta)\partial_t$ thus survives ensemble averaging without approximation: whenever the event tendency is globally linear, $F_B(U)=F(U)=(\lambda+i\omega)U$, both the cubic term and $\mathcal R_B$ vanish, and the first moment closes exactly into the linear PGLE,
\be\label{Eq_linear_PGLE}
(I-\epsilon\Delta)\partial_tu=(\lambda+i\omega)u.
\ee

For the nonlinear oscillatory law, the average of the cubic response differs from the cubic response of the average field by the fluctuation moments of $v=U-u$ ($\mathbb E[v]=0$). Expanding $U=u+v$ in the cubic term gives
\be\label{Eq_third_moment_expansion}
\mathbb E[|U|^2U]
=|u|^2u+2u\mathbb E[|v|^2]
+\overline u\mathbb E[v^2]+\mathbb E[|v|^2v],
\ee
which shows that even though centered marks drop out of the linear drift, parcel-event fluctuations feed back onto the mean evolution through the local variance $\mathbb E[|v|^2]$, phase-sensitive covariance $\mathbb E[v^2]$, and third-order skewness $\mathbb E[|v|^2v]$. In the regime where these fluctuation moments and the cutoff remainder $\mathcal R_B$ are small compared with the coherent amplitude $|u|^2$, we close the hierarchy at leading order via
\be\label{Eq_simple_approx}
\mathbb E[|U|^2U]\simeq |u|^2u,
\ee
which reduces Eq.~\eqref{Eq_PGLE_mean_field} to the deterministic pseudoparabolic Ginzburg--Landau equation (PGLE),
\be\label{Eq_PGLE}
\partial_tu=(\lambda+i\omega)u-\kappa|u|^2u
+\epsilon\Delta\partial_tu,
\ee
or, in its equivalent nonlocal convolution form on $\Torus$,
\be\label{Eq_PGLE_nonlocal}
\partial_tu=G_\epsilon*F(u).
\ee

This derivation endows the PGLE with a clear microphysical and statistical interpretation: it governs the leading ensemble-mean cloud--rain anomaly field when intermittent episodes of condensation, mixing, conversion, and fallout are represented by screened parcel jumps and the fluctuation correlations in Eq.~\eqref{Eq_third_moment_expansion} remain subordinate to the coherent oscillation. Crucially, the stochastic formulation does not merely motivate Eq.~\eqref{Eq_PGLE} and disappear; it identifies the exact moment corrections that the deterministic equation omits. Quantitative conditions on the parcel event rate $r_p$, element area $a_p$, and mark variance $b^2\mathbb E|\zeta|^2$ that keep the fluctuation variance $\mathbb E[|v|^2]$ small---together with the regularization of short-wavelength variance by finite parcel footprints---are derived in the companion study~\cite{chekroun_koren_liu_variance2026}. The consequences of Eq.~\eqref{Eq_third_moment_expansion} for the open-system liquid-water budget and for future LES-calibrated stochastic closures are examined in Secs.~\ref{Sec_liquid_water_budget} and \ref{Sec_concluding_perspective}.

Positioning this continuum jump closure alongside stochastic lattice cloud models~\cite{khouider2010multicloud,Frenkel2012,Frenkel2013,khouider2019new} highlights why the PGLE is directly suited to resolving cloud-field morphology. In the lattice framework of Ref.~\cite{khouider2019new}, discrete grid sites flip between clear and cloudy states at transition rates controlled by nearest-neighbor occupancy and bulk moisture, and passing to the deterministic mean-field limit collapses the spatial lattice into an ordinary differential equation for the domain-averaged cloud fraction. While that scalar coverage equation captures bistability and hysteretic transitions between high- and low-coverage regimes, it averages out spatial structure and therefore cannot describe the geometry of the cloud patterns themselves without reverting to stochastic lattice realizations. In the present framework, by contrast, jumps act on a continuous complex cloud--rain amplitude and are coupled nonlocally across space by the thermodynamic screening kernel $G_\epsilon$. Passing to the mean-field limit therefore preserves spatial coupling and yields a partial differential equation whose interplay of dynamic inertia (Sec.~\ref{Sec_dispersion_nonlinear}), global Noether-charge constraints (Sec.~\ref{Sec_Noether_balance}), and sideband instabilities (Secs.~\ref{Sec_Instab_Neimark} and \ref{Sec_Nonlin_Wave_Interact}) intrinsically generates and reorganizes two-dimensional cloud morphologies---connecting coherent \emph{undulatus} bands to defect-rich \emph{asperitas}-like wave networks (Secs.~\ref{Sec_banded_solutions}--\ref{Sec_defect_analysis}), consolidating cellular interference patterns into parallel convective rolls (Sec.~\ref{Sec_spontaneous_stabilization}), and building open-cell-like and actinoform radial structures out of unstructured random anomalies (Sec.~\ref{Sec_random_organization}).

\subsection{What the PGLE describes physically}
\label{Sec_PGLE_interpretation}

Under the parcel-event assumptions and the cubic moment closure of Sec.~\ref{Sec_mean_field_linear}, the PGLE governs the ensemble-mean cloud--rain anomaly field $u=\rho+iq$. In the direct water-anomaly representation of Eq.~\eqref{Eq_water_anomalies}, its real and imaginary parts satisfy the coupled system
\bea\label{PGLE_system}
(I-\epsilon\Delta)\partial_t\rho
&=(\lambda-\alpha|u|^2)\rho-(\omega-\beta|u|^2)q,\\
(I-\epsilon\Delta)\partial_tq
&=(\omega-\beta|u|^2)\rho+(\lambda-\alpha|u|^2)q,
\eea
where every term has units of inverse time ($[\epsilon]=\mathsf L^2$ and $u$ is dimensionless). For a spatially uniform state $u=a e^{i\vartheta}$, Eq.~\eqref{PGLE_system} reduces to $\dot a=(\lambda-\alpha a^2)a$ and $\dot\vartheta=\omega-\beta a^2$, whereas for a heterogeneous field the Fourier relation $(1+\epsilon|\kvec|^2)\partial_t\widehat u_{\kvec}=\widehat{F(u)}_{\kvec}$ expresses dynamic inertia by slowing the adjustment of finer spatial scales. The reduced coefficients $(\lambda,\omega,\alpha,\beta)$ summarize the net feedback of condensation, collision--coalescence, evaporation, and fallout across an oscillation cycle rather than isolated microphysical rate constants. In particular, the cubic damping $-\alpha|u|^2u$ acts as an amplitude-dependent restoring rate $\alpha|u|^2$ that pulls both $C-C_0$ and $P-P_0$ back toward the background state and therefore changes sign with the anomaly, while the off-diagonal factor $\omega-\beta|u|^2$ couples the timing of the two water categories. If $u$ instead denotes a normal-form coordinate derived from a parent cloud--rain model, the physical anomalies are recovered locally through the associated coordinate map, $(\rho,q)^{\mathsf T}=\mathsf P(\operatorname{Re}u,\operatorname{Im}u)^{\mathsf T}+O(|u|^2)$, with a nonsingular real matrix $\mathsf P$ that reduces to the identity in the direct representation of Eq.~\eqref{Eq_water_anomalies}.

This water-path closure complements the droplet-population theory of Sec.~\ref{Sec_PGLE_physics} by shifting from condensational activation to the precipitating oscillatory regime. In Eq.~\eqref{Eq_hover_parcel}, the prognostic variable $\phi=\langle r^2\rangle_{\x,t}$ follows directly from the diffusional growth law in Eq.~\eqref{Eq_droplet_growth} and determines the projected droplet surface-area density,
\be\label{Eq_droplet_area}
A_{\rm drop}(\x,t)=\pi N\phi(\x,t),
\ee
which controls optical extinction and optical depth for a droplet number concentration $N$~\cite{khain_pinsky_2018}. Converting $\phi$ into liquid-water mass concentration, however, requires the third radius moment,
\bea\label{Eq_LWC_mapping}
\mathrm{LWC}(\x,t)&=\frac{4\pi}{3}\rho_w N\langle r^3\rangle_{\x,t}
=\frac{4\pi}{3}\rho_w N r_e(\x,t)\phi(\x,t),\\
r_e(\x,t)&=\frac{\langle r^3\rangle_{\x,t}}{\langle r^2\rangle_{\x,t}},
\eea
where $\rho_w$ is liquid-water density and $r_e$ is the effective radius ($r_e=\sqrt{\phi}$ only for a monodisperse spectrum). Because predicting water mass and precipitation from $\phi$ alone would require closing the evolving drop-size distribution $r_e(\x,t)$ and vertical column structure across the collision--coalescence threshold, the PGLE adopts the screened environmental operator $(I-\epsilon\Delta)\partial_t$ as a closure at the water-path scale and couples it directly to the Stuart--Landau cloud--rain oscillation law.

Accordingly, the model variables $\rho$ and $q$ must not be confused with the total nonnegative cloud-water and rain-water paths (or liquid-water content) themselves, while remaining directly meaningful for interpreting cloud-field organization. Rather than absolute water inventories, $\rho$ and $q$ represent dimensionless, signed oscillatory departures about an activated cloudy background $(C_0,P_0)$---where negative values denote local cloud thinning or rain depletion relative to $(C_0,P_0)$ within the nonnegative physical validity range $C=C_0+C_*\rho\ge 0$ and $P=P_0+P_*q\ge 0$ (whose column mass budget is analyzed in Sec.~\ref{Sec_liquid_water_budget}). Because optical depth and cloud reflectance vary monotonically with cloud-water path at fixed droplet concentration, spatial maps of $\rho=\operatorname{Re}u$ capture the visible morphology of thicker cloud crests ($\rho>0$) alternating with thinner or clearing troughs ($\rho<0$) across bands, rolls, and cellular walls, while $q=\operatorname{Im}u$ tracks the phase-lagged precipitation anomalies that sustain the cycle. At the same time, the modulus $|u|=\sqrt{\rho^2+q^2}$ maps the spatial envelope of vigorous cloud--rain cycling---distinguishing actively oscillating cells or wave trains from quiescent background regions and defect cores where $|u|\to 0$---and the phase $\arg u$ measures how synchronized or disrupted the condensation--precipitation--recovery cycle is across neighboring cloud elements.

\section{Dispersion and Nonlinear Effects}
\label{Sec_dispersion_nonlinear}

\subsection{Dispersion relation, phase velocity, and group velocity}
\label{Sec_linear_dispersion}

Before analyzing how nonlinear interactions reorganize a cloud field, we characterize how the screened operator $(I-\epsilon\Delta)\partial_t$ shapes the propagation and growth of elementary wave disturbances. In the linearized PGLE, Eq.~\eqref{Eq_linear_PGLE},
\be\label{Eq_linear_PGLE_disp}
(I-\epsilon\Delta)\partial_t u=(\lambda+i\omega)u,
\ee
a plane-wave perturbation $u(\x,t)=u_0 e^{i(\kvec\cdot\x-\Omega t)}$ with wavevector $\kvec$, wavenumber $k=|\kvec|$, and complex angular frequency $\Omega$ satisfies $-i\Omega(1+\epsilon k^2)=\lambda+i\omega$, which yields the complex dispersion relation
\be\label{Eq_complex_dispersion}
\Omega(\kvec)=-\frac{\omega}{1+\epsilon k^2}+i\frac{\lambda}{1+\epsilon k^2}.
\ee
The imaginary part, $\operatorname{Im}\Omega(\kvec)=\lambda/(1+\epsilon k^2)$, governs modal growth or decay, while the real part,
\be\label{Eq_base_freq}
\operatorname{Re}\Omega(\kvec)=-\frac{\omega}{1+\epsilon k^2},
\ee
sets the oscillation frequency of the linear wave. Along the wavevector direction, individual crests and troughs propagate at the phase velocity
\be\label{Eq_phase_velocity}
v_p(k)=\frac{\operatorname{Re}\Omega}{k}=-\frac{\omega}{k(1+\epsilon k^2)},
\ee
whereas a modulated wavepacket centered near wavenumber $k$ propagates at the group velocity
\be\label{Eq_group_velocity}
v_g(k)=\frac{\d(\operatorname{Re}\Omega)}{\d k}=\frac{2\omega\epsilon k}{(1+\epsilon k^2)^2}.
\ee

Equations~\eqref{Eq_complex_dispersion}--\eqref{Eq_group_velocity} make the kinematic role of dynamic inertia explicit. In the unscreened limit $\epsilon\to 0$, all spatial scales oscillate at the same local frequency $\omega$, the group velocity vanishes, and wavepackets do not disperse. When environmental screening is active ($\epsilon>0$), the factor $Q_{\kvec}=1+\epsilon k^2$ slows both the temporal growth rate and the oscillation frequency at shorter wavelengths. As a result, the group velocity vanishes both in the long-wave limit $k\to 0$ and in the short-wave limit $k\sqrt{\epsilon}\gg 1$, attaining its maximum at the intermediate scale $k\sqrt{\epsilon}=1/\sqrt{3}$ Set by the screening length $L_{\rm scrn}=\sqrt{\epsilon}$. Atmospheric wave disturbances, such as convective inertia--gravity waves, are well known to organize cumulus cloud fields into coherent bands and streets~\citep{plank1966wind,plank1969size,da2011cloud,dror_2020}. Here, dispersion arises even in the absence of external gravity-wave forcing because the shared supersaturation halo couples the phase timing of neighboring parcels across the scale $L_{\rm scrn}$. How this scale-dependent timing interacts with parcel forcing to shape the spatial amplitude spectrum is examined in Sec.~\ref{Sec_dynamic_inertia_relaxation}, while its coupling to nonlinear amplitude saturation is developed next.

\subsection{Nonlinear frequency shift, twist, and phase lag}
\label{Sec_dyn_insights}

When the cubic cloud--rain feedback in Eq.~\eqref{Eq_PGLE} is restored with $\kappa=\alpha+i\beta$ ($\alpha>0$), the oscillation frequency becomes coupled to the excursion amplitude. To isolate this mechanism, we seek a spatially monochromatic solution of the nonlinear PGLE in the form $u(\x,t)=A(t)e^{i(\kvec\cdot\x+\Omega_0(\kvec)t)}$, where $A(t)=R(t)e^{i\phi(t)}$ and $\Omega_0(\kvec)=\omega/(1+\epsilon|\kvec|^2)$ denotes the linear base frequency magnitude from Eq.~\eqref{Eq_base_freq} (written in the $+i\Omega_0 t$ phase convention used for the nonlinear analysis below). Substituting this ansatz into Eq.~\eqref{Eq_PGLE} and separating real and imaginary parts yields the coupled amplitude--phase system
\bea\label{Eq_R_phi_ODEs}
(1+\epsilon|\kvec|^2)\dot R&=\lambda R-\alpha R^3,\\
(1+\epsilon|\kvec|^2)(\dot\phi+\Omega_0(\kvec))&=\omega-\beta R^2.
\eea
Because $(1+\epsilon|\kvec|^2)\Omega_0(\kvec)=\omega$, the linear frequency cancels from the phase equation, leaving the purely nonlinear phase drift
\be\label{Eq_phi_dot}
\dot\phi(t)=-\frac{\beta R(t)^2}{1+\epsilon|\kvec|^2}.
\ee
For $\lambda>0$, integrating the Bernoulli equation for $R(t)$ and substituting the result into Eq.~\eqref{Eq_phi_dot} yields the exact transient monochromatic wave
\be\label{Eq_Mono_general}
u(\x,t)=R(t)\exp\!\left[i\left(\kvec\cdot\x+\frac{\omega}{1+\epsilon|\kvec|^2}t+\phi(t)\right)\right],
\ee
whose evolving modulus and nonlinear phase shift are given by
\bea\label{Eq_R_phi_exact}
R(t)&=\sqrt{\frac{\lambda/\alpha}{1+\bigl(\lambda/(\alpha R_0^2)-1\bigr)e^{-2\lambda t/(1+\epsilon|\kvec|^2)}}},\\
\phi(t)&=\phi_0-\frac{\beta}{2\alpha}\ln\left|e^{2\lambda t/(1+\epsilon|\kvec|^2)}+\frac{\lambda}{\alpha R_0^2}-1\right|,
\eea
where $A(0)=R_0 e^{i\phi(0)}$ specifies the initial amplitude and phase, and the constant $\phi_0$ is related to the initial data through
\be\label{Eq_phi0}
\phi_0=\phi(0)+\frac{\beta}{2\alpha}\ln\left|\frac{\lambda}{\alpha R_0^2}\right|
=\phi(0)-\frac{\beta}{\alpha}\ln\left(\frac{R_0}{R_*}\right),
\ee
with $R_*=\sqrt{\lambda/\alpha}$ denoting the saturated limit-cycle radius.

Equation~\eqref{Eq_phi0} has a direct geometric interpretation in terms of the \emph{twist} parameter $\eta=\beta/\alpha$ and the isochron structure of the cloud--rain oscillator developed in \ref{App_twist_isochrons}~\cite{guckenheimer1975isochrons,Tantet_al_Hopf,Chekroun_al22SciAdv}. For a uniform oscillation ($\kvec=0$), curves of constant asymptotic phase---the isochrons---are logarithmic spirals defined by the invariant phase coordinate $\Theta=\vartheta-(\beta/\alpha)\ln(R/R_*)$ in Eq.~\eqref{Eq_PGLE_isochron}. Equation~\eqref{Eq_phi0} shows that the same logarithmic spiral governs each screened spatial mode $\kvec$: as the wave amplitude relaxes from $R_0$ toward $R_*$ at the screened rate $2\lambda/(1+\epsilon|\kvec|^2)$, the ratio $\beta/\alpha$ converts the radial excursion $R_0/R_*$ into a permanent asymptotic phase offset $\phi_0-\phi(0)=-(\beta/\alpha)\ln(R_0/R_*)$. In the long-time regime $t\to\infty$, expanding $\phi(t)$ in Eq.~\eqref{Eq_R_phi_exact} gives
\bes
\phi(t)\approx\phi_0-\frac{\beta\lambda}{\alpha(1+\epsilon|\kvec|^2)}t
-\frac{\beta}{2\alpha}\left(\frac{\lambda}{\alpha R_0^2}-1\right)e^{-2\lambda t/(1+\epsilon|\kvec|^2)},
\ees
so that Eq.~\eqref{Eq_Mono_general} converges to the family of saturated plane-wave solutions
\be\label{Eq_u_plane_asymp}
u(\x,t)=R_*\exp\bigl(i(\kvec\cdot\x+\Omega_{\rm eff}(\kvec)t+\phi_0)\bigr),
\ee
with $R_*=\sqrt{\lambda/\operatorname{Re}\kappa}$, $\phi_0\in[0,2\pi)$, and the effective nonlinear frequency
\be\label{Eq_effective_Omega}
\Omega_{\rm eff}(\kvec)=\frac{\omega-\beta\lambda/\alpha}{1+\epsilon|\kvec|^2}.
\ee
The asymptotic frequency in Eq.~\eqref{Eq_effective_Omega} combines the linear oscillatory tendency $\omega$ with the twist-induced shift $-\beta R_*^2=-(\beta/\alpha)\lambda$, both attenuated by the screening factor $1+\epsilon|\kvec|^2$.

\begin{figure}[htbp]
\centering
\includegraphics[width=.48\textwidth,height=0.25\textwidth]{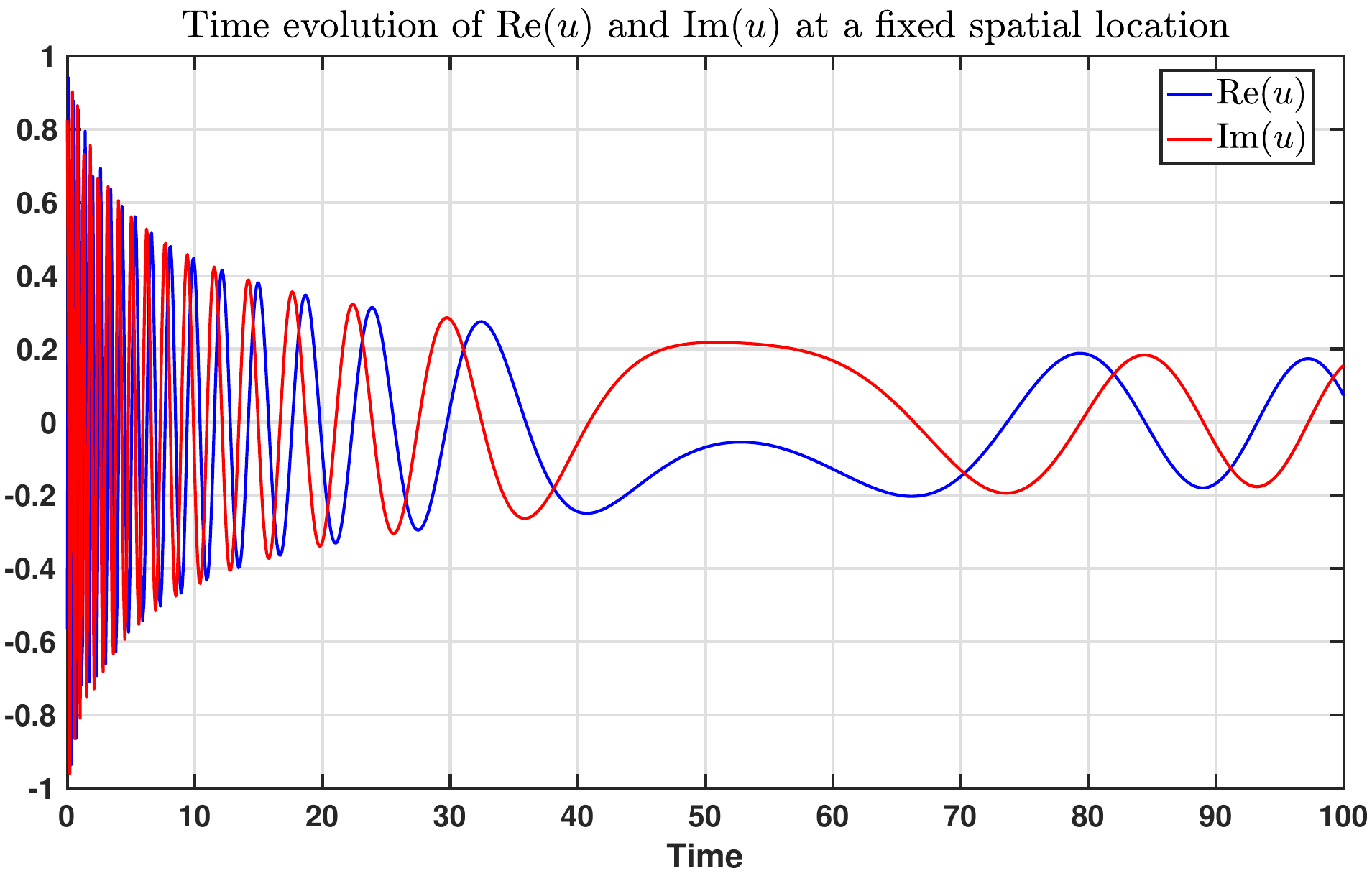} 
\caption{{\bf Nonlinear shift in frequency and phase at a fixed spatial location.} Time series of the normalized cloud-water anomaly $\rho=\operatorname{Re}u$ (blue) and rain-water anomaly $q=\operatorname{Im}u$ (red) evaluated from Eq.~\eqref{Eq_Mono_general} at a fixed position for $\lambda=10^{-2}$, $\omega=5$, $\alpha=1$, $\beta=100$, $\epsilon=10^{-2}$, $R_0=1$, $\phi_0=0$, and $\kvec=(13,-15)$. As the amplitude adjusts toward saturation, the twist term $-\beta R(t)^2$ continuously modifies the instantaneous frequency in Eq.~\eqref{Eq_instantaneous_phase_rate}, warping the oscillation period and the temporal lag between cloud-water and rain-water peaks.}\label{Fig_lag_variance}
\vspace{-2ex}
\end{figure}

During transients, this amplitude--frequency coupling directly modulates the timing between cloud-water and rain-water anomalies. Writing the total phase of Eq.~\eqref{Eq_Mono_general} as $\Psi(\x,t)=\kvec\cdot\x+\Omega_0(\kvec)t+\phi(t)$, the observables $\rho=\operatorname{Re}u=R(t)\cos\Psi$ and $q=\operatorname{Im}u=R(t)\sin\Psi$ remain in phase quadrature ($\Delta\Psi=\pi/2$), but their instantaneous angular frequency evolves in time according to
\be\label{Eq_instantaneous_phase_rate}
\partial_t\Psi(\x,t)=\Omega_0(\kvec)+\dot\phi(t)=\frac{\omega-\beta R(t)^2}{1+\epsilon|\kvec|^2}.
\ee
In a reduced cloud--rain description, Eq.~\eqref{Eq_instantaneous_phase_rate} encodes the state-dependent cycle period inherited from delayed microphysical feedback, where larger cloud-water and rain-water excursions alter the depletion and replenishment timescales~\cite{chekroun2020efficient,Chekroun_al22SciAdv}. Over any interval on which $\partial_t\Psi$ maintains a definite sign, successive quadrature crossings of $\rho$ and $q$ separated by a time lag $\delta t$ satisfy $\int_t^{t+\delta t}\partial_s\Psi(\x,s)\,\d s=\pi/2$, yielding the local lag estimate $\delta t\approx\pi/(2|\partial_t\Psi(\x,t)|)$ when the amplitude varies slowly. If $R(t)$ passes through $(\omega/\beta)^{1/2}$ during its radial adjustment, the instantaneous phase rate $\partial_t\Psi$ vanishes and reverses sign, producing a pronounced stretching of the oscillation period and a transient reversal of the lead--lag order between $\rho$ and $q$, as illustrated around $t\approx 50$ in Fig.~\ref{Fig_lag_variance}.

Beyond single-wave transients, the twist mechanism summarized by Eqs.~\eqref{Eq_phi0}--\eqref{Eq_instantaneous_phase_rate} serves as a central organizing thread throughout the remainder of this study. For a single traveling carrier, the competition between the linear frequency $\omega$, the twist shift $\lambda\beta/\alpha$, and the wavenumber screening factor $1+\epsilon|\kvec|^2$ governs sideband instability and defines the zero-frequency locus $\Omega_{\rm eff}=0$ in Sec.~\ref{Sec_instab} (Eqs.~\eqref{Eq_stab_cond2}--\eqref{Eq_single_frequency_intervals} and Fig.~\ref{Fig_stab_chart}), as well as the finite-wavenumber modulational instability of a spatially uniform oscillation when $\beta\Omega_*<0$ in Sec.~\ref{Sec_PGLE_modulation} (Eq.~\eqref{Eq_uniform_modulation_band} and Appendix~\ref{App_PGLE_modulation}). For interacting wave trains, the twist ratio $|\beta/\alpha|$ controls whether conjugate sideband coupling can overcome linear damping in Sec.~\ref{Sec_detuning_resonant_gain} (Eq.~\eqref{Eq_gain_damping_ratio}). Finally, in two-dimensional fields where the amplitude $R(\x,t)$ varies in space, the local reaction-frequency gradient $\nabla(\omega-\beta R^2)=-2\beta R\nabla R$ (Eq.~\eqref{Eq_reaction_frequency_gradient}) and the screened isochron response (Eq.~\eqref{Eq_screened_isochron_response} in Appendix~\ref{App_phase_geometry}) convert spatial amplitude contrasts into differential phase winding, driving e.g.~the transition from coherent undulatus-like bands to defect-rich asperitas-like wave patterns in Secs.~\ref{Sec_banded_solutions}--\ref{Sec_defect_analysis}.

\subsection{Dynamic inertia vs diffusive relaxation}
\label{Sec_dynamic_inertia_relaxation}

To appreciate why the pseudoparabolic operator is suited to cloud-field organization, it is instructive to contrast the PGLE with the standard complex Ginzburg--Landau equation (CGLE)~\cite{cross1993pattern,aranson2002world},
\be\label{Eq_CGLE_comparison}
\partial_tu=(\lambda+i\omega)u-\kappa|u|^2u
+(D_r+iD_i)\Delta u,\qquad D_r>0,
\ee
which couples the same local Stuart--Landau tendency $F(u)=(\lambda+i\omega)u-\kappa|u|^2u$ across space through a complex Laplacian with amplitude diffusivity $D_r$ and dispersive stiffness $D_i$ (both with units $\mathsf L^2\mathsf T^{-1}$). In the CGLE, spatial coupling acts directly on the state $u$, so diffusion progressively smooths out spatial gradients while leaving the cubic interaction purely local. In the PGLE, Eq.~\eqref{Eq_PGLE}, spatial coupling originates instead from eliminating the shared supersaturation bath (Sec.~\ref{Sec_screened_origin}), which places the Helmholtz operator $I-\epsilon\Delta$ on the rate of change $\partial_t u$ and filters the entire reaction tendency through the screened Green function, $\partial_t u=G_\epsilon*F(u)$. 

The physical difference is immediately apparent in a linearly stable regime ($\lambda=-\gamma$ with damping rate $\gamma>0$).  For a Fourier mode of wavevector $\kvec$ and wavenumber $k=|\kvec|$, let $Q_k\equiv 1+\epsilon k^2=\widehat G_\epsilon(\kvec)^{-1}$ denote the Fourier symbol of the screening operator $I-\epsilon\Delta$, i.e., the inverse of the Green-function multiplier in Eq.~\eqref{Eq_Green_symbol}. In the CGLE, the modal decay rate $\gamma+D_r k^2$ grows with wavenumber, so its amplitude-relaxation time $(\gamma+D_r k^2)^{-1}$ vanishes as $k^{-2}$ and fine-scale structure is rapidly erased; in the PGLE, by contrast, the modal decay rate is attenuated by screening to $\gamma/Q_k$, so the relaxation time $Q_k/\gamma=(1+\epsilon k^2)/\gamma$ increases at wavelengths shorter than the screening length $L_{\rm scrn}=\sqrt{\epsilon}$, resulting in fine-scale spatial structures that adjust slowly and persist rather than being diffusively smoothed out.

Correspondingly, whereas the global CGLE amplitude balance dissipates gradients through a diffusive sink $-D_r\int_{\Torus}|\nabla u|^2\,\d\x$, the PGLE embeds the gradient norm $\frac{\epsilon}{2}\int_{\Torus}|\nabla u|^2\,\d\x$ inside the time derivative of the conserved-sector energy $\mathcal E_\epsilon$ (Eq.~\eqref{Eq_PGLE_energy} in Sec.~\ref{Sec_amplitude_balances}). This slow short-wavelength adjustment is the physical essence of \emph{dynamic inertia} for clouds: a cloud element cannot abruptly alter its fine-scale boundaries without readjusting its surrounding supersaturation halo across the scale $L_{\rm scrn}$ \cite{chekroun_koren_hovering2026}, allowing sharp cloud edges, localized phase defects, and steep gradient filaments to persist while the broader cloud--rain cycle evolves. 
At the same time, because $(I-\epsilon\Delta)^{-1}$ acts on both the real and imaginary parts of $F(u)$, screening directly couples the nonlinear frequency twist $\beta$ to spatial phase gradients without requiring a separate dispersive diffusivity $D_i$.

This contrast between dynamic inertia and diffusive smoothing also governs how the two models filter unresolved fluctuations about a stable reference state ($\lambda=-\gamma<0$). Let $P_{\rm PGLE}(k)$ and $P_{\rm CGLE}(k)$ denote the stationary modal variance $\mathbb E|\widehat U_{\kvec}|^2$ (with units of area $\mathsf L^2$ for dimensionless amplitude) at wavenumber $k$, and let $\mathcal J(k)$ (units $\mathsf L^2\mathsf T^{-1}$) denote the variance injected into that mode per unit time by stochastic forcing. Because modal variance decays at twice the linear amplitude relaxation rate---namely $2\gamma/Q_k$ in the PGLE and $2(\gamma+D_r k^2)$ in the CGLE---balancing injection against relaxation yields
\be\label{Eq_spectral_comparison}
P_{\rm PGLE}(k)=\frac{Q_k\,\mathcal J(k)}{2\gamma},
\quad
P_{\rm CGLE}(k)=\frac{\mathcal J(k)}{2(\gamma+D_rk^2)}.
\ee
In the parcel-event model of Sec.~\ref{Sec_spatial_jump_process}, each event impulse is distributed spatially by $G_\epsilon$ and its Fourier amplitude is weighted by $\widehat G_\epsilon(\kvec)=Q_k^{-1}$, so the variance injected per unit time into mode $k$ is doubly screened, $\mathcal J(k)=\mathcal D_*/Q_k^2$, where the injection amplitude $\mathcal D_*$ (units $\mathsf L^2\mathsf T^{-1}$) is determined self-consistently by the parcel event frequency, element area, and mark variance, as derived in the companion study~\cite{chekroun_koren_liu_variance2026}. Substituting $\mathcal J(k)=\mathcal D_*/Q_k^2$ into Eq.~\eqref{Eq_spectral_comparison} shows that one factor of $Q_k$ cancels in the PGLE between the screened event footprint ($\propto Q_k^{-2}$) and slow inertial relaxation ($\propto Q_k$), leaving a Lorentzian spectrum $P_{\rm PGLE}(k)=\mathcal D_*/[2\gamma(1+\epsilon k^2)]$ with a $k^{-2}$ high-wavenumber tail, whereas the same screened forcing in the CGLE decays as $k^{-6}$. A linearly stable CGLE driven instead by unscreened space--time white noise ($\mathcal J(k)=\mathrm{const}$) also produces a $k^{-2}$ tail~\cite{staliunas2001spectra}, showing that identical snapshot spectra can reflect opposite physical mechanisms: in the parcel-driven PGLE, the screening length $L_{\rm scrn}$ simultaneously sets the spatial reach of an event's thermodynamic footprint and the prolonged lifetime of fine spatial structure.

The linear stochastic-diffusion model of cloud regimes~\cite{stechmann2016cloud} illustrates this distinction directly. That model evolves a scalar total-water variable $\mathsf q$ under $\partial_t\mathsf q=D_{\mathsf q}\Delta\mathsf q-\tau^{-1}\mathsf q+F_0+\sigma\dot W$, with eddy diffusivity $D_{\mathsf q}$, bulk relaxation time $\tau$, mean source $F_0$, and space--time white noise $\sigma\dot W$, diagnosing cloud cover by thresholding $\mathsf q$ at saturation. With a grid cutoff $\Delta x$ understood, its stationary spectrum is proportional to $(1+D_{\mathsf q}\tau k^2)^{-1}$, which matches the static Lorentzian profile of $P_{\rm PGLE}(k)$ when $\epsilon=D_{\mathsf q}\tau$ and shares the same $\log(L/\Delta x)$ pointwise variance growth on a two-dimensional grid~\cite{stechmann2016cloud,chekroun_koren_liu_variance2026}. Dynamically, however, its fluctuations at wavenumber $k$ flicker and decay over the fast diffusive time $\tau/(1+\epsilon k^2)$ rather than persisting over the slow inertial time $(1+\epsilon k^2)/\gamma$, so temporal persistence separates the two descriptions where a single-time spectrum cannot. Moreover, while linear diffusive-threshold models capture bulk cloud-fraction statistics without resolving cellular geometry~\cite{stechmann2016cloud}, the PGLE couples dynamic inertia to nonlinear cloud--rain oscillations: in the supercritical regime ($\lambda>0$), cubic interactions arrest scale coarsening and generate a persistent negative spatial-correlation lobe in the oscillation amplitude $|u|$ from random initial anomalies (Sec.~\ref{Sec_random_organization}), capturing how organized mesoscale cellular morphology endures even as individual cloud elements continually form and dissipate~\cite{koren2013adaptive}.

\section{Liquid-Water and Amplitude Budgets}
\label{Sec_budgets}

This section collects the exact budgets of the model: the integrated liquid-water budget that constrains the event increments, and the amplitude balance of the deterministic PGLE with its phase-symmetry interpretation. Here ``energy'' denotes a quadratic measure of normalized cloud--rain oscillation amplitude, not thermodynamic energy or water mass; its relation to phase symmetry is developed in Sec.~\ref{Sec_Noether_balance}.

\subsection{Liquid-water budget and unresolved reservoirs}
\label{Sec_liquid_water_budget}

Because the PGLE resolves suspended cloud water $C$ and precipitating rain water $P$ while leaving water vapor implicit, its physical water budget is that of an open liquid system. In a full warm-cloud budget, condensation and evaporation exchange mass between vapor and liquid, whereas collision--coalescence transfers mass internally between cloud droplets and rain and cancels from the total liquid-water tendency, leaving vapor phase changes and precipitation fallout as the net column sources and sinks~\cite{frierson2004precipitation}. In the direct water-path representation of Eq.~\eqref{Eq_water_anomalies} with fixed background paths $(C_0,P_0)$ and scales $(C_*,P_*)$  (Eq.~\eqref{Eq_water_anomalies}), the local liquid-water-path anomaly $\ell(\x,t)$ and the domain-integrated liquid-water mass $M_l(t)$ are related to $u=\rho+iq$ through the real-valued  linear projection $\mathscr L(z)=C_*\operatorname{Re}z+P_*\operatorname{Im}z$:
\bea\label{Eq_liquid_mass_definition}
\ell(\x,t)&=\mathscr L(u)=C_*\rho+P_*q,\\
M_l(t)&=\int_{\Torus}(C+P)\,\d\x
=|\Torus|(C_0+P_0)+\int_{\Torus}\ell(\x,t)\,\d\x.
\eea
Applying $\mathscr L$ to the deterministic PGLE, Eq.~\eqref{Eq_PGLE}, yields the local and domain-integrated liquid-water balances
\bea\label{Eq_liquid_local_budget}
(I-\epsilon\Delta)\partial_t\ell&=S_l(u),\\
S_l(u)&=\mathscr L(F(u))=(\lambda-\alpha|u|^2)(C_*\rho+P_*q)\\
&\quad +(\omega-\beta|u|^2)(P_*\rho-C_*q),
\eea
and, by periodicity of $\Torus$, or equivalently because $\partial_t\ell=G_\epsilon*S_l(u)$ with $\int_{\Torus}G_\epsilon\,\d\x=1$,
\be\label{Eq_liquid_integrated_budget}
\frac{\d M_l}{\d t}=\int_{\Torus}S_l(u)\,\d\x.
\ee
Here $S_l(u)$, with units of water path per unit time, represents the net unscreened column source combining vapor condensation, evaporation, and rain fallout, while environmental screening $G_\epsilon*S_l(u)$ redistributes this liquid-water tendency laterally across the scale $L_{\rm scrn}=\sqrt{\epsilon}$ without altering its domain integral.

Equations~\eqref{Eq_liquid_local_budget} and \eqref{Eq_liquid_integrated_budget} reveal how spatial pattern geometry governs the domain liquid-water inventory. Because $F(-u)=-F(u)$ is odd under a half-cycle phase shift $u\mapsto -u$, the source $S_l(u)$ integrates to zero over any periodic wave or cellular pattern whose spatial phase distribution is symmetric between opposite halves of the cloud--rain cycle. In particular, for any monochromatic wave solution of Sec.~\ref{Sec_dyn_insights} with $\kvec\neq 0$, the modulus $|u(\x,t)|=R(t)$ is spatially uniform while $\int_{\Torus}u(\x,t)\,\d\x=0$, so
\beas
\int_{\Torus}S_l(u)\,\d\x
&=\bigl(\lambda-\alpha R(t)^2\bigr)\left[C_*\int_{\Torus}\rho\,\d\x+P_*\int_{\Torus}q\,\d\x\right]\\
&+\bigl(\omega-\beta R(t)^2\bigr)\left[P_*\int_{\Torus}\rho\,\d\x-C_*\int_{\Torus}q\,\d\x\right]\\
&=0.
\eeas
and $\d M_l/\d t=0$ identically at all times. Thus, condensation in growing cloud crests is balanced exactly across the domain by evaporation and fallout in decaying rain troughs, maintaining the total liquid inventory at $M_l=|\Torus|(C_0+P_0)$---consistent with large-domain LES where out-of-phase subregions cancel out in the domain-integrated water mass \cite[Fig.~8]{dagan2018organization}.

When $u(\x,t)$ is a superposition of multiple waves, $|u(\x,t)|^2$ varies with $\x$, so $(\lambda-\alpha|u|^2)$ and $(\omega-\beta|u|^2)$ cannot be factored out of the spatial integral. Instead, integrating $S_l(u)=\mathscr L\bigl((\lambda+i\omega)u-\kappa|u|^2u\bigr)$ gives
\bes
\frac{\d M_l}{\d t}
=\mathscr L\!\left((\lambda+i\omega)\int_{\Torus}u\,\d\x-\kappa\int_{\Torus}|u|^2u\,\d\x\right).
\ees
If the pattern has no resonant triad $\kvec_1+\kvec_2=\kvec_3$ (or has half-cycle translation symmetry $u(\x+\boldsymbol\delta,t)=-u(\x,t)$) and starts with zero spatial mean, then $\int_{\Torus}|u|^2u\,\d\x=0$ and $\d M_l/\d t=0$. Conversely, if the pattern contains a resonant triad $\kvec_1+\kvec_2=\kvec_3$ (such as the $k_0=5$ carrier and $k_0+q=10$ sideband in Fig.~\ref{Fig_Space_Time}B, where $5+5=10$), then $\int_{\Torus}|u|^2u\,\d\x\neq 0$, which excites the $\kvec=\mathbf 0$ spatial-mean mode and causes $M_l(t)$ to oscillate in time around $|\Torus|(C_0+P_0)$ while remaining bounded with zero long-time average drift.
Physically, such resonant triads correspond to phase-asymmetric spatial patterns in which large amplitude excursions $|u|^2$ correlate preferentially with one phase of the cloud--rain cycle rather than canceling across opposite halves, yielding $\int_{\Torus}|u|^2u\,\d\x\neq 0$ and driving a net domain-mean water-mass tendency $\d M_l/\d t\neq 0$.

Prior to the mean-field closure, taking the expectation of the exact jump balance in Eq.~\eqref{Eq_mean_exact_cutoff} for the realization mass $\mathsf M_l[U_t]=\int_{\Torus}(C_t+P_t)\,\d\x$ gives
\be\label{Eq_liquid_exact_mean_budget}
\frac{\d\mathbb E[\mathsf M_l[U_t]]}{\d t}
=\int_{\Torus}\mathscr L\bigl(\mathbb E[F_B(U_t)]\bigr)\,\d\x,
\ee
which shows via Eq.~\eqref{Eq_third_moment_expansion} that unresolved parcel fluctuations can also drive a net domain water-mass tendency whenever the local fluctuation covariance $2u\mathbb E[|v|^2]+\bar u\mathbb E[v^2]$ or third-order skewness $\mathbb E[|v|^2v]$ breaks phase symmetry.

At the level of an individual parcel episode centered at $\y$, the total dimensionless increment
\be\label{Eq_liquid_event_increment}
\eta=\frac{F_B(U_{t-}(\y))}{r_p}+b\zeta
\ee
changes the domain liquid-water mass of a realization by
\be\label{Eq_liquid_event_mass}
\delta\mathsf M_l=a_p\mathscr L(\eta)=a_p\bigl(C_*\operatorname{Re}\eta+P_*\operatorname{Im}\eta\bigr).
\ee
Equation~\eqref{Eq_liquid_event_mass} partitions the complex increment plane $(\operatorname{Re}\eta,\operatorname{Im}\eta)$ into distinct microphysical regimes. Pure conversion episodes (autoconversion and accretion), which transfer mass from suspended cloud droplets to rain without vapor exchange or immediate fallout, must lie along the zero-sum line $\mathscr L(\eta)=0$ in the fourth quadrant ($\operatorname{Im}\eta/\operatorname{Re}\eta=-C_*/P_*<0$, $\operatorname{Re}\eta<0$), conserving $\mathsf M_l$ event by event. Departures from this line quantify open-system column exchanges: condensation and cloud evaporation act parallel to the real axis ($\delta\mathsf M_l=a_p C_*\operatorname{Re}\eta\gtrless 0$), whereas rain evaporation and surface precipitation fallout act along the negative imaginary axis ($\delta\mathsf M_l=a_p P_*\operatorname{Im}\eta<0$). As a parcel progresses through its Stuart--Landau cycle, the systematic tendency $F(u)$ rotates continuously through these quadrants---from net vapor condensation ($\dot\rho>0$) into cloud-to-rain conversion ($\dot\rho<0,\dot q>0$) and subsequent fallout ($\dot q<0$)---while the mark law $\Pi(\d\zeta)$ encodes how episode fluctuations partition between anti-correlated conversion scatter along $\mathscr L(\zeta)=0$ and open-system vapor or fallout variance.

For cloud physics, the resulting take-home message is twofold. Pointwise and event by event, the PGLE does not conserve liquid-water mass---nor should it, because a precipitating cloud layer is an open liquid system that continually draws water from the unresolved vapor reservoir through condensation and exports it through evaporation and surface fallout, with only internal collision--coalescence episodes conserving $C+P$ along the line $\mathscr L(\eta)=0$.
At the mesoscale domain scale, however, because the screened response $G_\epsilon$ is mass-conserving ($\int_{\Torus}G_\epsilon\,\d\x=1$) and the oscillatory source $S_l(u)$ is odd under a half-cycle phase shift, any monochromatic wave, parallel roll, or triad-free pattern with zero initial mean maintains an exactly constant domain-integrated liquid-water inventory $M_l(t)=|\Torus|(C_0+P_0)$ through a dynamic equilibrium between growing and decaying cloud regions, while multi-mode cellular or modulated fields with active triads oscillate in a bounded fashion around $|\Torus|(C_0+P_0)$ without secular drift.
 What the PGLE regulates structurally is therefore not a closed parcel water inventory, but the spatial intensity and gradient structure of the cloud--rain cycle itself: as shown next in Secs.~\ref{Sec_amplitude_balances} and \ref{Sec_Noether_balance}, the rotational sector of the equation conserves an exact $U(1)$ Noether charge $\mathcal E_\epsilon=\mathcal{N}_\epsilon$ combining oscillation variance $|u|^2$ with screening-weighted spatial gradients $\epsilon|\nabla u|^2$, and the balance between linear growth and cubic saturation locks this global oscillatory activity onto an invariant level even as water continuously fluxes through individual clouds.

\subsection{Amplitude and spectral balances}
\label{Sec_amplitude_balances}

Let $V=|\Torus|$ and use orthonormal Fourier coefficients
\[
\widehat u_{\kvec}=V^{-1/2}\int_{\Torus}
u(\x)e^{-i\kvec\cdot\x}\,\d\x,
\qquad Q_{\kvec}=1+\epsilon|\kvec|^2.
\]
Thus $\sum_{\kvec}|\widehat u_{\kvec}|^2=\int_{\Torus}|u|^2\,\d\x$. To obtain the integrated balance, multiply Eq.~\eqref{Eq_PGLE} by $\bar u$ and integrate over the periodic domain:
\bea\label{Eq_PGLE_energy_step}
\int_{\Torus}\bar u\,(I-\epsilon\Delta)\partial_tu\,\d\x&=(\lambda+i\omega)\int_{\Torus}|u|^2\,\d\x\\
&\qquad -(\alpha+i\beta)\int_{\Torus}|u|^4\,\d\x.
\eea
Periodicity removes the boundary term in integration by parts. The real parts of the two terms on the left are therefore
\bea\label{Eq_PGLE_energy_parts}
\operatorname{Re}\int_{\Torus}\bar u\,\partial_tu\,\d\x
&=\frac12\frac{\d}{\d t}\int_{\Torus}|u|^2\,\d\x,\\
-\operatorname{Re}\int_{\Torus}\bar u\,\Delta\partial_tu\,\d\x
&=\operatorname{Re}\int_{\Torus}\nabla\bar u\cdot\nabla\partial_tu\,\d\x\\
&=\frac12\frac{\d}{\d t}\int_{\Torus}|\nabla u|^2\,\d\x.
\eea
On the right, both integrals are real, so the terms multiplied by $i\omega$ and $i\beta$ have zero real part. Consequently,
\bea\label{Eq_PGLE_energy}
\mathcal E_\epsilon&=\frac12\int_{\Torus}
\bigl(|u|^2+\epsilon|\nabla u|^2\bigr)\,\d\x,\\
\frac{\d\mathcal E_\epsilon}{\d t}
&=\lambda\int_{\Torus}|u|^2\,\d\x
-\alpha\int_{\Torus}|u|^4\,\d\x.
\eea
Thus the mixed derivative contributes the gradient term to $\mathcal E_\epsilon$, while linear growth and cubic saturation determine its rate of change. This balance describes the amplitude field and is distinct from the liquid-water budget in Eq.~\eqref{Eq_liquid_integrated_budget}.

For the linear PGLE, the modal power evolves as
\be\label{Eq_linear_modal_power}
|\widehat u_{\kvec}(t)|^2
=|\widehat u_{\kvec}(0)|^2
\exp\!\left(\frac{2\lambda t}{Q_{\kvec}}\right).
\ee
 In the unforced linear equation, each initial spectral component evolves at its own scale-dependent rate. Sustained spectral levels therefore require a balance with forcing or nonlinear transfer. In the nonlinear equation the exact balance is
\bea\label{Eq_deterministic_spectral_balance}
\frac{\d|\widehat u_{\kvec}|^2}{\d t}
=\frac{2}{Q_{\kvec}}\Bigl[\lambda|\widehat u_{\kvec}|^2
-\operatorname{Re}\bigl\{\kappa\widehat u_{\kvec}^{\,*}
\widehat{|u|^2u}_{\kvec}\bigr\}\Bigr].
\eea
The cubic term couples Fourier modes and combines redistribution with amplitude saturation. Under an ensemble average, it involves fourth-order statistics, so the nonlinear spectrum requires further closure or direct computation.

\subsection{Phase symmetry, Noether charge, and dynamical constraints on pattern selection}
\label{Sec_Noether_balance}

The quadratic functional $\mathcal E_\epsilon$ in Eq.~\eqref{Eq_PGLE_energy} is more than a convenient Lyapunov-type diagnostic: it is the conserved Noether charge of the conservative sector of the PGLE, and its production--dissipation balance imposes exact global constraints on the pattern-formation dynamics analyzed in Sec.~\ref{Sec_modeling_results}. The deterministic PGLE is equivariant under the global $U(1)$ phase rotation $u\mapsto e^{i\theta}u$. In normal-form coordinates this symmetry represents a uniform shift in oscillation timing across the cloud field, whereas in the direct water-anomaly representation of Eq.~\eqref{Eq_water_anomalies} it rotates the normalized anomaly pair $(\rho,q)$ and therefore generally alters the linear liquid-water combination $\ell=C_*\rho+P_*q$ in Eq.~\eqref{Eq_liquid_mass_definition}. Consequently, phase invariance governs the organization of oscillatory activity rather than the physical water inventory of Sec.~\ref{Sec_liquid_water_budget}. While classical Noether theory links continuous symmetries to conservation laws in variational systems~\cite{noether1918invariante} and recent extensions formulate analogous symmetry identities for dissipative gradient flows~\cite{white2025beyond}, the PGLE splits naturally into a conservative Hamiltonian flow and a dissipative gradient flow that are both governed by the same dynamic-inertia metric operator $A_\epsilon=I-\epsilon\Delta$. This decomposition parallels the splitting of the standard CGLE into nonlinear Schr\"odinger and reaction--diffusion (Allen--Cahn) sectors~\cite{cross1993pattern,sulem1999nonlinear,aranson2002world}, with the crucial difference that here all spatial coupling resides in the operator $A_\epsilon$ acting on the time derivative rather than in separate diffusive and dispersive Laplacians.

To make the conservative sector explicit in Eq.~\eqref{Eq_PGLE}, we set $\lambda=\alpha=0$ on the periodic domain $\Torus$ ($\epsilon>0$) while retaining the linear frequency $\omega$ and twist parameter $\beta$. We obtain then
\be\label{Eq_PGLE_conservative_sector}
A_\epsilon\partial_tu=i(\omega-\beta|u|^2)u.
\ee
Factoring out the base oscillation via $u(\x,t)=e^{i\omega t}\psi(\x,t)$ transforms Eq.~\eqref{Eq_PGLE_conservative_sector} into 
\be
i(I-\epsilon\Delta)\partial_t\psi+\epsilon\omega\Delta\psi-\beta|\psi|^2\psi=0, 
\ee
which is a pseudoparabolic, Benjamin--Bona--Mahony-type regularization of the nonlinear Schr\"odinger equation~\cite{benjamin1972model,sulem1999nonlinear}. Indeed, just as the Benjamin--Bona--Mahony operator $(I-\epsilon\partial_x^2)\partial_t$ regularizes unbounded short-wave dispersion in long-wave hydrodynamics~\cite{benjamin1972model}, $A_\epsilon\partial_t$ regularizes the Schr\"odinger dispersion at wavelengths shorter than $L_{\rm scrn}=\sqrt{\epsilon}$. 

Endowing the space of smooth periodic fields with the real-valued  $L^2$-inner product  $(v,w)_{\mathbb R}=\operatorname{Re}\int_{\Torus}\bar v w\,\d\x$ and denoting its variational gradient by $\operatorname{grad}$, Eq.~\eqref{Eq_PGLE_conservative_sector} takes the noncanonical Hamiltonian form~\cite{olver1993applications,morrison1998hamiltonian}
\bea\label{Eq_PGLE_Hamiltonian}
H[u]&=\int_{\Torus}\left(-\frac{\omega}{2}|u|^2+\frac{\beta}{4}|u|^4\right)\d\x,\\
J_\epsilon&=-iA_\epsilon^{-1},\qquad
\partial_tu=J_\epsilon\operatorname{grad}H[u].
\eea
Because $A_\epsilon$ is positive, self-adjoint, and invertible on $L^2(\Torus)$, the Poisson (cosymplectic) operator $J_\epsilon=-i(I-\epsilon\Delta)^{-1}$ is constant and skew-adjoint with respect to $(\cdot,\cdot)_{\mathbb R}$, satisfying the Jacobi identity automatically \cite{zafferi2023generic}. Both $H[u]$ and $A_\epsilon$ are invariant under global phase rotations, and the quadratic functional
\bea\label{Eq_PGLE_phase_charge}
\mathcal{N}_\epsilon[u]&=\frac12\int_{\Torus}\bar u A_\epsilon u\,\d\x\\
&=\frac12\int_{\Torus}\bigl(|u|^2+\epsilon|\nabla u|^2\bigr)\,\d\x=\mathcal E_\epsilon[u],\\
\operatorname{grad}\mathcal{N}_\epsilon[u]&=A_\epsilon u,\qquad
J_\epsilon\operatorname{grad}\mathcal{N}_\epsilon[u]=-iu,
\eea
is the infinitesimal Hamiltonian generator of $u\mapsto e^{-i\theta}u$. By Noether's theorem for infinite-dimensional Hamiltonian systems~\cite{olver1993applications,sulem1999nonlinear}, $U(1)$ invariance of $H[u]$ implies $(\operatorname{grad}\mathcal{N}_\epsilon, J_\epsilon\operatorname{grad}H)_{\mathbb R}=0$, so both $\mathcal{N}_\epsilon[u]$ and $H[u]$ are exact invariants of Eq.~\eqref{Eq_PGLE_conservative_sector}. Unlike the standard nonlinear Schr\"odinger equation, where the $U(1)$ wave-action charge is the $L^2$ norm $\frac{1}{2}\int|u|^2\,\d\x$ and $\int|\nabla u|^2\,\d\x$ resides in the Hamiltonian~\cite{sulem1999nonlinear}, the pseudoparabolic operator $A_\epsilon\partial_t$ embeds the screening-weighted gradient norm $\frac{\epsilon}{2}\int|\nabla u|^2\,\d\x$ directly into the conserved Noether charge $\mathcal{N}_\epsilon$~\cite{benjamin1972model}.

The complementary growth--saturation sector ($\omega=\beta=0$, with $\lambda,\alpha>0$), namely $A_\epsilon\partial_t u=(\lambda-\alpha|u|^2)u$, is a pseudoparabolic Allen--Cahn equation driven by the Landau potential
\be\label{Eq_PGLE_Landau_potential}
V[u]=\int_{\Torus}\left(-\frac{\lambda}{2}|u|^2+\frac{\alpha}{4}|u|^4\right)\d\x.
\ee
Rather than a parabolic reaction--diffusion flow, this sector is a gradient system $\partial_t u=-A_\epsilon^{-1}\operatorname{grad}V[u]$ whose steepest descent is measured with respect to the Sobolev dynamic-inertia inner product $\langle v,w\rangle_{H^1_\epsilon}=(v,A_\epsilon w)_{\mathbb R}$. Combining both sectors expresses the full PGLE as a metric-coupled Hamiltonian and gradient flow on the Sobolev space $H^1_\epsilon(\Torus)$,
\be\label{Eq_PGLE_charge_forcing}
\partial_tu=J_\epsilon\operatorname{grad}H[u]
-A_\epsilon^{-1}\operatorname{grad}V[u],
\ee
in which $H[u]$ and $V[u]$ are purely local potential integrals and all spatial coupling enters through $A_\epsilon^{-1}$. Taking the (real) inner product  of Eq.~\eqref{Eq_PGLE_charge_forcing} with $\operatorname{grad}\mathcal{N}_\epsilon=A_\epsilon u$ annihilates the conservative vector field $J_\epsilon\operatorname{grad}H[u]$ and recovers the exact global balance in Eq.~\eqref{Eq_PGLE_energy}. 

In practical terms, the Noether charge $\mathcal N_\epsilon=\mathcal E_\epsilon$ serves as a global budget that separates conservative spatial reshaping from amplitude growth and saturation. Because $\omega$ and $\beta$ enter only through the Hamiltonian vector field $J_\epsilon\operatorname{grad}H[u]$, they make zero direct contribution to $\d\mathcal E_\epsilon/\d t$ in Eq.~\eqref{Eq_PGLE_energy}: the rotational dynamics cannot create or destroy total oscillatory activity on its own, but instead redistributes $\mathcal N_\epsilon$ between bulk oscillation amplitude and fine-scale spatial gradients---an insight used in Secs.~\ref{Sec_banded_solutions} and \ref{Sec_defect_analysis}  below to isolate how phase twist drives wave breakup and filament formation. On the other hand, net production and dissipation of $\mathcal N_\epsilon$ are governed entirely by the growth--saturation potential $V[u]$, which caps the global mean-square amplitude at the isolated-wave saturation level and separates overall amplitude equilibration from ongoing spatial reorganization. This is illustrated in the roll-consolidation and random-emergence experiments of Secs.~\ref{Sec_spontaneous_stabilization} and \ref{Sec_random_organization} below.

\section{Instability and Nonlinear Transitions}\label{Sec_Instab_Neimark}

This section analyzes the stability of a \emph{single} exact periodic state of the PGLE---first a traveling plane wave ($k_0\neq 0$) on a one-dimensional periodic domain, following the sideband framework of the CGLE~\cite{holmes1985bounded,doelman1993traveling}, and then the spatially homogeneous oscillation ($k_0=0$) in any spatial dimension. We show below that, although the PGLE lacks a separate diffusive dispersion coefficient $D_i$, the interplay of dynamic inertia ($\epsilon>0$) and phase twist ($\beta\neq 0$) suffices to destabilize a traveling carrier through paired sidebands $k_0\pm q$ (Eq.~\eqref{Eq_stab_cond2}), with non-harmonic modulations ($q\neq nk_0$) opening broad instability regions on both sides of the zero-frequency locus $\Omega_0=0$ (Fig.~\ref{Fig_stab_chart}). Tracking a destabilized wave into saturation, we also show that nonlinear evolution reorganizes the field into a persistent modulated state whose power spectrum forms a sharp combination-tone comb $\omega_n=|\Omega_c+n\Omega_m|$ (Figs.~\ref{Fig_Space_Time}--\ref{Fig_PSD}), separating the linear sidebands that initiate breakup from the frequencies selected after saturation. Specializing to the $k_0=0$ limit further reveals that a spatially uniform cloud--rain cycle spontaneously loses synchrony across a finite wavenumber band $0<\epsilon|\kvec|^2<-2\beta A^2/\Omega_*$ whenever $\beta\Omega_*<0$ (Eq.~\eqref{Eq_uniform_modulation_band}), providing a dynamic-inertia analog of Turing pattern onset from an oscillating reference state.

\subsection{Our perturbation approach}
 We perturb the exact traveling waves associated with the asymptotic form in Eq.~\eqref{Eq_u_plane_asymp}. Their amplitude is constant, $R(t)=A$, and their phase advances uniformly at frequency $\Omega_0$:
\be \label{Eq_A_Omega}
|A|^2  = \frac{\lambda}{\text{Re}(\kappa)}, \quad  \Omega_0 = (1 + \epsilon |\mathbf{k}|^2)^{-1} \left(\omega - \frac{\lambda \text{Im}(\kappa)}{\text{Re}(\kappa)} \right). 
\ee
We start with an unperturbed traveling wave solution
\bes
u_0(x, t) = A \exp(i (k_0 x + \Omega_0 t)), \; x \in (0,2\pi),
\ees
where $A$ and $\Omega_0$ are given by Eq.~\eqref{Eq_A_Omega}. Here, we assume thus implicitly that $\phi(0)$ is chosen such that $\phi_0=0$ (Eq.~\eqref{Eq_phi0}). 

First, we linearize the PGLE around the plane wave solution, $u_0$. Let the full solution be $u=u_0+\delta u$, where $\delta u$ is a small perturbation. Substituting this into the PGLE (Eq.~\eqref{Eq_PGLE}), we keep only terms that are linear in $\delta u$.  The nonlinear term is $-\kappa|u|^2u$. We can write $|u|^2=u\bar{u}=(u_0+\delta u)(\bar{u}_0+\bar{\delta u})$.
Expanding to first order in the perturbation, we get:
$$ |u|^2 \approx |u_0|^2+u_0\bar{\delta u}+\bar{u}_0\delta u.$$
The full nonlinear term becomes:
$$ -\kappa|u|^2u \approx -\kappa(|u_0|^2+u_0\bar{\delta u}+\bar{u}_0\delta u)(u_0+\delta u).$$
Keeping only terms up to first order:
\beas
 -\kappa|u|^2u=&-\kappa(|u_0|^2 u_0+|u_0|^2\delta u+u_0^2\bar{\delta u}+|u_0|^2\delta u)\\
 &= -\kappa(|u_0|^2u_0+2|u_0|^2\delta u+u_0^2\bar{\delta u}).
 \eeas
The zeroth-order term, $-\kappa|u_0|^2u_0$, cancels with the corresponding term from the left side of the PGLE since $u_0$ is a solution. This leaves us with the linearized PGLE for the perturbation $\delta u$:
\be\label{Eq_linearized_PGLE}
 \partial_t\delta u=(\lambda+i\omega)\delta u-\kappa(2|u_0|^2\delta u+u_0^2\bar{\delta u})+\epsilon\Delta\partial_t\delta u.
 \ee

\subsection{Instability Regions}\label{Sec_instab}
Write $u=(1+V)u_0$, so that $\delta u=Vu_0$, and expand the modulation as
\be\label{Eq_modulational}
V(x,t)=\sum_{q\in\mathbb Z}V_q(t)e^{iqx}.
\ee
On the $2\pi$-periodic domain, the carrier wavenumber $k_0$ and modulation wavenumber $q$ are integers; the expansion allows all integer modulations. Each $q\ne0$ couples the sidebands $k_0+q$ and $k_0-q$. Define
\be\label{Eq_single_sideband_factors}
Q_\pm=1+\epsilon(k_0\pm q)^2,\qquad Q_0=1+\epsilon k_0^2.
\ee
Substituting $\delta u=Vu_0$ into Eq.~\eqref{Eq_linearized_PGLE} gives the exact paired linear system
\begin{widetext}
\be\label{Eq_system_perturb}
\partial_t\begin{pmatrix} V_q\\ \overline V_{-q}\end{pmatrix}
=M_q\begin{pmatrix} V_q\\ \overline V_{-q}\end{pmatrix},\qquad
M_q=\begin{pmatrix}
(\lambda+i\omega-2\kappa A^2)/Q_+-i\Omega_0 &-\kappa A^2/Q_+\\
-\overline\kappa A^2/Q_- &(\lambda-i\omega-2\overline\kappa A^2)/Q_-+i\Omega_0
\end{pmatrix}.
\ee
\end{widetext}
The rotating-frame terms $\mp i\Omega_0$ lie outside the screening denominators: differentiating $Vu_0$ produces $ [I-\epsilon(\partial_x+ik_0)^2](\partial_tV+i\Omega_0V)$ in the shifted spatial frame. Equivalently, the frequency inside each numerator is $\omega-\Omega_0Q_\pm$. Both components of an eigenvector evolve with the same factor $e^{\sigma t}$; consequently $V_{-q}$ itself carries the conjugate exponent.

 With $\kappa=\alpha+i\beta$ and $A^2=\lambda/\alpha$, define $Y=\omega-2\beta A^2$. Write the diagonal entries of $M_q$ as $-d_++ib_+$ and $-d_-+ib_-$, where $d_\pm=\lambda/Q_\pm>0$, $b_+=Y/Q_+-\Omega_0$, and $b_-=-Y/Q_-+\Omega_0$. The positive off-diagonal product is $g_q^2=|\kappa|^2A^4/(Q_+Q_-)$. The characteristic equation takes the factored form
\be\label{Eq_single_pair_characteristic}
(\sigma+d_+-ib_+)(\sigma+d_--ib_-)=g_q^2.
\ee
At the stability boundary the perturbation neither grows nor decays, so its exponent is purely imaginary, $\sigma=i\delta_q$. The imaginary part of Eq.~\eqref{Eq_single_pair_characteristic} then gives
\be\label{Eq_single_neutral_imaginary}
d_+(\delta_q-b_-)+d_-(\delta_q-b_+)=0,
\ee
and therefore
\be\label{Eq_chi_Omega}
\delta_q=\frac{d_+b_-+d_-b_+}{d_++d_-}
=\Omega_0\frac{Q_--Q_+}{Q_++Q_-}.
\ee
Thus the neutral modulation frequency is a damping-weighted balance of the two diagonal frequencies in the carrier frame. Substitution into the real part gives
\be\label{Eq_single_neutral_gain}
g_q^2=d_+d_-\left[1+\frac{(b_+-b_-)^2}{(d_++d_-)^2}\right].
\ee
The frequency mismatch raises the coupling needed to overcome damping. The exact eigenvalue calculation in \ref{App_single_carrier_stability} establishes which side of this boundary grows: the pair is unstable precisely when
\be\label{Eq_stab_cond1}
|\kappa|^2A^4>\lambda^2+
\left[Y-\frac{2\Omega_0Q_+Q_-}{Q_++Q_-}\right]^2.
\ee
This criterion follows from linearization about a single exact traveling wave.

\begin{figure*}[t]
\centering
\includegraphics[width=0.98\textwidth]{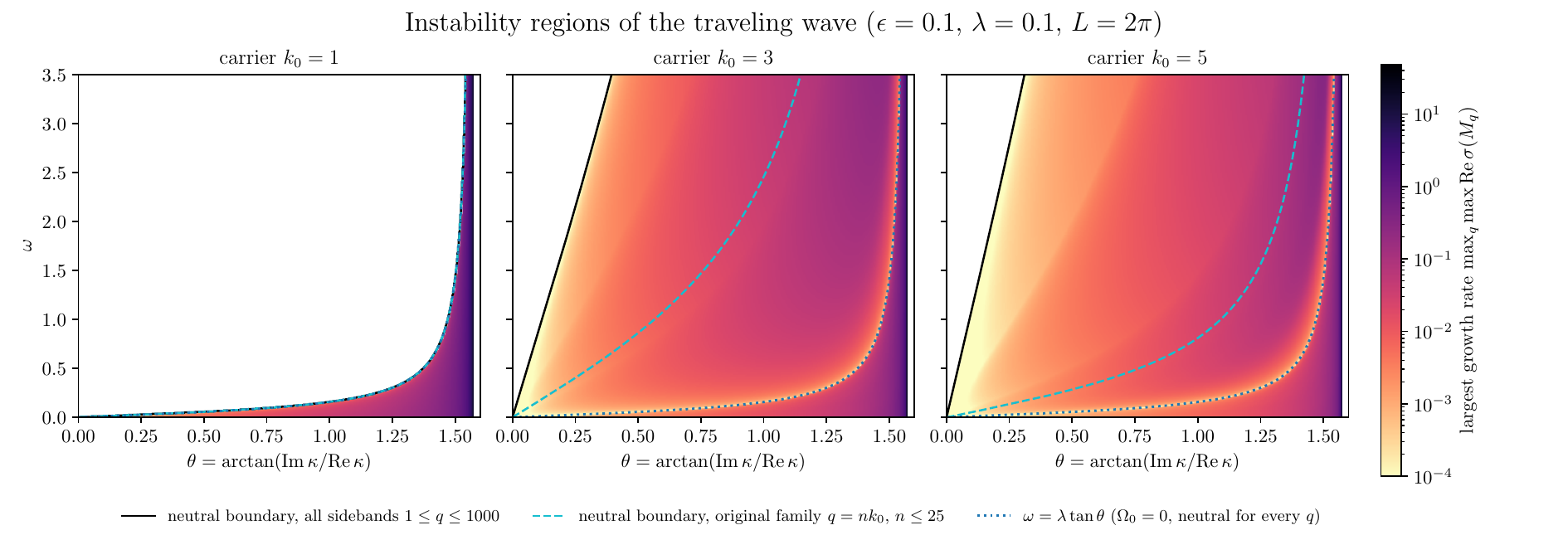}
\caption{{\bf Scale-selective sideband instability of traveling waves.} Largest positive growth rate from Eq.~\eqref{Eq_system_perturb} in the $(\theta,\omega)$ plane, where $\theta=\arctan(\beta/\alpha)$ is the twist angle, for carriers $k_0=1,3,5$ at $\lambda=0.1$, $\epsilon=0.1$, and domain length $L=2\pi$. The blue dotted curve $\omega=\lambda\tan\theta$ marks where the carrier frequency $\Omega_0=(\omega-\lambda\beta/\alpha)/Q_0$ in Eq.~\eqref{Eq_A_Omega} vanishes. Dynamic inertia flips the instability geometry as the carrier wavenumber increases: at $k_0=1$, growing sidebands are confined below $\omega=\lambda\tan\theta$ ($h_q>1$ for all $q\ge 1$), whereas at $k_0=3$ and $5$, long-wavelength modulations ($h_q<1$ in Eq.~\eqref{Eq_single_frequency_intervals}) open a broad unstable region extending above $\omega=\lambda\tan\theta$ up to the solid black neutral boundary (computed over all integer sidebands $1\le q\le 1000$). The cyan dashed curve marks the upper neutral boundary of the restricted harmonic family $q=nk_0$ ($1\le n\le 25$), so only the region below the cyan dashed curve is unstable to harmonic perturbations; the large shaded wedge between the cyan dashed and solid black curves demonstrates that non-harmonic modulations ($q\neq nk_0$) dominate the destabilization of shorter waves. White regions are linearly stable; positive growth rates are shown on a logarithmic scale clipped at $10^{-4}$.}
\label{Fig_stab_chart}
\end{figure*}

For plotting, set $\eta=\beta/\alpha$ and
\bes
h_q=\frac{2Q_+Q_{-}}{Q_0(Q_{+}+Q_{-})}. 
\ees
Since $\Omega_0=(\omega-\lambda\eta)/Q_0$, the exact instability criterion in Eq.~\eqref{Eq_stab_cond1} takes the compact dimensionless form
\be\label{Eq_stab_cond2}
\boxed{\left[(1-h_q)\omega+(h_q-2)\lambda\eta\right]^2<\lambda^2\eta^2,}
\ee
where equality defines the neutral curve for a specified nonzero modulation wavenumber $q$. A traveling wave is spectrally unstable if at least one admissible pair satisfies the strict inequality (the $q=0$ mode, whose exponents are $0$ by phase invariance and $-2\lambda/Q_0$, is excluded). Notably, when $\beta=0$ ($\eta=0$) the strict inequality cannot hold: without amplitude-dependent timing shifts, every traveling plane wave is linearly stable regardless of screening. Dynamic inertia and nonlinear twist must therefore act together to destabilize a coherent wave train.

In the $(\theta,\omega)$ plane of Fig.~\ref{Fig_stab_chart}, where $\theta=\arctan\eta$ measures the twist angle for $\eta>0$, Eq.~\eqref{Eq_stab_cond2} factors into two explicit frequency windows separated by the sign of $h_q-1$:
\be\label{Eq_single_frequency_intervals}
\boxed{\begin{aligned}
h_q<1:\quad &\lambda\eta<\omega<\frac{3-h_q}{1-h_q}\lambda\eta,\\
h_q>1:\quad &\frac{3-h_q}{1-h_q}\lambda\eta<\omega<\lambda\eta,
\end{aligned}}
\ee
with neutral endpoints and every pair neutral at $h_q=1$ while Eq.~\eqref{Eq_stab_cond2} remains valid for either sign of $\eta$. Figure~\ref{Fig_stab_chart} plots the union of these intervals over integer sidebands $1\le q\le 1000$, revealing three striking consequences of dynamic inertia for cloud-wave stability. First, the sign of $h_q-1$ is governed purely by the screening geometry,
\be\label{Eq_single_screening_sign}
h_q-1=\frac{\epsilon q^2(1+\epsilon q^2-3\epsilon k_0^2)}
{(1+\epsilon k_0^2)[1+\epsilon(k_0^2+q^2)]},
\ee
which flips sign across the threshold $q^2=3k_0^2-\epsilon^{-1}$. For $\epsilon=0.1$ and $k_0=1$, every nonzero integer $q$ satisfies $h_q>1$, confining instability strictly below the line $\omega=\lambda\eta$ (Fig.~\ref{Fig_stab_chart}, left). For shorter carriers, however, long-wavelength modulations satisfy $h_q<1$ ($q=1,\ldots,4$ for $k_0=3$; $q=1,\ldots,8$ for $k_0=5$) and open a broad upper instability lobe above $\omega=\lambda\eta$, while shorter modulations ($q\ge 5$ and $q\ge 9$, respectively) remain unstable below it (Fig.~\ref{Fig_stab_chart}, middle and right). Even though all carriers share the same saturated amplitude $A=\sqrt{\lambda/\alpha}$, identical local cloud--rain feedback can thus sustain a coherent wave at one spatial scale while destabilizing it at another.

Second, the blue dotted locus $\omega=\lambda\tan\theta$ marks where $\Omega_0=0$---the balance at which the linear frequency $\omega$ and twist shift $-\lambda\eta$ cancel so the carrier becomes stationary---rather than a stability boundary: for $k_0=3$ and $5$, growing sidebands populate both sides of this line. Away from it, the color map measures the fastest amplification rate, which surges as $\theta\to\pi/2$ because the diverging twist $\beta/\alpha=\tan\theta$ converts small amplitude perturbations into large asymptotic phase shifts (Eq.~\eqref{Eq_PGLE_isochron}). Third, comparing the full stability boundary (solid black) against the restricted harmonic family $q=nk_0$ (cyan dashed) exposes a decisive physical point: restricting perturbations to integer multiples of the carrier wavenumber misses the vast majority of the unstable domain for $k_0=3$ and $5$. A banded cloud pattern therefore loses coherence primarily through non-harmonic spatial modulations that break translational periodicity with the carrier. While Fig.~\ref{Fig_stab_chart} maps where such departures initiate and how fast they grow, which modes survive saturation is a nonlinear question that we trace next.

\subsection{Sidebands and transitions to quasi-periodic states}\label{Sec_sidebands}

While the linear stability boundary in Fig.~\ref{Fig_stab_chart} locates where a traveling cloud--rain wave first loses coherence, it does not determine the spatial scale or temporal spectrum that emerges after nonlinear saturation. To trace this transition, we perturb the $k_0=5$ plane wave ($A=\sqrt{22}$, $\Omega_0=1.4667$) at $\lambda=2.2$, $\epsilon=0.1$, $\omega=22/3$, and $\kappa=0.1+0.1i$ on a $2\pi$-periodic domain through the sideband expansion
\be\label{Eq_single_wave_sideband_field}
u=u_0+\sum_{q\in\mathbb Z}A V_q(t)e^{i[(k_0+q)x+\Omega_0t]}.
\ee
For the seeded modulation $q=5$, the sideband matrix $M_5$ in Eq.~\eqref{Eq_system_perturb} yields the unstable eigenvalue $\sigma_5=0.16811-1.23588i$, so the perturbation grows exponentially while drifting in phase relative to the carrier. Initializing the growing eigenvector at norm $10^{-3}$ isolates this linear mechanism at early times and reveals a two-stage transition in Fig.~\ref{Fig_Space_Time}. During the early stage, the measured growth rate converges to $\operatorname{Re}\sigma_5$ as the timestep is refined, and the off-carrier amplitude reaches $10\%$ of the carrier near $t=27.3$ (robust to halving $\Delta t$ or doubling spatial resolution), confirming that linear sideband instability quantitatively governs the onset of wave breakup. Moreover, because the carrier $k_0=5$ and upper sideband $k_0+q=10$ satisfy the resonant triad $5+5=10$, cubic coupling immediately excites the $\kvec=0$ spatial-mean mode $\langle\operatorname{Re}u\rangle_x$ (Fig.~\ref{Fig_Space_Time}B), illustrating how sideband destabilization generates the domain-mean liquid-water oscillations discussed in Sec.~\ref{Sec_liquid_water_budget}.

\begin{figure}[t]
\centering
\includegraphics[width=0.48\textwidth]{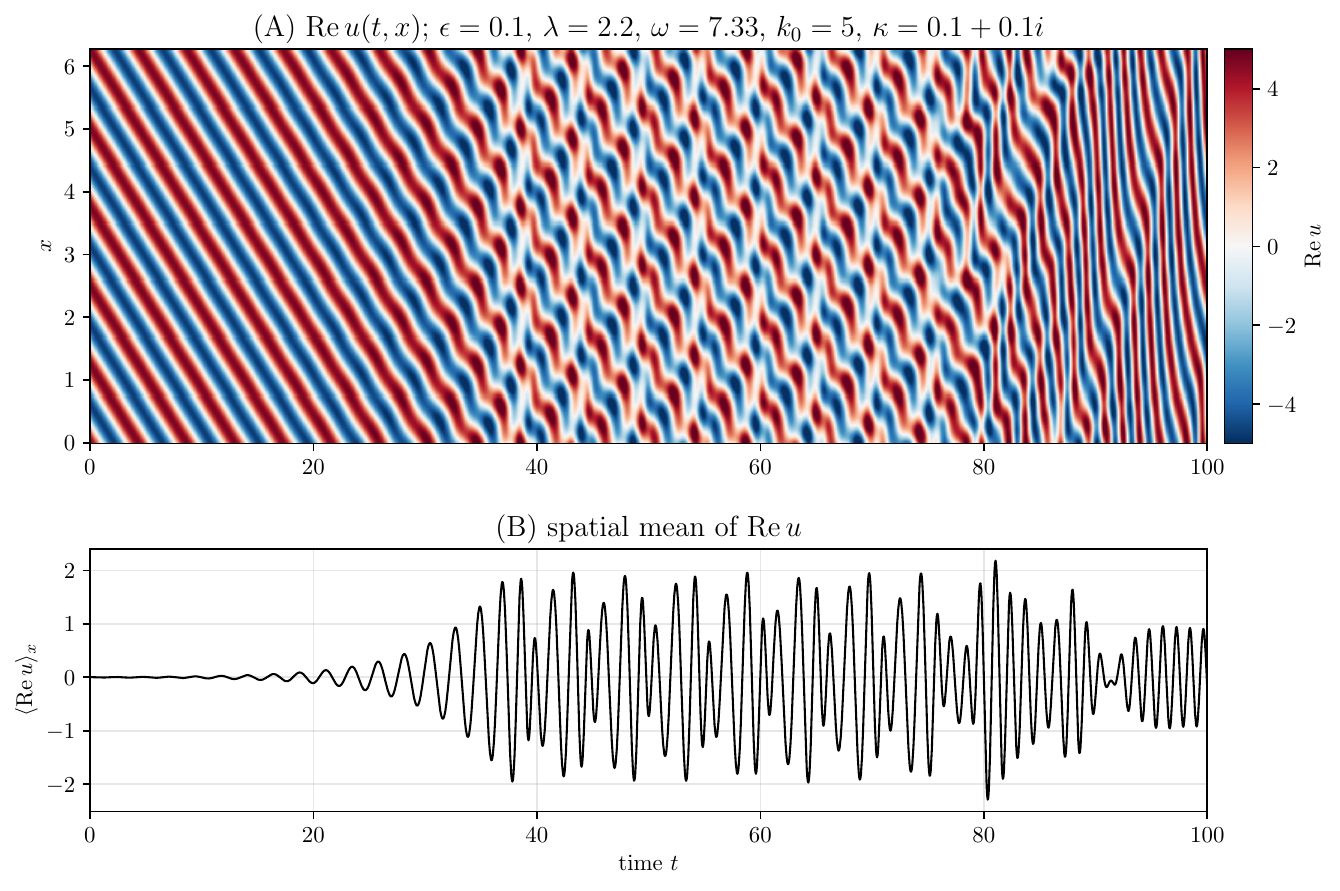}
\caption{{\bf Nonlinear transition from a destabilized traveling wave to a modulated state.} (A) Space--time evolution of $\operatorname{Re}u$ and (B) its spatial mean $\langle\operatorname{Re}u\rangle_x$ (the $\kvec=0$ Fourier mode), illustrating how linear sideband growth initiates wave breakup and nonlinear mode coupling reorganizes the field into a persistent modulated state (whose late-time spectrum is analyzed in Fig.~\ref{Fig_PSD}). Parameters are $\lambda=2.2$, $\epsilon=0.1$, $\omega=22/3$, $\kappa=0.1+0.1i$, $k_0=5$, and $L=2\pi$, initialized from $u=Ae^{ik_0x}[1+ae^{i5x}+\overline b e^{-i5x}]$ where $(a,b)$ is the growing eigenvector of $M_5$ in Eq.~\eqref{Eq_system_perturb} normalized to $\sqrt{|a|^2+|b|^2}=10^{-3}$. Integrated with $256$ grid points and forward Euler ($\Delta t=0.01$, without dealiasing); panels display $0\le t\le 100$ of a run to $t=2000$.}
\label{Fig_Space_Time}
\end{figure}

After the transient breakup ($t\gtrsim 30$), nonlinear saturation transfers power downscale to longer wavelengths, where competing persistent states coexist: depending on the transient perturbation and numerical timestep, both forward-Euler and fourth-order Runge--Kutta integrations settle onto either a nearly monochromatic $k=1$ traveling wave or the $k=2$-dominated modulated state shown in Fig.~\ref{Fig_Space_Time}. Remarkably, the linear sideband theory of Eq.~\eqref{Eq_system_perturb} accounts for this dichotomy directly: at these parameters the $k=1$ plane wave is linearly stable, whereas the $k=2$ plane wave is weakly unstable to its $q=1$ sidebands (with growth rate $0.015$), sustaining persistent temporal modulation around the saturated $k=2$ carrier.
The late-time power spectrum of $\operatorname{Re}u(\pi,t)$ over $100\le t\le 2000$ (Fig.~\ref{Fig_PSD}) provides a decisive demonstration of nonlinear frequency reorganization. Low-order combinations $|p\Omega_0+q\operatorname{Im}\sigma_5|$ ($|p|,|q|\le 3$) built from the linear carrier and initiating sideband frequencies match only one of the ten detected spectral peaks within the two-bin tolerance. By contrast, fitting just the three strongest peaks ($n=-1,0,1$) to the two-frequency combination law
\be\label{Eq_saturated_combination_tones}
\omega_n=|\Omega_c+n\Omega_m|
\ee
with $\Omega_c=3.6038$ and $\Omega_m=0.9074$ predicts all seven remaining peaks---including the folded low-frequency difference tone $|\Omega_c-4\Omega_m|\simeq 0.0258$ and the high-frequency sideband $\Omega_c+5\Omega_m\simeq 8.1408$---within $0.02$ frequency bins. This same combination-tone comb persists unchanged across the split windows $100\le t\le 1050$ and $1050\le t\le 2000$ (yielding $\Omega_m=0.9070$ and $0.9074$ with residuals below $0.12$ bins). Here the saturated carrier $\Omega_c=3.6038$ sits close to the exact $k=2$ plane-wave frequency $(\omega-\beta A^2)/(1+4\epsilon)=3.667$ (shifted slightly by finite-amplitude modulation and the $O(\Delta t)$ phase error of forward Euler), while the incommensurate ratio $\Omega_c/\Omega_m\simeq 3.972$ rules out simple $4{:}1$ frequency locking at the resolved precision and points to a two-frequency quasi-periodic state whose formal continuation as an invariant torus would require bifurcation tracking~\cite{guckenheimer2013nonlinear} not pursued here. Together, Figs.~\ref{Fig_Space_Time} and \ref{Fig_PSD} establish a clean separation between initiation and saturation: linear sideband instability explains why a regular cloud--rain wave breaks up, while nonlinear wave coupling selects a new, lower-wavenumber carrier and modulation frequency that organize the long-time variability into a tmulti-timescale  oscillation reminiscent of the coexisting fast wave-driven and slow precipitation-driven periods resolved in LES~\cite{dagan2018organization}.

\begin{figure}[t]
\centering
\includegraphics[width=0.48\textwidth]{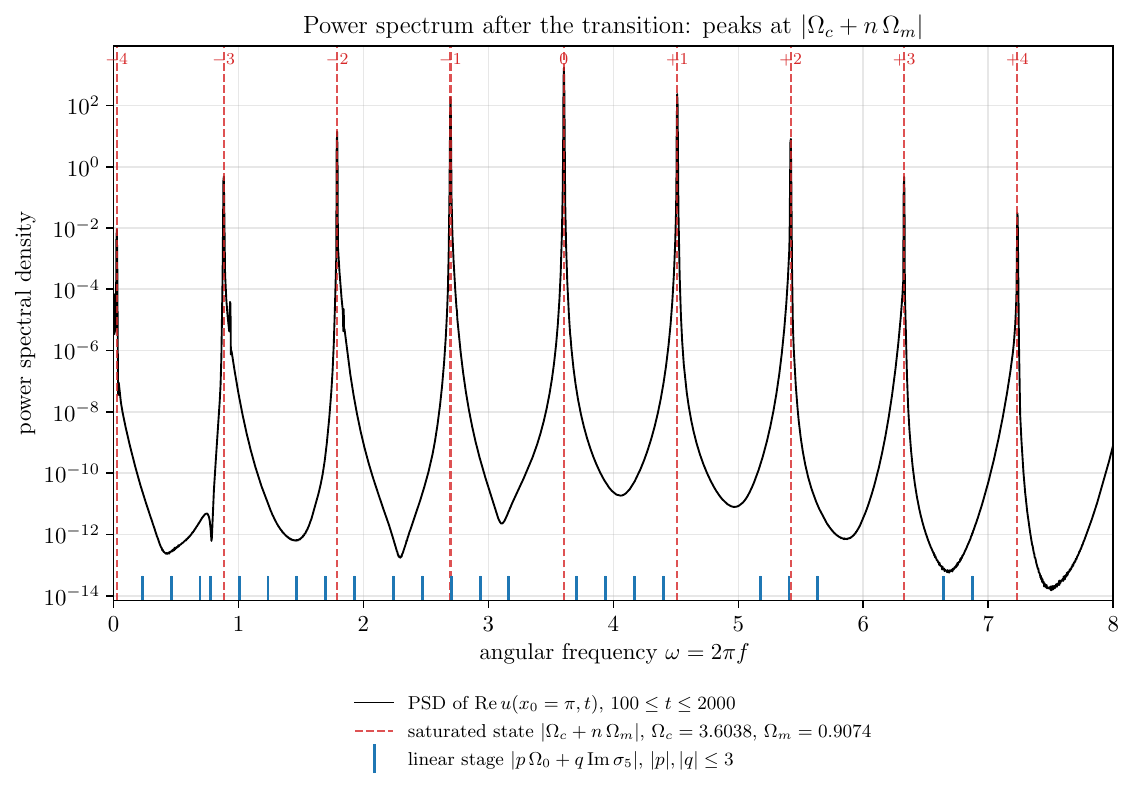}
\caption{{\bf Nonlinear frequency reorganization in the saturated modulated state.} Power spectral density of $\operatorname{Re}u(\pi,t)$ from the trajectory of Fig.~\ref{Fig_Space_Time} ($100\le t\le 2000$, sampled every $0.05$ after mean removal and Hann windowing; angular-frequency units). Red dashed lines show the two-frequency combination comb $|\Omega_c+n\Omega_m|$ ($n=-4,\ldots,4$) with $\Omega_c=3.6038$ and $\Omega_m=0.9074$ fitted solely to the three strongest peaks ($n=-1,0,1$); this single pair predicts all seven remaining detected peaks (including $n=5$ beyond the plotted range) within $0.02$ frequency bins. By contrast, low-order combinations $|p\Omega_0+q\operatorname{Im}\sigma_5|$ ($|p|,|q|\le 3$, blue ticks) of the linear carrier and seeded sideband frequencies match only one peak within the two-bin tolerance ($0.0066$), demonstrating that nonlinear saturation selects a new carrier--modulation pair rather than preserving the initiating frequencies. }
\label{Fig_PSD}
\end{figure}

 The contrast between the two frequency sets separates initiation from saturation. Linear sideband growth explains why the original carrier loses stability, whereas nonlinear evolution establishes a different frequency pair organizing the late signal. {The carrier $\Omega_c$ lies near the frequency $(\omega-\beta A^2)/(1+4\epsilon)=3.667$ of a $k=2$ plane wave, consistent with the $k=2$-dominated state of Fig.~\ref{Fig_Space_Time}. Its displacement from that value can reflect both the finite modulation and the $O(\Delta t)$ phase error of the forward-Euler integrator; their individual contributions have not been separated.} The ratio $\Omega_c/\Omega_m\simeq3.972$ excludes simple $4{:}1$ locking at the resolved precision and is consistent with a two-frequency quasiperiodic interpretation. The resolved evidence is the persistent combination-tone structure generated by nonlinear reorganization. Establishing an invariant two-frequency torus and its bifurcation type would additionally require continuation and the associated bifurcation conditions~\cite{guckenheimer2013nonlinear}. In the cloud--rain interpretation, the PGLE provides a route from a regular cycle to richer temporal variability through the spatial interaction of oscillations.

\subsection{Modulational instability of the homogeneous oscillation}
\label{Sec_PGLE_modulation}

The preceding analysis describes how sidebands destabilize a traveling wave that already carries a spatial phase gradient. Setting its carrier wavenumber $k_0$ to zero asks a complementary question: can spatial structure develop from a cloud--rain cycle whose amplitude and timing are initially uniform? This is a special case of the same stability problem, with $Q_+=Q_-$ in Eq.~\eqref{Eq_system_perturb}, but it makes the feedback between amplitude and phase particularly transparent. We develop this limit below in any spatial dimension and distinguish its oscillatory instability from classical Turing pattern formation \cite{turing1952morphogenesis,cross1993pattern}.

For $\lambda>0$ and $\alpha>0$, the PGLE admits
\be\label{Eq_uniform_cycle}
u_*(t)=A e^{i\Omega_*t},\qquad
A^2=\lambda/\alpha,\qquad \Omega_*=\omega-\beta A^2.
\ee
We assume $\Omega_*\ne0$. The amplitude returns toward $A$ after a uniform disturbance, whereas a uniform phase shift simply changes the timing of the same cycle. Spatial modulation asks whether neighboring regions can maintain that common timing and amplitude.

Write $u=e^{i\Omega_*t}(A+v)$ and resolve the real and imaginary parts of $v$ into spatial Fourier modes. For each mode, denote these coefficients by $a_{\kvec}$ and $b_{\kvec}$. To first order, $a_{\kvec}$ is the Fourier coefficient of the amplitude displacement, while $b_{\kvec}/A$ is that of the phase displacement relative to the uniform cycle. With $r_{\kvec}=\epsilon|\kvec|^2$ and $Q_{\kvec}=1+r_{\kvec}$, linearization gives
\be\label{Eq_uniform_modulation_matrix}
\partial_t\begin{pmatrix}a_{\kvec}\\b_{\kvec}\end{pmatrix}
=\frac{1}{Q_{\kvec}}
\begin{pmatrix}
-2\lambda & \Omega_*r_{\kvec}\\
-2\beta A^2-\Omega_*r_{\kvec} & 0
\end{pmatrix}
\begin{pmatrix}a_{\kvec}\\b_{\kvec}\end{pmatrix}.
\ee
The $-2\beta A^2$ term converts an amplitude disturbance into a timing disturbance. Screening supplies the reverse coupling, proportional to $\Omega_*r_{\kvec}$: a spatial phase modulation can then feed back on amplitude. This feedback is absent for the uniform mode, whose exponents are $-2\lambda$ and $0$.

For nonzero wavenumber the matrix has negative trace. An exponentially growing mode therefore occurs precisely when its determinant is negative:
\be\label{Eq_uniform_modulation_condition}
\Omega_*r_{\kvec}\bigl(2\beta A^2+\Omega_*r_{\kvec}\bigr)<0.
\ee
Equivalently, a band of unstable modes exists when
\be\label{Eq_uniform_modulation_band}
\beta\Omega_*<0,\qquad
0<\epsilon|\kvec|^2<-\frac{2\beta A^2}{\Omega_*},
\ee
provided the periodic domain contains a wavenumber in this interval. Appendix~\ref{App_PGLE_modulation} gives the exponents and their limiting behavior. The growth vanishes at both ends of the continuous band, so the fastest modulation has nonzero wavenumber. This wavelength selection concerns departures from the uniform cycle; the eventual pattern requires nonlinear saturation.

The spatial symmetry breaking is Turing-like, but the reference state and feedback mechanism differ from the classical diffusion-driven destabilization of a stationary equilibrium~\cite{turing1952morphogenesis,cross1993pattern}. Here the reference state is a sustained oscillation, and dynamic inertia couples its amplitude-dependent timing back to spatial amplitude variations. Neighboring regions can therefore lose synchrony even though a uniform amplitude disturbance decays. Equation~\eqref{Eq_uniform_modulation_band} identifies the spatial scales at which that feedback grows. This completes the single-carrier analysis of Sections~\ref{Sec_instab} and \ref{Sec_sidebands} by treating its spatially uniform limit. Section~\ref{Sec_sideband_instab} next considers amplification supplied by two distinct carriers, while Section~\ref{Sec_random_organization} examines growth from the zero-amplitude state, whose real modal rate $\lambda/Q_{\kvec}$ has the same sign at every wavenumber. These three starting states separate loss of synchrony, interaction between existing waves, and emergence from initially weak anomalies.

\section{Nonlinear Wave Interactions}\label{Sec_Nonlin_Wave_Interact}

Whereas Sec.~\ref{Sec_Instab_Neimark} perturbs a single exact traveling or uniform wave, a realistic cloud--rain field typically superposes several wave trains with distinct wavenumbers and orientations. Because the sum of two or more exact PGLE waves is no longer an exact solution, multi-wave fields introduce collective interactions absent from single-carrier stability. We show below that overlapping carriers immediately force new spatial and temporal harmonics at combination wavevectors $\kvec=\kvec_1+\kvec_2-\kvec_3$ through the cubic superposition residual $\mathcal R$ (Eqs.~\eqref{Eq_multimode_Fourier} and \eqref{Eq_wave_residual}), reshaping wave crests and damping peaks even when all modes are linearly stable. We also show that the cross-carrier product $2u_1u_2$ opens a mixed two-sideband amplification channel coupling perturbations with $\kvec_p+\kvec_q=\kvec_1+\kvec_2$, governed by an exact gain--detuning threshold (Eq.~\eqref{Eq_gain_detuning_threshold}). At the isolated-wave amplitude $A^2=\lambda/\alpha$, this cross-carrier channel can overcome damping only when the twist ratio exceeds the critical threshold $|\beta/\alpha|>\sqrt{5}/2$ (Eq.~\eqref{Eq_gain_damping_ratio}), while dynamic inertia selects which sideband pairs satisfy the resonance bandwidth in Eq.~\eqref{Eq_resonance_bandwidth}. Together, these results explain why weak-twist wave superpositions ($\beta/\alpha<\sqrt{5}/2$) either preserve ordered \emph{undulatus} bands or consolidate into parallel rolls, whereas strong-twist superpositions ($\beta/\alpha>\sqrt{5}/2$) trigger off-axis amplification and shatter into defect-rich \emph{asperitas}-like patterns in Sec.~\ref{Sec_modeling_results}.

\subsection{Mode coupling and phase matching}\label{Sec_Fourier_nonlinearity}

Using the orthonormal Fourier convention of Section~\ref{Sec_amplitude_balances}, the PGLE gives
\begin{widetext}
\be\label{Eq_multimode_Fourier}
Q_{\kvec}\,\partial_t\widehat u_{\kvec}
=(\lambda+i\omega)\widehat u_{\kvec}
-\frac{\kappa}{V}
\sum_{\kvec_1+\kvec_2-\kvec_3=\kvec}
\widehat u_{\kvec_1}\widehat u_{\kvec_2}
\overline{\widehat u_{\kvec_3}},
\qquad Q_{\kvec}=1+\epsilon|\kvec|^2,\qquad V=|\Torus|.
\ee
\end{widetext}
Three input components therefore generate an output at
$\kvec=\kvec_1+\kvec_2-\kvec_3$. If their phases evolve as
$e^{i\Omega_j t}$, the forcing has frequency
$\Omega_1+\Omega_2-\Omega_3$. These are Fourier matching relations inherited from the product $|u|^2u$.  They determine which spatial and temporal harmonics the cubic product generates. Their amplitudes evolve under the growth and nonlinear saturation terms in Eq.~\eqref{Eq_PGLE_energy}. Resonance refers to a generated forcing remaining close in frequency to the response it excites; off-resonant forcing is also present.

For two distinct carrier waves, write
\be\label{Eq_wave_carriers}
u_0=u_1+u_2,\qquad
u_j=Ae^{i(\kvec_j\cdot\x+\Omega_jt)},\qquad A>0.
\ee
Each wave separately solves the PGLE when
$A^2=\lambda/\alpha$ and
$\Omega_j=(\omega-\beta A^2)/Q_j$, with
$Q_j=1+\epsilon|\kvec_j|^2$ and $\lambda>0$.
Their sum generally does not. Its residual contains changes to the carrier amplitudes and new components at
$2\kvec_1-\kvec_2$ and $2\kvec_2-\kvec_1$
(Appendix~\ref{App_wave_interactions}). The residual therefore generates additional wavelengths directly, providing the forcing against which amplification is analyzed below.

\subsection{A two-sideband amplification mechanism}\label{Sec_sideband_instab}

 Sections~\ref{Sec_instab} and \ref{Sec_sidebands} linearized about a single exact traveling wave. Its two sidebands couple through the square of that carrier, and their growth initiates the transition examined in Figs.~\ref{Fig_Space_Time}--\ref{Fig_PSD}. We now ask how amplification changes when two distinct waves coexist. The cubic law contains a mixed product $2u_1u_2$ that couples perturbations whose wavevectors sum to $\kvec_1+\kvec_2$, with a driving frequency $\Omega_1+\Omega_2$. This opens an interaction channel associated with the pair of carriers. Because their superposition also generates new harmonics, we isolate this channel with prescribed carrier amplitudes and frequencies over the interval of interest. The resulting gain--detuning criterion explains when coupling to that background overcomes the damping of the retained perturbation pair.

To isolate one coupling channel, take sidebands
$\kvec_p=\kvec_1+\mathbf s$ and
$\kvec_q=\kvec_2-\mathbf s$, where $\mathbf s$ is an admissible modulation wavevector. Their sum equals that of the carriers. Retain the spatially uniform part of $2|u_0|^2$ and the mixed-carrier contribution $2u_1u_2$ in the linearized equation. This approximation neglects scattering by the carrier interference pattern and coupling through $u_1^2$ and $u_2^2$. It is useful when these other channels are weak or sufficiently detuned over the interval considered; the full coupling system is given in Appendix~\ref{App_wave_interactions}.

With sideband carrier frequencies $\nu_p,\nu_q$ satisfying
$\nu_p+\nu_q=\Omega_1+\Omega_2$, the amplitudes
$(a_p,\overline a_q)$ obey
\begin{widetext}
\be\label{Eq_wave_pair_matrix}
\frac{\d}{\d t}
\begin{pmatrix}a_p\\\overline a_q\end{pmatrix}
=M_{pq}\begin{pmatrix}a_p\\\overline a_q\end{pmatrix},
\qquad
M_{pq}=
\begin{pmatrix}
\Lambda/Q_p-i\nu_p & -2\kappa A^2/Q_p\\
-2\overline\kappa A^2/Q_q & \overline\Lambda/Q_q+i\nu_q
\end{pmatrix},
\qquad
\Lambda=\lambda+i\omega-4\kappa A^2=X_0+iY_0.
\ee
\end{widetext}
Here $X_0=\lambda-4\alpha A^2$ and
$Y_0=\omega-4\beta A^2$. The nonlinear correction to both diagonal entries is essential: the carriers modify the damping and frequency of the perturbations as well as coupling them. Both entries of the state vector share the same eigenvalue $\sigma$; the conjugate eigenproblem describes the complementary block.

For $X_0<0$, the uncoupled sidebands decay. The analytical calculation in \ref{App_wave_neutral_crossing} proves that the retained pair has an eigenvalue with positive real part if and only if
\begin{widetext}
\be\label{Eq_wave_pair_threshold}
\boxed{\quad
4|\kappa|^2A^4>
X_0^2+\left[Y_0-
\frac{Q_pQ_q}{Q_p+Q_q}(\Omega_1+\Omega_2)\right]^2.
\quad}
\ee
\end{widetext}
Equality gives neutral stability. The left-hand side measures the strength of the conjugate-sideband coupling; the right-hand side combines damping and frequency detuning. Screening enters the detuning and the growth rates through $Q_p$ and $Q_q$. The criterion is unchanged by how the fixed carrier-frequency sum is split between $\nu_p$ and $\nu_q$, and applies without modification when $Q_p=Q_q$: its derivation requires no division by their difference. Equation~\eqref{Eq_wave_pair_threshold} is the exact instability condition for the retained matrix~\eqref{Eq_wave_pair_matrix}. As an independent numerical check of the analytical result, it agrees with direct matrix eigenvalues across $10{,}000$ parameter checks, with equal screening also checked separately. The approximation lies in reducing the full coupling system to this pair; the criterion itself follows analytically from that reduced system.

For $X_0>0$, the real part of the matrix trace is $X_0(Q_p^{-1}+Q_q^{-1})>0$, so at least one sideband eigenmode grows for any coupling strength or detuning. For $X_0=0$, stability follows directly from the characteristic equation~\eqref{Eq_wave_characteristic}; \ref{App_wave_neutral_crossing} distinguishes these cases from the damped-sideband regime.

For carriers at the isolated-wave amplitude $A^2=\lambda/\alpha$, the damping is $X_0=-3\lambda$. A nonzero imaginary cubic coefficient can both strengthen the coupling and change its detuning; amplification requires the full inequality, not phase matching alone. In particular, $|\beta/\alpha|>\sqrt5/2$ is necessary for this retained pair to grow, although it is not sufficient. The parameter dependence and the nonsingular eigenvalue formulas are derived in Appendix~\ref{App_wave_interactions}.

\subsection{The Instability Condition: Detuning and Resonant Gain}
\label{Sec_detuning_resonant_gain}

The threshold in Eq.~\eqref{Eq_wave_pair_threshold} can be expressed as a competition between mutual sideband amplification, decay, and frequency mismatch. For $X_0<0$, define the uncoupled decay rates $d_p,d_q$, the coupling strength $g_{pq}$, and the detuning $\Delta_{pq}$ by
\begin{widetext}
\bea\label{Eq_sideband_gain_detuning}
d_p=-\frac{X_0}{Q_p}>0,\qquad
 d_q=-\frac{X_0}{Q_q}>0,\qquad
 g_{pq}=\frac{2|\kappa|A^2}{\sqrt{Q_pQ_q}},\\
\Delta_{pq}=Y_0\left(\frac{1}{Q_p}+\frac{1}{Q_q}\right)
-(\Omega_1+\Omega_2).
\eea
\end{widetext}
All four quantities have units of inverse time. The squared coupling $g_{pq}^2$ is the product of the off-diagonal coupling magnitudes in Eq.~\eqref{Eq_wave_pair_matrix}; $g_{pq}$ is not itself the exponential growth rate. The detuning compares the sum of the uncoupled sideband frequencies, shifted by the carrier background, with the carrier-frequency sum. The instability condition becomes
\be\label{Eq_gain_detuning_threshold}
\boxed{g_{pq}^{\,2}>d_pd_q
\left[1+\left(\frac{\Delta_{pq}}{d_p+d_q}\right)^2\right].}
\ee
Frequency mismatch therefore raises the coupling needed to overcome decay. At exact resonance, $\Delta_{pq}=0$, the threshold is lowest: $g_{pq}^{\,2}>d_pd_q$. The largest eigenvalue real part at resonance is
\be\label{Eq_resonant_growth}
\operatorname{Re}\sigma_+
=-\frac{d_p+d_q}{2}
+\sqrt{\frac{(d_p-d_q)^2}{4}+g_{pq}^{\,2}}.
\ee
For equal decay rates $d_p=d_q=d$, this reduces to $g_{pq}-d$, making the balance between amplification and decay explicit.

When $g_{pq}^{\,2}>d_pd_q$, Eq.~\eqref{Eq_gain_detuning_threshold} also defines a finite resonance bandwidth:
\be\label{Eq_resonance_bandwidth}
|\Delta_{pq}|<(d_p+d_q)
\sqrt{\frac{g_{pq}^{\,2}}{d_pd_q}-1}.
\ee
Exact frequency matching is unnecessary, but the mismatch must remain within this interval. Through $Q_p,Q_q$ and the carrier frequencies, screening makes the condition selective in wavenumber. Minimizing detuning lowers the threshold; identifying the fastest-growing pair additionally requires the eigenvalues, since its decay and coupling rates also depend on wavelength.

For any carrier amplitude in the damped-sideband regime ($X_0=\lambda-4\alpha A^2<0$, Eq.~\eqref{Eq_wave_pair_matrix}), substituting $d_p$, $d_q$, and $g_{pq}$ from Eq.~\eqref{Eq_sideband_gain_detuning} into the ratio $g_{pq}^{\,2}/(d_pd_q)$ gives
\be\label{Eq_gain_damping_general}
\frac{g_{pq}^{\,2}}{d_pd_q}
=\frac{4|\kappa|^2A^4}{X_0^2}
=\frac{4(\alpha A^2)^2}{(\lambda-4\alpha A^2)^2}
\left[1+\left(\frac{\beta}{\alpha}\right)^2\right],
\ee
because the factor $(Q_pQ_q)^{-1}$ enters both $g_{pq}^{\,2}$ and $d_pd_q$, and cancels out.

At fixed carrier amplitude, environmental screening therefore alters frequency compatibility and the amplification timescale rather than boosting coupling relative to decay. Specializing to the isolated-wave amplitude $A^2=\lambda/\alpha$ (for which $X_0=-3\lambda$), Eq.~\eqref{Eq_gain_damping_general} reduces to
\be\label{Eq_gain_damping_ratio}
\frac{g_{pq}^{\,2}}{d_pd_q}
=\frac{4}{9}\left[1+\left(\frac{\beta}{\alpha}\right)^2\right].
\ee
Because $d_p,d_q>0$ and $\bigl(\Delta_{pq}/(d_p+d_q)\bigr)^2\ge 0$ in Eq.~\eqref{Eq_gain_detuning_threshold}, instability requires $g_{pq}^{\,2}/(d_pd_q)>1$ as a necessary condition, attained at exact resonance $\Delta_{pq}=0$. Substituting Eq.~\eqref{Eq_gain_damping_ratio} gives then $\frac{4}{9}[1+(\beta/\alpha)^2]>1$, recovering the threshold $|\beta/\alpha|>\sqrt{5}/2$ of Sec.~\ref{Sec_sideband_instab}. Because $\beta$ simultaneously strengthens the coupling and shifts the detuning $\Delta_{pq}$, increasing $|\beta/\alpha|$ above $\sqrt{5}/2$ need not increase growth monotonically, with detuning deciding whether the available coupling actually produces instability.

In the cloud--rain interpretation, this mechanism allows coupled oscillations to amplify a spatial modulation when their nonlinear interaction maintains sufficiently compatible phases despite decay. It describes resonant amplification within the retained two-sideband approximation; the full evolution also involves the carrier residual and the other coupling channels identified in Appendix~\ref{App_wave_interactions}.

\subsection{Interpretation for cloud--rain organization}
\label{Sec_wave_cloud_interpretation}

Translated into the cloud--rain observables of Sec.~\ref{Sec_PGLE_interpretation}, each spatial Fourier mode $\kvec$ represents a propagating wave of cloud-water thickening ($\rho=\operatorname{Re}u$) and phase-lagged rain-water variability ($q=\operatorname{Im}u$) whose phase velocity sets how rapidly the condensation--precipitation--recovery cycle sweeps across the cloud layer. The analysis of Secs.~\ref{Sec_Fourier_nonlinearity}--\ref{Sec_detuning_resonant_gain} shows that when multiple such wave trains coexist, the cubic feedback $F(u)$ and the screened operator $(I-\epsilon\Delta)\partial_t$ act together as a selective spatiotemporal filter that governs whether the cloud field remains ordered or breaks into complex two-dimensional structure.

In particular, the gain--detuning threshold in Eq.~\eqref{Eq_gain_detuning_threshold} imposes a dual condition---one energetic, the other temporal---for two coexisting cloud--rain patterns $(\kvec_1,\Omega_1)$ and $(\kvec_2,\Omega_2)$ to amplify an off-axis modulation pair $(\kvec_p,\kvec_q)$ with $\kvec_p+\kvec_q=\kvec_1+\kvec_2$. First, at the isolated-wave amplitude $A^2=\lambda/\alpha$, Eq.~\eqref{Eq_gain_damping_ratio} requires the twist ratio to exceed $|\beta/\alpha|>\sqrt{5}/2$: without sufficiently strong amplitude-dependent frequency shifts, the conservative Hamiltonian coupling (Sec.~\ref{Sec_Noether_balance}) cannot overcome linear amplitude damping even at exact resonance. Second, even when $|\beta/\alpha|>\sqrt{5}/2$, the screened frequency mismatch $\Delta_{pq}$ in Eq.~\eqref{Eq_sideband_gain_detuning} must remain within the finite resonance bandwidth of Eq.~\eqref{Eq_resonance_bandwidth}. Physically, if the background-shifted oscillation frequencies of the sidebands drift too rapidly relative to $\Omega_1+\Omega_2$, the cross-carrier phase slips before net amplification can occur. Dynamic inertia enters this temporal selection through the screening factors $Q_{\kvec}=1+\epsilon|\kvec|^2$,  albeit in an indirect fashion.  Although $Q_p$ and $Q_q$ have been canceled out to form the two-carrier gain-to-damping ratio in Eq.~\eqref{Eq_gain_damping_ratio}, they shift both the modal frequencies in $\Delta_{pq}$ (Eq.~\eqref{Eq_sideband_gain_detuning}) and the growth timescale in Eq.~\eqref{Eq_resonant_growth}, thereby tuning which spatial wavelengths and orientations fall inside the amplification band.

Taken together, these results separate two complementary routes to cloud-field reorganization across the parameter regimes numerically explored in Sec.~\ref{Sec_modeling_results} below. Whenever $|\beta/\alpha|<\sqrt{5}/2$, the retained two-sideband amplification channel is closed; reorganization proceeds purely through the deterministic superposition residual $\mathcal R$ (Eq.~\eqref{Eq_wave_residual}) and full Fourier mode coupling (Eq.~\eqref{Eq_multimode_Fourier}), which reshape wave crests, saturate peak excursions, and can either preserve coherent \emph{undulatus} bands ($\beta/\alpha=0.25$ in Sec.~\ref{Sec_banded_solutions}) or consolidate a cellular interference pattern into parallel convective rolls ($\beta/\alpha=1$ in Sec.~\ref{Sec_spontaneous_stabilization}). Once $|\beta/\alpha|>\sqrt{5}/2$, however, resonant two-sideband amplification unlocks alongside residual harmonic forcing, enabling compatible off-axis modulations to grow exponentially and drive the transition from ordered bands to defect-rich \emph{asperitas}-like wave networks ($\beta/\alpha=2$ in Secs.~\ref{Sec_complex_patterns}--\ref{Sec_defect_analysis}). When more than two carriers are present, pairwise channels share modes and interact simultaneously with carrier depletion and higher harmonics; the full Fourier system in Eq.~\eqref{Eq_multimode_Fourier} (and Appendix~\ref{App_wave_interactions}) governs that joint nonlinear evolution, while Eq.~\eqref{Eq_wave_pair_threshold} identifies the resonant instability channels that initiate it.
\section{Modeling Results}\label{Sec_modeling_results}

We now test how the analytical mechanisms established in Secs.~\ref{Sec_dispersion_nonlinear}--\ref{Sec_Nonlin_Wave_Interact}---dynamic inertia, the Hamiltonian--gradient decomposition of the Noether charge, phase-twist sensitivity, and resonant sideband coupling---organize two-dimensional cloud--rain fields beyond the linear regime. The simulations integrate the deterministic PGLE, Eq.~\eqref{Eq_PGLE}, on a periodic square domain $\Torus=[0,L)^2$ of side $L=20$ (with parameters, coordinates, and time reported in dimensionless model units, and cloud photographs used for morphological comparison rather than dimensional calibration). Throughout the experiments, $\rho=\operatorname{Re}u$ represents the signed cloud-water path anomaly, whereas $|u|$ measures the joint cloud--rain oscillation amplitude (Sec.~\ref{Sec_PGLE_interpretation}). We examine three fundamental pathways of cloud-field organization: the twist-driven transformation of coherent \emph{undulatus}-like bands into defect-rich \emph{asperitas}-like wave networks (Secs.~\ref{Sec_banded_solutions}--\ref{Sec_defect_analysis}), the spontaneous consolidation of cellular interference patterns into parallel convective rolls (Sec.~\ref{Sec_spontaneous_stabilization}), and the emergence of cellular and branching actinoform-like structures from unstructured random anomalies (Sec.~\ref{Sec_random_organization}).

\subsection{Undulatus-like banded solutions}\label{Sec_banded_solutions}

In cloud morphology, the variety \emph{undulatus} denotes quasi-regular wave bands across a cloud layer, whereas \emph{asperitas} designates a strongly convoluted, locally roughened cloud base~\cite{wmo2017undulatus,wmo2017asperitas,harrison2017asperitas}. While atmospheric realizations of asperitas can arise from hydro-thermodynamic instabilities such as droplet settling and evaporation in sheared layers~\cite{ravichandran2022asperitas}, the PGLE isolates an intrinsic oscillatory mechanism that connects regular bands to convoluted wave patterns as the amplitude--frequency twist $\beta=\operatorname{Im}\kappa$ increases. To probe this transition, we initialize the field with a superposition of $n$ nonlinear plane waves,
\bea\label{eq_combined_wave}
S(\x,t)&=\sum_{j=1}^{n}A e^{i(\kvec_j\cdot\x+\Omega_jt)},
\qquad A^2=\frac{\lambda}{\alpha},\\
\Omega_j&=\frac{\omega-\beta A^2}{1+\epsilon|\kvec_j|^2},
\qquad u(\x,0)=S(\x,0).
\eea
Each summand is an exact monochromatic solution of the PGLE with saturated amplitude $A$ and twist-shifted frequency $\Omega_j$ (Eq.~\eqref{Eq_u_plane_asymp}), so the linear sum $S(\x,t)$ serves as an analytical reference that tracks independent carrier propagation while omitting nonlinear mode coupling. We initialize four carriers with the wavevectors listed in Table~\ref{Table_wave_vectors} at fixed growth, base frequency, screening, and saturation parameters ($\lambda=5$, $\omega=3$, $\epsilon=0.1$, $\alpha=10$) and vary the twist parameter $\beta$.

\begin{table}[tb]
\centering
\caption{Wavevectors of the four-wave ($n=4$) initial condition in Eq.~\eqref{eq_combined_wave}. Here $a=2\pi/L$ is the fundamental wavenumber and $L=20$.}
\label{Table_wave_vectors}
\begin{tabular}{ll}
\toprule
Wavevector & Components\\
\midrule
$\kvec_1$ & $a(10,-18)$\\
$\kvec_2$ & $a(12,-18)$\\
$\kvec_3$ & $a(14,-14)$\\
$\kvec_4$ & $a(16,-12)$\\
\bottomrule
\end{tabular}
\end{table}

At weak twist ($\beta=2.5$), the PGLE solution preserves the diagonal orientation and large-scale envelope of the analytical reference $S(\x,t)$ while sharpening the individual crests and reducing their peak amplitude (Fig.~\ref{Fig_undulatus_vs_analytic} and Fig.~\ref{Fig_CS_PGLE_complex_kappa}C,D). At strong twist ($\beta=20$), the identical initial wave superposition breaks up into a labyrinthine network of bent crests, line terminations, and intersecting wave ridges reminiscent of asperitas (Fig.~\ref{Fig_CS_PGLE_complex_kappa}A,B). 

The Hamiltonian--gradient structure established in Sec.~\ref{Sec_Noether_balance} (Eq.~\eqref{Eq_PGLE_charge_forcing}) explains why this comparison cleanly isolates the role of phase twist: because $\omega$ and $\beta$ reside entirely in the conservative Hamiltonian $H[u]$ (Eq.~\eqref{Eq_PGLE_Hamiltonian}), they drop out of the global production--dissipation law for the Noether charge $\mathcal{N}_\epsilon=\mathcal E_\epsilon$ in Eq.~\eqref{Eq_PGLE_energy}. Varying $\beta$ from $2.5$ to $20$ at fixed $(\lambda,\alpha,\epsilon)$ leaves the Landau potential $V[u]$ (Eq.~\eqref{Eq_PGLE_Landau_potential}) and the isolated-wave saturation amplitude $A=\sqrt{\lambda/\alpha}=1/\sqrt{2}$ completely unchanged. Thus, the morphological transition in Fig.~\ref{Fig_CS_PGLE_complex_kappa}B-C is therefore driven by conservative Hamiltonian phase-shearing $J_\epsilon\operatorname{grad}H[u]$ that redistributes the Noether charge $\mathcal{N}_\epsilon=\mathcal E_\epsilon$ from bulk oscillation amplitude $\frac{1}{2}\int_{\Torus}|u|^2\,\d\x$ into fine-scale spatial gradients $\frac{\epsilon}{2}\int_{\Torus}|\nabla u|^2\,\d\x$.

\begin{figure}[tb]
\centering
\includegraphics[width=\columnwidth]{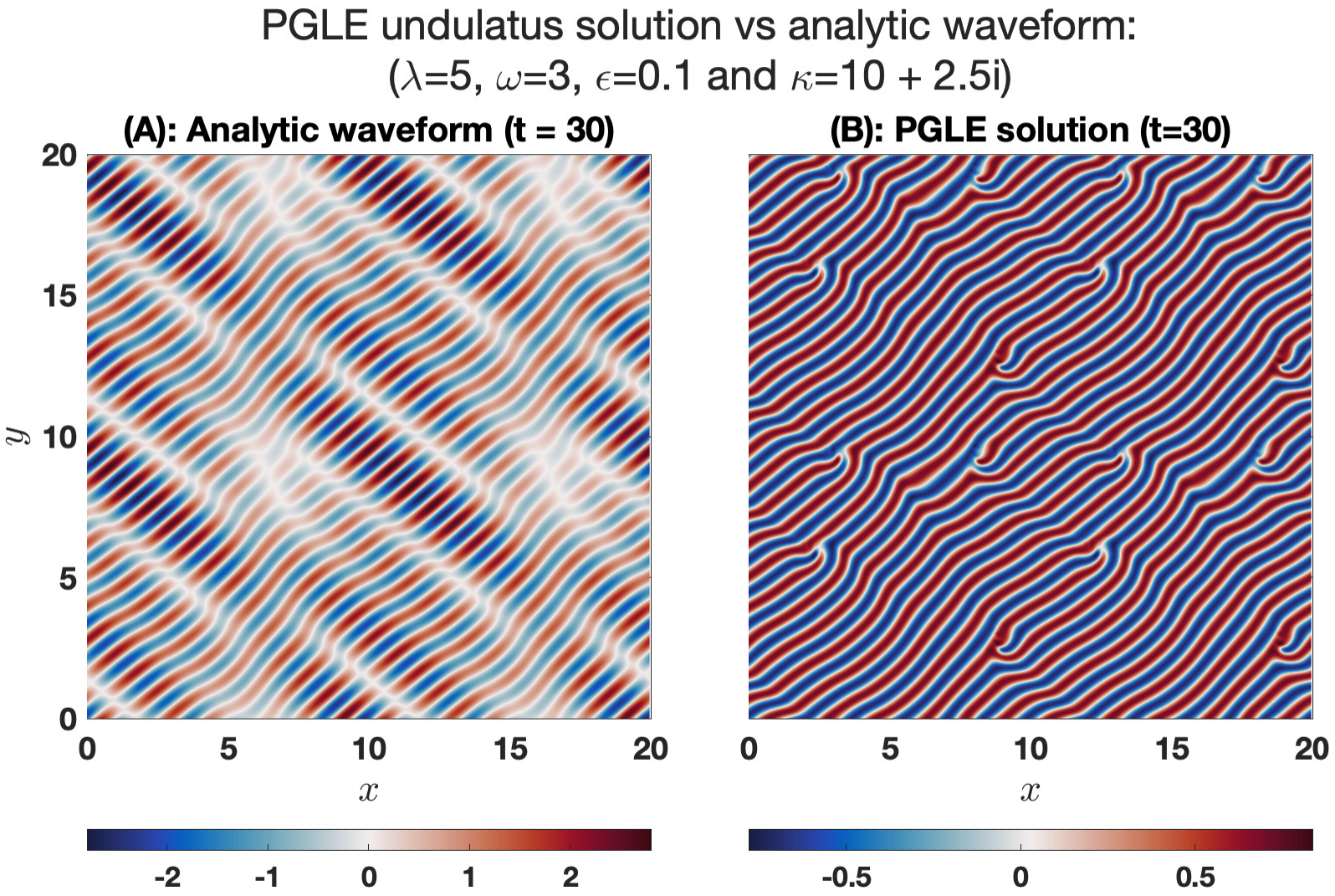}
\caption{{\bf Persistence of banded organization under nonlinear evolution.} Real parts of (A) the independently evolving wave superposition $S(\x,t)$ and (B) the PGLE solution at $t=30$, initialized from $S(\x,0)$. The four wavevectors are listed in Table~\ref{Table_wave_vectors}; $\lambda=5$, $\omega=3$, $\epsilon=0.1$, $\kappa=10+2.5i$, and $L=20$. The separate color scales show the amplitude reduction accompanying nonlinear saturation.}
\label{Fig_undulatus_vs_analytic}
\end{figure}
\begin{figure*}[tb]
\centering
\includegraphics[width=\textwidth]{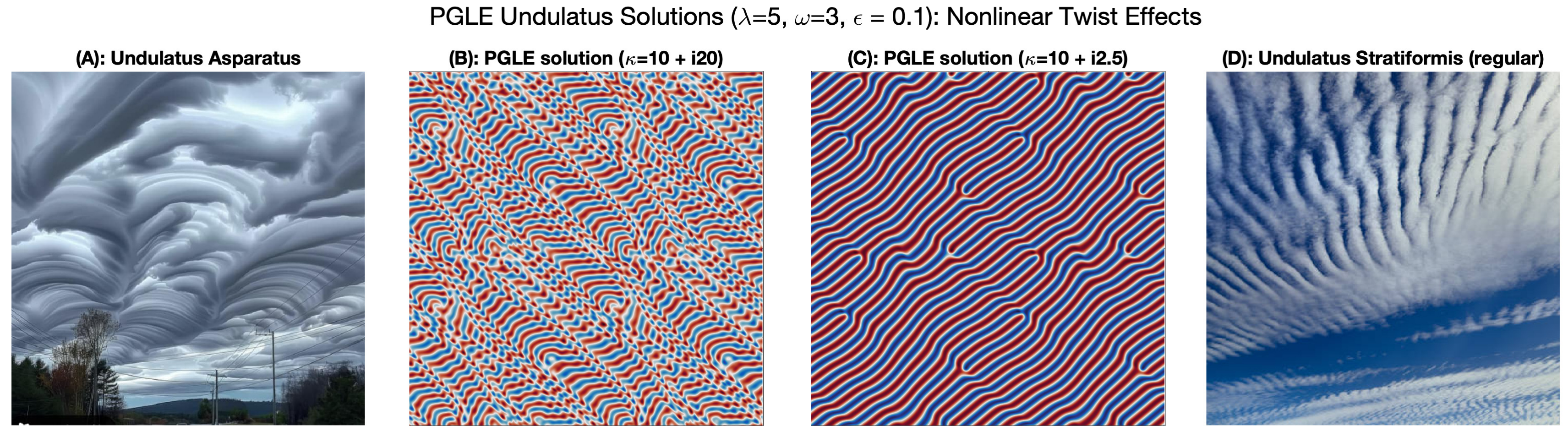}
\caption{{\bf Regular and irregular wave organization.} Cloud photographs illustrate (A) asperitas over Gorham, New Hampshire, in June 2024, and (D) regular undulatus bands. Panels (B,C) show the real part of the PGLE field for $\kappa=10+20i$ and $10+2.5i$, respectively, from the same four-wave initial pattern. Both runs use $\lambda=5$, $\omega=3$, $\epsilon=0.1$, and $L=20$. {The comparison highlights band interruptions within ordered undulatus (C,D) and the intricate wave geometry of the irregular state (A,B). In the model, phase defects identify band terminations, while stronger amplitude-dependent timing shifts disrupt the ordered pattern. Panels (B,C) display water-path anomalies, providing a dynamical interpretation of the photographed morphology.}}
\label{Fig_CS_PGLE_complex_kappa}
\end{figure*}

Table~\ref{Table_Corr_coef} quantifies the progressive loss of coherence across intermediate twist values through the spatial correlation between $\operatorname{Re}u(\x,t)$ and $\operatorname{Re}S(\x,t)$. Because $S(\x,t)$ already incorporates the single-carrier frequency shift $\Omega_j(\beta)$, any decay of correlation measures the cumulative effect of cross-carrier interactions in Eq.~\eqref{Eq_multimode_Fourier}. For $\beta=2.5$, the nonlinear field remains strongly locked to the reference ($0.76$ at $t=50$), whereas increasing $\beta$ accelerates decorrelation until, at $\beta=20$, memory of the independent-wave pattern is lost before $t=20$.

\begin{table*}[tb]
\centering
\caption{Spatial correlation between $\operatorname{Re}S(\x,t)$ and $\operatorname{Re}u(\x,t)$ at matched times. The analytic reference is defined by Eq.~\eqref{eq_combined_wave}, and every run starts from $u(\x,0)=S(\x,0)$. Parameters are $\epsilon=0.1$, $\omega=3$, $\lambda=5$, and $L=20$. Each coefficient compares the two spatial fields at the same time, measuring their departure from independent carrier evolution.}
\label{Table_Corr_coef}
\begin{tabular}{ccccccc}
\toprule
$\kappa$ & $t=5$ & $t=10$ & $t=20$ & $t=30$ & $t=40$ & $t=50$\\
\midrule
$10+2.5i$ & $0.88$ & $0.88$ & $0.86$ & $0.84$ & $0.80$ & $0.76$\\
$10+7.5i$ & $0.72$ & $0.71$ & $0.68$ & $0.65$ & $0.60$ & $0.56$\\
$10+10i$ & $0.58$ & $0.53$ & $0.42$ & $0.35$ & $0.18$ & $-0.059$\\
$10+12.5i$ & $0.42$ & $0.34$ & $0.27$ & $-0.028$ & $-0.041$ & $-0.026$\\
$10+15i$ & $0.26$ & $0.29$ & $0.022$ & $0.0022$ & $-0.0038$ & $-0.0063$\\
$10+20i$ & $0.21$ & $0.07$ & $-0.014$ & $0.0062$ & $0.0099$ & $0.00073$\\
\bottomrule
\end{tabular}
\end{table*}

\subsection{From Plane Waves to Complex Patterns}\label{Sec_complex_patterns}

The wave-interaction theory of Sec.~\ref{Sec_Nonlin_Wave_Interact} distinguishes the two dynamical mechanisms responsible for the progression in Table~\ref{Table_Corr_coef}. The first mechanism is direct harmonic forcing by the superposition residual $\mathcal R$ in Eq.~\eqref{Eq_wave_residual}, which operates at every value of $\beta$: wherever the four carriers overlap, the cubic product $|u|^2u$ excites combination wavevectors $\kvec=\kvec_1+\kvec_2-\kvec_3$ (Eq.~\eqref{Eq_multimode_Fourier}) and damps the peak excursion of the four-wave sum toward the saturation level $R_*$, accounting for the amplitude reduction and crest sharpening seen at $\beta=2.5$ in Fig.~\ref{Fig_undulatus_vs_analytic}. As $\beta$ increases to intermediate values such as $\beta=10$---where the two-carrier gain-to-damping ratio in Eq.~\eqref{Eq_gain_damping_ratio} remains subcritical, $g_{pq}^2/(d_pd_q)=8/9<1$---this residual forcing combines with stronger amplitude-induced phase shifts to gradually dephase the carriers and drive the correlation in Table~\ref{Table_Corr_coef} to zero by $t=50$.

The second mechanism is resonant two-sideband amplification (Sec.~\ref{Sec_detuning_resonant_gain}), which opens once the twist ratio exceeds $|\beta/\alpha|>\sqrt{5}/2\approx 1.12$. While $\kappa=10+2.5i$ yields $g_{pq}^2/(d_pd_q)=17/36<1$ and precludes exponential growth of the retained sideband pair, $\kappa=10+20i$ yields $g_{pq}^2/(d_pd_q)=20/9>1$, allowing any sideband pair whose screened frequency mismatch $\Delta_{pq}$ falls within the resonance bandwidth of Eq.~\eqref{Eq_resonance_bandwidth} to amplify exponentially against linear damping (Eq.~\eqref{Eq_gain_detuning_threshold}). In physical terms, strong amplitude--frequency coupling enables overlapping cloud--rain wave trains to mutually reinforce off-axis modulations whenever dynamic inertia places their frequencies in near-resonance, providing a collective instability route that parallels the interplay of precipitation cycles and gravity-wave modulation in cloud-resolving studies~\cite{feingold_2010,dagan2018organization}.

\subsection{Topological defects}\label{Sec_topo_defects}

A striking structural feature of both the ordered undulatus state ($\beta=2.5$, Fig.~\ref{Fig_CS_PGLE_complex_kappa}C,D) and the disrupted asperitas state ($\beta=20$, Fig.~\ref{Fig_CS_PGLE_complex_kappa}A,B) is the presence of localized band terminations and Y-shaped dislocations. Writing $u=R e^{i\vartheta}$ with amplitude $R=|u|$ and cycle phase $\vartheta=\arg u$, we distinguish a genuine topological dislocation from an ordinary cloud-water trough by evaluating the phase winding number around a positively oriented closed contour $\mathcal C$ avoiding zeros of $u$~\cite{shraiman1992spatiotemporal,aranson2002world}:
\be\label{Eq_winding_nu}
\nu=\frac{1}{2\pi}\oint_{\mathcal C}\nabla\vartheta\cdot\d\boldsymbol\ell\in\mathbb Z.
\ee
A nonzero integer charge $\nu=\pm 1$ means that the cloud--rain cycle advances or retards by one full period around $\mathcal C$, which by continuity requires the oscillation amplitude to vanish ($R=0$, so both water-path anomalies return to their reference values $\rho=q=0$) at a defect core inside $\mathcal C$. On the periodic torus $\Torus$, the net topological charge vanishes, so defects are created and annihilated in oppositely charged ($\pm 1$) pairs.

For $\kappa=10+2.5i$, Fig.~\ref{Fig_topo_defect1} shows that the isolated band breaks within the otherwise ordered undulatus pattern correspond precisely to $\nu=\pm 1$ phase singularities paired with sharp point-like minima of $|u|$. Once formed, the motion of such a defect core is constrained by dynamic inertia: for an idealized one-dimensional depression moving at speed $c$, substituting $u(x,t)=e^{i\Omega t}U(x-ct)$ into Eq.~\eqref{Eq_PGLE} yields the third-order profile equation
\be\label{Eq_main_traveling_profile}
(i\Omega-c\partial_z)(I-\epsilon\partial_z^2)U
=(\lambda+i\omega)U-\kappa|U|^2U,
\ee
expanded in \ref{App_moving_defects}, which requires any translating defect core to co-adjust its surrounding screened wave envelope. At $\kappa=10+20i$, by contrast, Fig.~\ref{Fig_topo_defect2} shows that these isolated point cores give way to a dense web of deep amplitude troughs (Fig.~\ref{Fig_topo_defect2}B) bordered by intense filaments of the cloud-anomaly gradient $|\nabla\operatorname{Re}u|$ (Fig.~\ref{Fig_topo_defect2}A).

\begin{figure}[tb]
\centering
\includegraphics[width=\columnwidth]{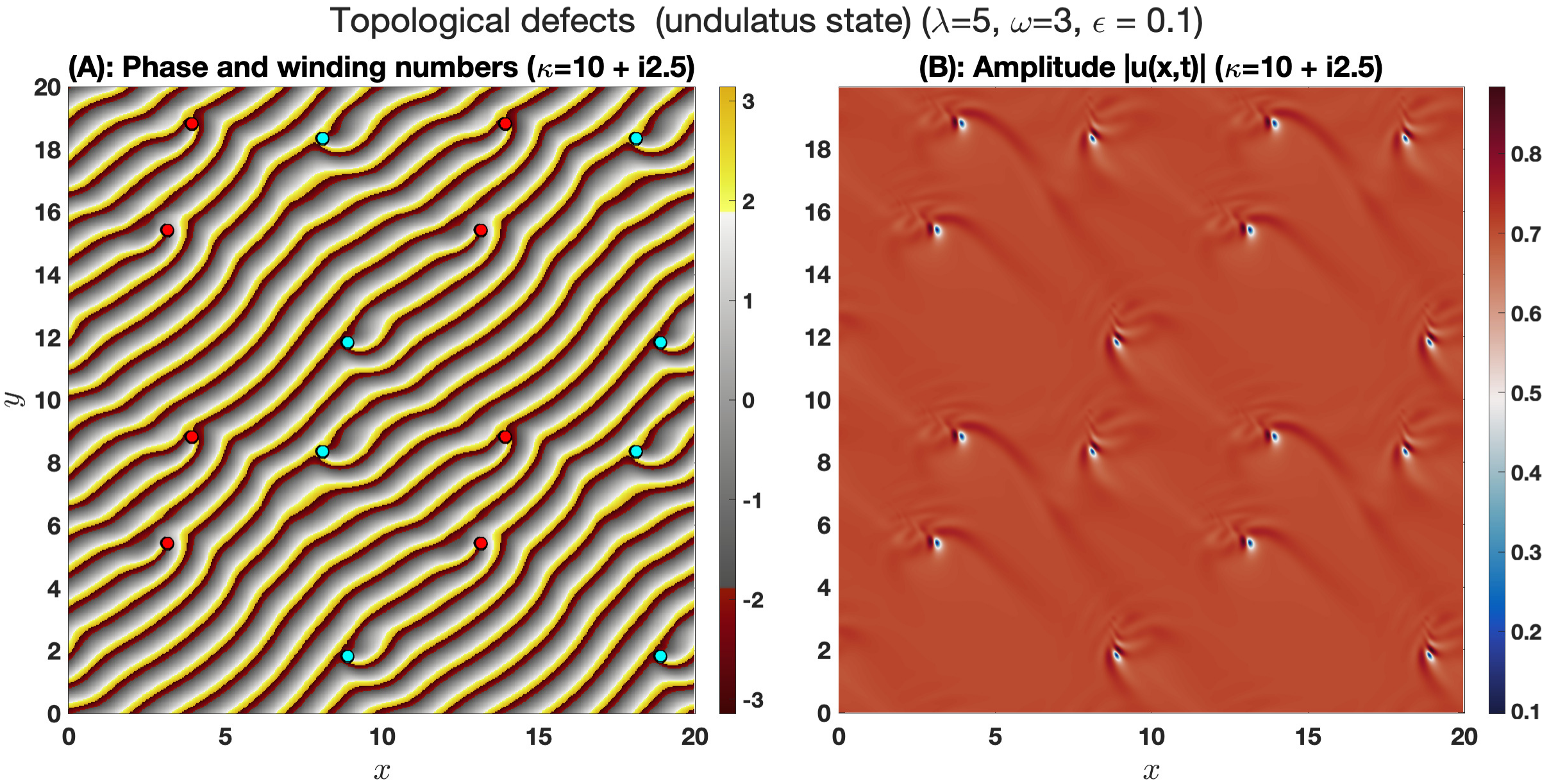}
\caption{{\bf Phase defects within an ordered banded state.} (A) Phase field with defects of charge $+1$ (red) and $-1$ (cyan), identified by the winding number in Eq.~\eqref{Eq_winding_nu}. (B) Corresponding amplitude $|u|$, with localized minima near the phase singularities. Parameters are $\lambda=5$, $\omega=3$, $\epsilon=0.1$, $\kappa=10+2.5i$, and $L=20$. Defects interrupt individual bands while preserving overall orientation in this $\beta=2.5$ run; Figs.~\ref{Fig_topo_defect2} and \ref{Fig_defect_count} describe the separate evolution at $\beta=20$.}
\label{Fig_topo_defect1}
\end{figure}
\begin{figure}[tb]
\centering
\includegraphics[width=\columnwidth]{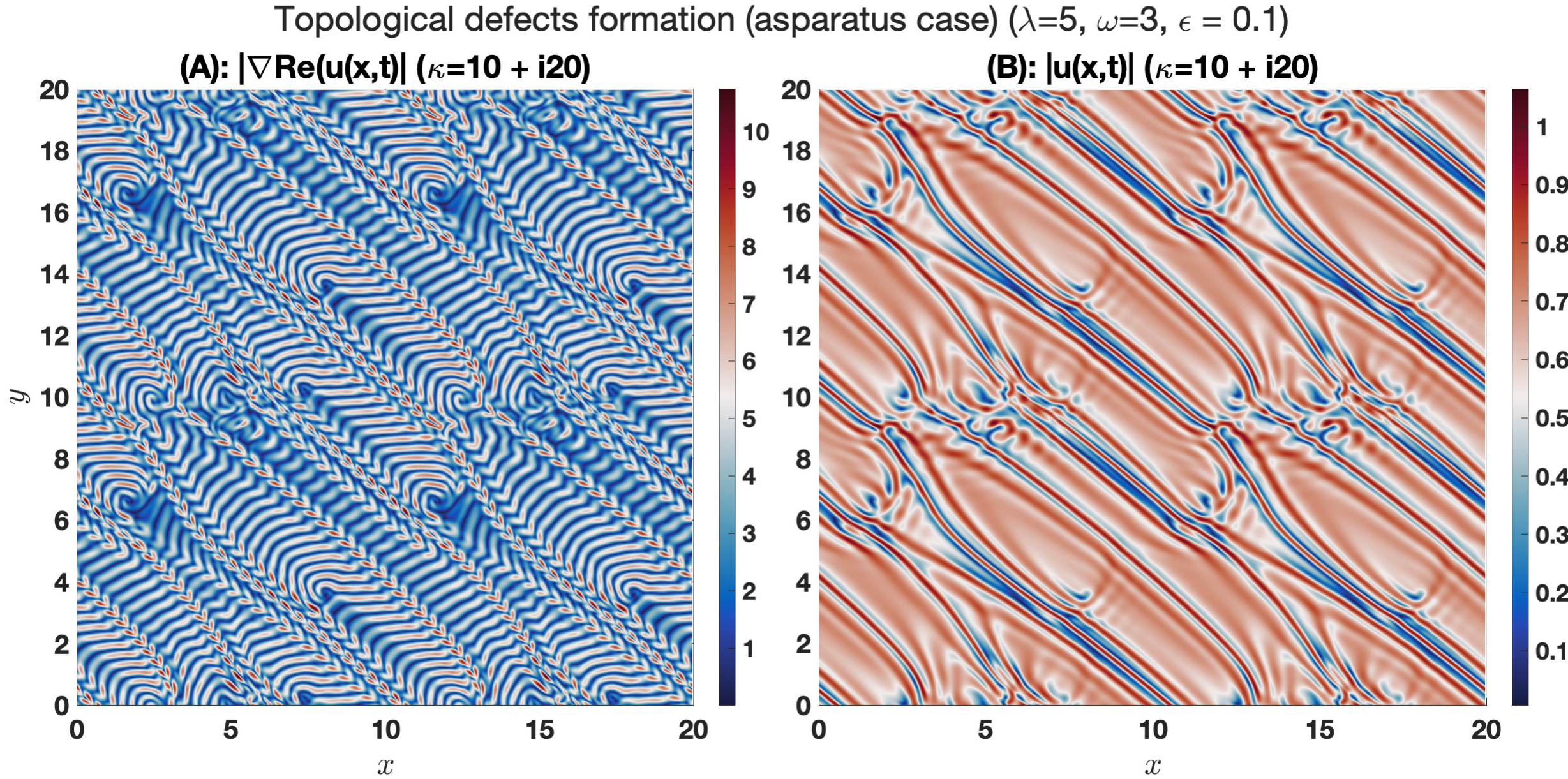}
\caption{{\bf Fine structure in the irregular wave state.} (A) Magnitude of the cloud-anomaly gradient, $|\nabla\operatorname{Re}u|$, and (B) joint oscillation amplitude $|u|$ at $t=9$ for $\kappa=10+20i$. The remaining parameters are $\lambda=5$, $\omega=3$, $\epsilon=0.1$, and $L=20$, as in Fig.~\ref{Fig_topo_defect1}, whose run uses $\beta=2.5$. Sharp contrasts and numerous amplitude depressions accompany the disrupted bands at $\beta=20$. Panel (B) is the same amplitude snapshot as Fig.~\ref{Fig_defect_count}B; that figure traces the evolution leading to this state through an earlier snapshot and the phase-defect count defined by Eq.~\eqref{Eq_winding_nu}.}
\label{Fig_topo_defect2}
\end{figure}

\subsection{Amplitude and phase near defect cores}\label{Sec_defect_analysis}

The screened polar equations derived in Appendix \ref{App_phase_geometry} (Eq.~\eqref{Eq_exact_screened_polar}) explain why increasing $\beta$ converts smooth wave interference into the sharp gradient filaments and defect cores of Fig.~\ref{Fig_topo_defect2}. In Eq.~\eqref{Eq_exact_screened_polar}, the local reaction tendency advances the phase at rate $\omega-\beta R^2$, whose spatial gradient is
\be\label{Eq_reaction_frequency_gradient}
\nabla(\omega-\beta R^2)=-2\beta R\nabla R.
\ee
Wherever carrier interference creates a spatial amplitude contrast between neighboring cloud elements ($R_1\neq R_2$), Eq.~\eqref{Eq_reaction_frequency_gradient} generates a differential phase drift $-\beta(R_2^2-R_1^2)$ that winds the phase surfaces and steepens $|\nabla u|$. From the perspective of the isochron geometry in \ref{App_twist_isochrons} (Eqs.~\eqref{Eq_PGLE_isochron} and \eqref{Eq_screened_isochron_response}), raising $\beta$ from $2.5$ to $20$ at fixed $\alpha=10$ increases the isochron twist parameter $\beta/\alpha$ eightfold, from $0.25$ to $2$: a given fractional amplitude disturbance $R_2/R_1$ therefore induces an eightfold larger asymptotic phase shift $-(\beta/\alpha)\log(R_2/R_1)$, amplifying the same shear-induced phase sensitivity that drives irregular open-cell oscillations in the nonspatial KTF delay model~\cite{guckenheimer1975isochrons,Tantet_al_Hopf,Chekroun_al22SciAdv}. In the spatial field, as this differential phase winding steepens $\nabla\vartheta$, the nonlocal screening response $\mathcal H=e^{-i\vartheta}(\mathcal Q^{-1}-I)F(u)$ in Eq.~\eqref{Eq_exact_screened_polar} depresses the local modulus $R$ where opposing phases meet, pulling $R$ down to zero to nucleate $\pm 1$ defect pairs. Viewed through the Noether-charge balance of Sec.~\ref{Sec_Noether_balance}, this twist-driven phase shearing is the exact mechanism by which the conservative Hamiltonian flow transfers $\mathcal{N}_\epsilon=\mathcal E_\epsilon$ out of the bulk amplitude integral $\frac{1}{2}\int_{\Torus}|u|^2\,\d\x$, creating the network of dark amplitude troughs in Fig.~\ref{Fig_topo_defect2}B, and concentrates it into the gradient integral $\frac{\epsilon}{2}\int_{\Torus}|\nabla u|^2\,\d\x$---producing the bright filamentary ridges of $|\nabla\operatorname{Re}u|$ in Fig.~\ref{Fig_topo_defect2}A---which persist without rapid diffusive erasure owing to dynamic inertia.

\paragraph*{Measuring nonlinear coherence and destabilization.} Tracking the total number of phase defects along the $\kappa=10+20i$ trajectory demonstrates that this transition from undulatus-like bands to asperitas-like turbulence occurs dynamically in time at fixed environmental parameters (Fig.~\ref{Fig_defect_count}). Up to $t=5$, the amplitude field retains broadly aligned diagonal bands punctured by only a few dozen isolated defects (Fig.~\ref{Fig_defect_count}A), and the total defect count relaxes from roughly $70$ to about $25$ (Fig.~\ref{Fig_defect_count}C). Near $t\simeq 5$, however, once the growing sidebands and twist-induced phase gradients reach finite amplitude, the pattern undergoes a rapid topological transition: defect pairs nucleate throughout the domain, driving the defect count above $700$ by $t=9$ and shattering the bands into the intricate web of Fig.~\ref{Fig_defect_count}B (identical to Fig.~\ref{Fig_topo_defect2}B). Together with Table~\ref{Table_Corr_coef}, Fig.~\ref{Fig_defect_count} shows that strong amplitude--phase twist and screened wave coupling are sufficient to trigger a spontaneous regime transition from ordered cloud bands to a defect-dominated wave state without external stochastic forcing.

\begin{figure}[tb]
\centering
\includegraphics[width=\columnwidth]{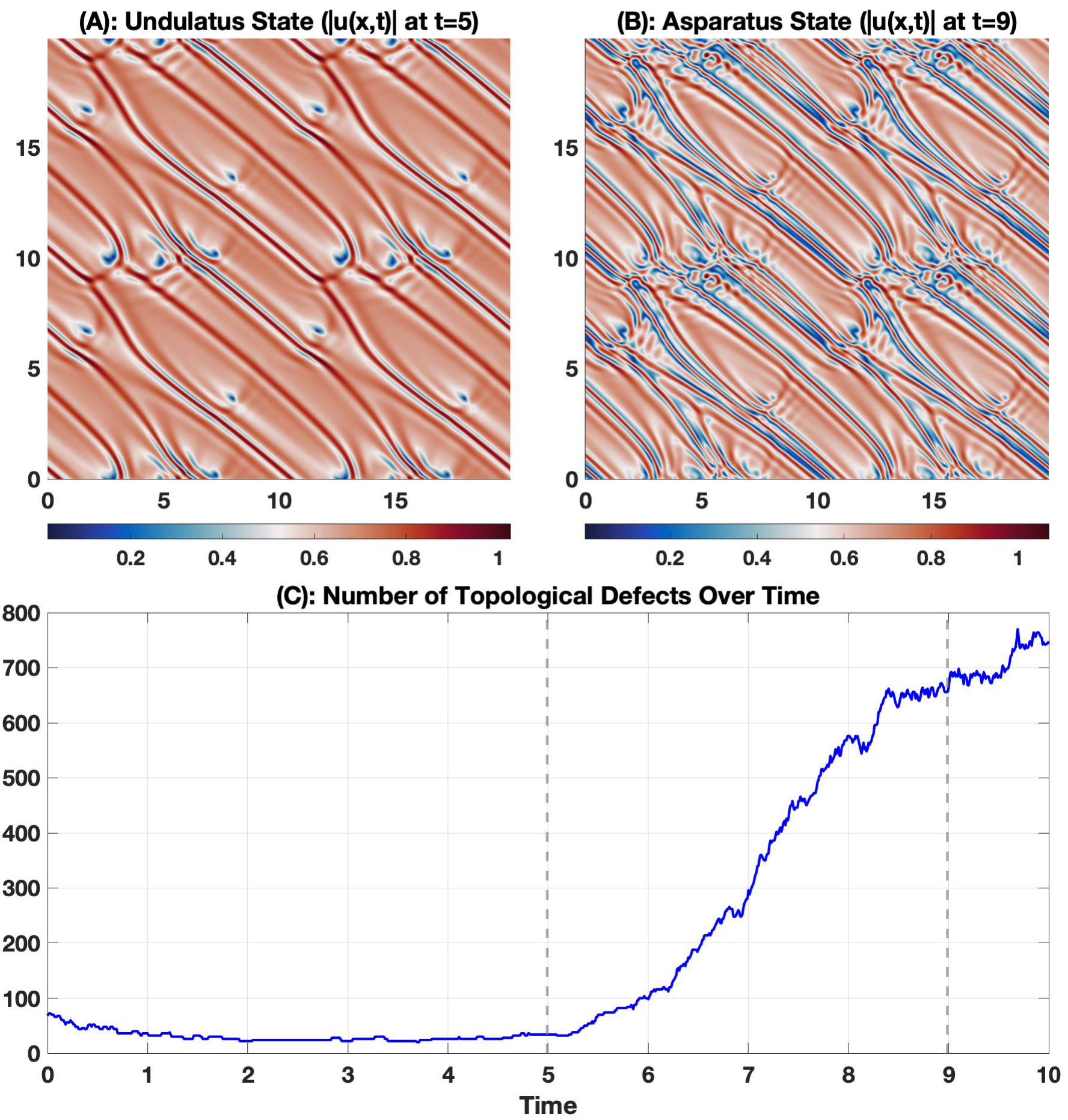}
\caption{{\bf Evolution toward the irregular state of Fig.~\ref{Fig_topo_defect2}.} Amplitude $|u|$ at (A) $t=5$ and (B) $t=9$ in the same run with fixed $\kappa=10+20i$; panel (B) reproduces Fig.~\ref{Fig_topo_defect2}B. Panel (C) gives the phase-defect count along this evolution, with vertical dashed lines marking the snapshots. The initial decline is followed by a sustained rise after $t\simeq5$, accompanying the transition from broadly aligned bands to the densely interrupted state. Both snapshots use $\lambda=5$, $\omega=3$, $\epsilon=0.1$, and $L=20$. Amplitude-dependent timing (Eq.~\eqref{Eq_reaction_frequency_gradient}) and the gain--detuning balance (Eq.~\eqref{Eq_gain_detuning_threshold}) provide the dynamical interpretation developed in the text, while phase winding (Eq.~\eqref{Eq_winding_nu}) identifies the topological changes.}
\label{Fig_defect_count}
\end{figure}

\subsection{Cellular-to-roll consolidation}\label{Sec_spontaneous_stabilization}

The PGLE can also operate in the opposite direction, spontaneously consolidating a two-dimensional cellular pattern into regular parallel rolls (Fig.~\ref{Fig_biphasic}). Initializing Eq.~\eqref{eq_combined_wave} with three symmetrically oriented zero-phase carriers---$\kvec_1=a(10,0)$, $\kvec_2=a(-5,8)$, and $\kvec_3=a(-5,-9)$ ($a=2\pi/L$, $L=20$) at $\lambda=2$, $\omega=3$, $\epsilon=0.1$, and $\kappa=2+2i$---produces a hexagonal cellular interference pattern in $\operatorname{Re}u$ at $t=0$ (Fig.~\ref{Fig_biphasic}A). Over a long quasi-stationary plateau ($15\lesssim t\lesssim 85$), the cellular array gradually deforms into aligned diagonal segments while the spatial mean amplitude $\langle|u|\rangle$ hovers near $0.94$, before abruptly reorganizing around $t\approx 90$ into defect-free parallel roll bands whose mean amplitude locks onto the isolated-wave saturation value $R_*=\sqrt{\lambda/\alpha}=1$ (Fig.~\ref{Fig_biphasic}B--G).

\begin{figure}[tb]
\centering
\includegraphics[width=\columnwidth]{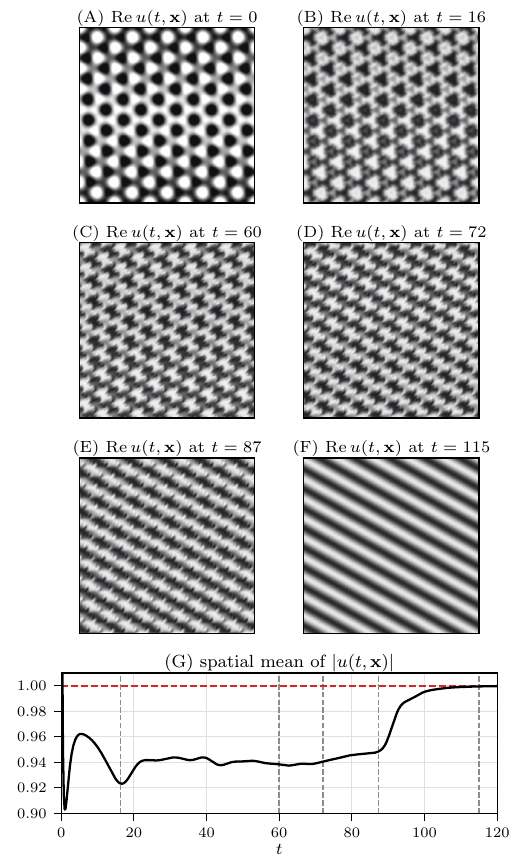}
\caption{{\bf Spontaneous organization into regular bands.} (A--F) Real-part snapshots at the indicated times, starting from the three-wave superposition specified in the text. (G) Spatial mean of $|u|$; the red dashed line is the isolated-wave amplitude $\sqrt{\lambda/\alpha}=1$, and vertical lines mark the later snapshots. The transition from a cellular interference pattern to bands accompanies the approach to this amplitude. Parameters are $\lambda=2$, $\omega=3$, $\epsilon=0.1$, $\kappa=2+2i$, and $L=20$. {The cellular-to-banded transition connects the isolated-wave saturation balance (Eq.~\eqref{Eq_wave_carriers}) with the integrated amplitude balance (Eq.~\eqref{Eq_PGLE_energy});  the decreasing spatial modulation and increasingly parallel bands document consolidation during the displayed interval.}}
\label{Fig_biphasic}
\end{figure}

Both the depressed amplitude plateau ($\langle|u|\rangle\approx 0.94$) during the cellular stage and its rise to $R_*=1$ upon roll formation follow directly from the Noether-charge balance in Sec.~\ref{Sec_Noether_balance} (Eqs.~\eqref{Eq_PGLE_energy} and \eqref{Eq_PGLE_charge_forcing}).  Applying the Cauchy--Schwarz inequality $\int_{\Torus}|u|^2\,\d\x\le V^{1/2}(\int_{\Torus}|u|^4\,\d\x)^{1/2}$ (with $V=|\Torus|$) to Eq.~\eqref{Eq_PGLE_energy} yields
\be\label{Eq_Noether_absorbing_bound}
\frac{\d\mathcal E_\epsilon}{\d t}\le\lambda\int_{\Torus}|u|^2\,\d\x\left(1-\frac{\alpha}{\lambda V}\int_{\Torus}|u|^2\,\d\x\right),
\ee
which guarantees that $\mathcal E_\epsilon$ strictly decreases whenever the spatial mean-square amplitude $\langle|u|^2\rangle=V^{-1}\int_{\Torus}|u|^2\,\d\x$ exceeds $R_*^2=\lambda/\alpha$.

Furthermore, on any quasi-stationary or asymptotic regime where global charge production balances cubic dissipation ($\d\mathcal E_\epsilon/\d t\simeq 0$), Eq.~\eqref{Eq_PGLE_energy} enforces $\langle|u|^4\rangle=R_*^2\langle|u|^2\rangle$. Expressing the spatial variance of $|u|^2$ as $\operatorname{Var}_x(|u|^2)=\langle|u|^4\rangle-\langle|u|^2\rangle^2\ge 0$ yields the exact amplitude-deficit identity
\be\label{Eq_stationary_amplitude_deficit}
\langle|u|^2\rangle=R_*^2-\frac{\operatorname{Var}_x(|u|^2)}{\langle|u|^2\rangle}\le R_*^2,
\ee
with equality $\langle|u|\rangle=R_*$ attained if and only if the oscillation modulus $|u(\x,t)|\equiv R_*$ is spatially uniform. During the three-wave cellular stage in Fig.~\ref{Fig_biphasic}A--E, interference nodes and antinodes maintain a finite spatial variance $\operatorname{Var}_x(|u|^2)>0$, which by Eq.~\eqref{Eq_stationary_amplitude_deficit} depresses $\langle|u|\rangle$ below $R_*=1$. Because $\beta/\alpha=1$ here lies below the two-carrier sideband instability threshold $|\beta/\alpha|>\sqrt{5}/2$ in Eq.~\eqref{Eq_gain_damping_ratio} ($g_{pq}^2/(d_pd_q)=8/9<1$), resonant sideband amplification is closed; instead, cubic mode competition in Eq.~\eqref{Eq_deterministic_spectral_balance} progressively suppresses two of the three carrier orientations, eliminating $\operatorname{Var}_x(|u|^2)$ and driving the cloud field to a single constant-modulus roll wave with $|u|\equiv R_*=1$.

\begin{figure*}[t]
\centering
\includegraphics[width=0.98\textwidth]{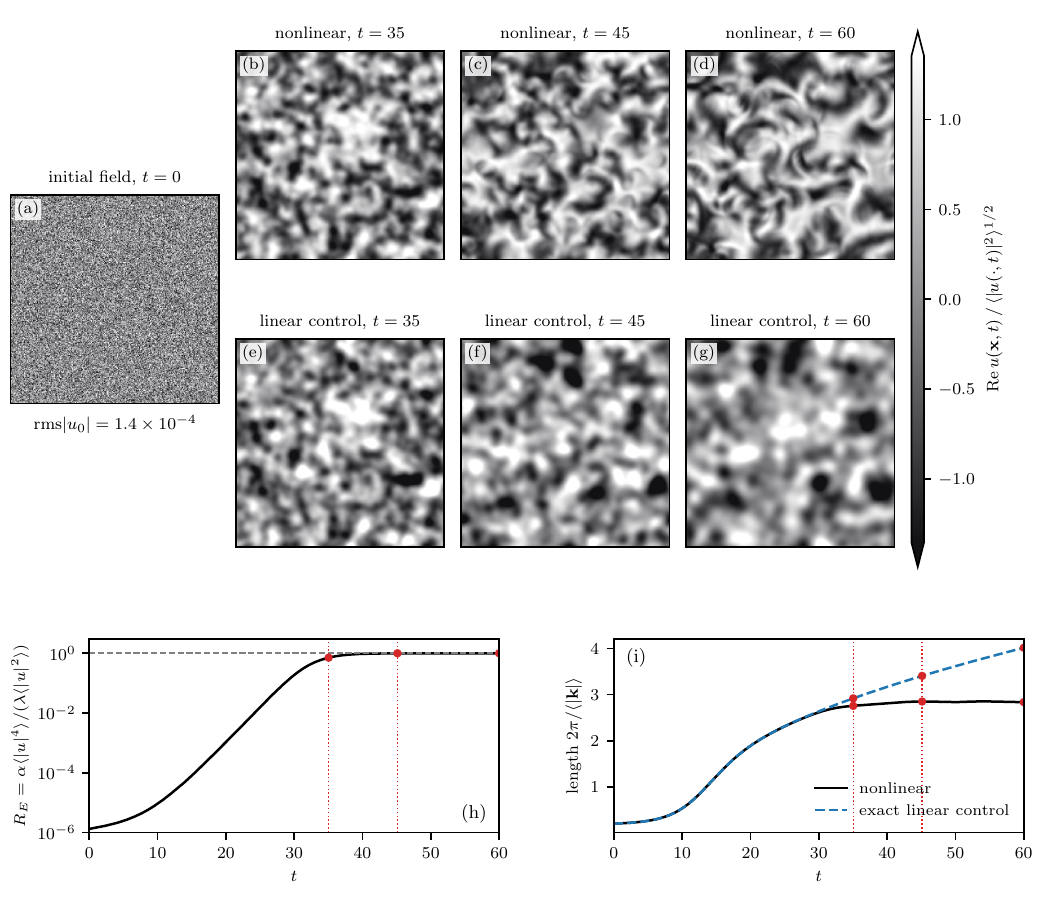}
\caption{{\bf Emergence of cellular and spiral cloud-water patterns from random anomalies.} (a) Real part of the initial field after Fourier projection of $10^{-4}(\xi+i\zeta)$, with independent standard normal values at the grid points. (b--d) Nonlinear PGLE cloud-water anomaly $\operatorname{Re}u$ at $t=35$, $45$, and $60$, developing sharp interlocking cells and curved spiral filaments, contrasted with (e--g) the exact linear evolution from the same projected initial field, which coarsens into diffuse blobs. Snapshot times are rounded ($t=35.04,\ 45.12,\ 60$); each map shows $\operatorname{Re}u$ normalized by the spatial RMS amplitude of $u$ on the common gray scale $\pm 1.355$ (the 99th percentile of the normalized field at $t=60$, with values beyond this range clipped; at most $6.5\%$ of pixels). (h) Growth--saturation ratio $R_E$ in Eq.~\eqref{Eq_cellular_energy_ratio}; the dashed line marks global Noether-charge balance. (i) Power-weighted characteristic length, Eq.~\eqref{Eq_cellular_length}, showing arrested coarsening in the nonlinear evolution relative to the linear control (red markers indicate snapshot times). Parameters: $\lambda=0.3$, $\omega=3$, $\epsilon=0.01$, $\kappa=10$, and $L=20$ in model units.}
\label{Fig_open_cells_formation}
\end{figure*}

\subsection{Emergence of coherent patterns from pure randomness}\label{Sec_random_organization}

Whereas the preceding experiments initialized the PGLE with finite-amplitude wave superpositions, Figs.~\ref{Fig_open_cells_formation}--\ref{Fig_cellular_radial} demonstrate that cellular and spiral actinoform-like cloud-water patterns emerge spontaneously from small, spatially uncorrelated Gaussian anomalies ($u_0\sim 10^{-4}$) without any prescribed carriers or twist. Here, we illustrate this feature for $\lambda=0.3$, $\omega=3$, $\epsilon=0.01$, and real $\kappa=10$ on $[0,20)^2$; see Appendix~\ref{App_cellular_numerics} for numerical and diagnostic details. Comparing the nonlinear PGLE trajectory against the exact linear evolution from the identical initial seed (Eq.~\eqref{Eq_cellular_linear_control}) isolates the precise role of cubic saturation in building mesoscale structure.

Rewriting the global Noether-charge balance of Eq.~\eqref{Eq_PGLE_energy} (Sec.~\ref{Sec_Noether_balance}) in terms of the dimensionless growth--saturation ratio $R_E(t)$ plotted in Fig.~\ref{Fig_open_cells_formation}h,
\be\label{Eq_cellular_energy_ratio}
R_E(t)=\frac{\alpha\langle|u|^4\rangle}
{\lambda\langle|u|^2\rangle},\qquad
\frac{\d\mathcal E_\epsilon}{\d t}=\lambda V\langle|u|^2\rangle\bigl[1-R_E(t)\bigr],
\ee
separates the emergence into two distinct dynamical stages. For $t\lesssim 30$, the amplitude is small ($R_E\ll 1$), so the potential gradient flow $-A_\epsilon^{-1}\operatorname{grad}V[u]$ in Eq.~\eqref{Eq_PGLE_charge_forcing} is governed by linear screened amplification $\exp[2\lambda t/(1+\epsilon|\kvec|^2)]$ (Eq.~\eqref{Eq_linear_modal_power}), which preferentially amplifies low wavenumbers and causes the nonlinear field and its linear control to coarsen identically. Between $t=30$ and $t=45$, the norm ratio drops below $0.9$ (at $t=30$), the normalized spectral total-variation distance exceeds $0.05$ (near $t=36$), and $R_E(t)$ rises through $0.5$ (near $t=33$) to approach unity by $t\approx 45$ (Fig.~\ref{Fig_open_cells_formation}h). At $R_E(t)\simeq 1$, net global production of the Noether charge $\mathcal{N}_\epsilon=\mathcal E_\epsilon$ ceases ($\dot{\mathcal E}_\epsilon\simeq 0$), locking the field onto its global saturation manifold. Crucially, however, arresting global charge growth does not freeze the spatial pattern: over $t\in[40,60]$, nonlinear mode coupling in Eq.~\eqref{Eq_deterministic_spectral_balance} redistributes power across wavenumbers at nearly constant $\mathcal E_\epsilon$, arresting the power-weighted characteristic length $\ell_u(t)$ (Eq.~\eqref{Eq_cellular_length}) at a plateau near $2.84$ while the uninhibited linear control continues to coarsen from $3.17$ to $4.01$ (Fig.~\ref{Fig_open_cells_formation}i) and its cloud-water anomalies broaden into diffuse blobs rather than the sharp, interlocking curved filaments of the nonlinear field (Fig.~\ref{Fig_open_cells_formation}b--g).

This post-saturation spectral redistribution imprints two robust geometric signatures on the cloud--rain field. First, the radial autocorrelation $\rho_{|u|}(r)=C(r)/C(0)$ of the detrended oscillation amplitude locks onto a nearly stationary profile with a persistent negative lobe at separations $r\approx 0.6$--$0.7$ model units, deepening from $-0.09$ in $t\in[30,40]$ to $-0.12$ in $t\in[40,60]$, whereas the linear control broadens monotonically without anticorrelation (Fig.~\ref{Fig_cellular_correlation}d,e). Physically, cubic saturation enforces a characteristic exclusion scale at which vigorously oscillating cloud--rain cells ($\operatorname{Re}u$ shown in Fig.~\ref{Fig_cellular_correlation}a--c) are systematically flanked by quieter amplitude troughs, providing a deterministic mechanism for the preferred mesoscale coherence window of cloud--rain oscillations and the endurance of cellular statistics as individual clouds cycle through growth and decay~\cite{koren2013adaptive,dagan2018organization}.

\begin{figure*}[t]
\centering
\includegraphics[width=0.92\textwidth]{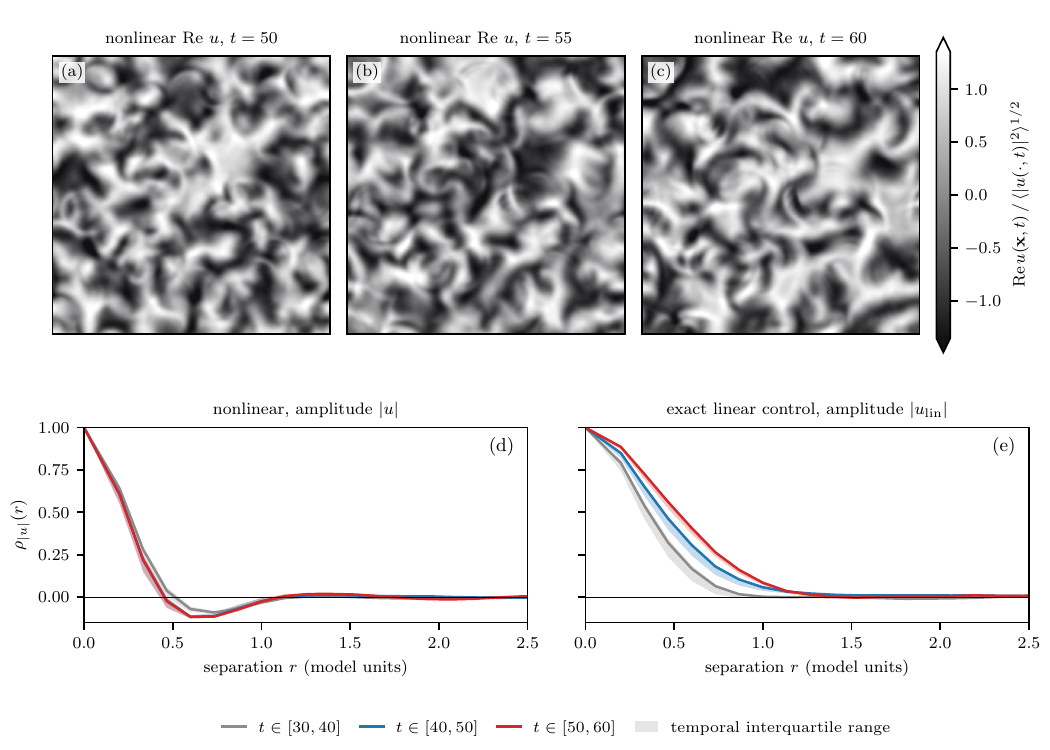}
\caption{{\bf Persistent cellular exclusion scale in the saturated cloud--rain field.} (a--c) Nonlinear cloud-water anomaly $\operatorname{Re}u$ at $t=50$, $55$, and $60$ (saved at $t=49.92,\ 54.96,\ 60$), normalized by the spatial RMS amplitude of $u$ on the gray scale $\pm 1.355$ of Fig.~\ref{Fig_open_cells_formation} (at most $1.1\%$ of pixels clipped). (d,e) Radial spatial autocorrelation $\rho_{|u|}(r)$ of the detrended oscillation amplitude for nonlinear evolution (d) and its exact linear control (e), showing framewise medians (solid curves) and temporal interquartile ranges (shading) over the indicated windows (Appendix~\ref{App_cellular_numerics}). While individual cloud-water structures evolve continuously across (a--c), the nonlinear amplitude correlation in (d) collapses onto a stationary curve with a pronounced negative exclusion lobe near $r\approx 0.6$--$0.7$ model units, absent in the broadening linear control (e).}
\label{Fig_cellular_correlation}
\end{figure*}

Second, the saturated cloud-water field develops a striking spiral-arm geometry around topological defects (Fig.~\ref{Fig_cellular_radial}). Because the uncorrelated complex initial field contains scattered phase singularities where $\vartheta=\arg u$ winds by $\pm 2\pi$ (Eq.~\eqref{Eq_winding_nu}), cubic saturation amplifies $|u|$ toward $R_*=\sqrt{\lambda/\alpha}$ everywhere away from those zeros while leaving the $\nu=\pm 1$ defect cores intact. Around each core, the cloud-water contours $\operatorname{Re}u=|u|\cos\vartheta$ must wind through a full cycle, and the screened phase coupling governed by Eq.~\eqref{Eq_exact_screened_polar} bends those contours into outward-radiating spiral arms of either handedness---quantified by the spiral index $S(c)$ in Eq.~\eqref{Eq_cellular_spiral_index}, which attains $S=+0.52$ and $-0.48$ around the prominent cores enlarged in Fig.~\ref{Fig_cellular_radial}b,c---evoking the radiating arms of marine actinoform cloud systems~\cite{garay2004actinoform}.

\begin{figure*}[t]
\centering
\includegraphics[width=0.98\textwidth]{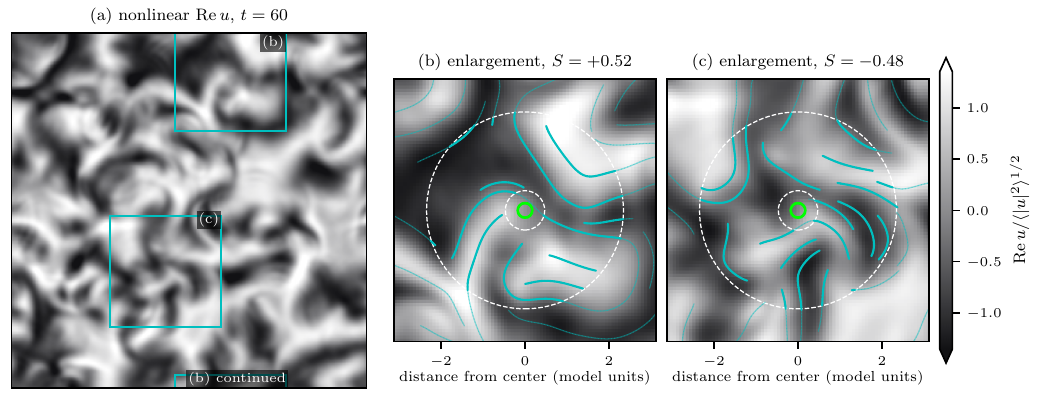}
\caption{{\bf Actinoform-like spiral arms around phase defects emerging from randomness.} (a) Nonlinear cloud-water anomaly $\operatorname{Re}u$ at $t=60$, grown from spatially uncorrelated anomalies of amplitude $10^{-4}$ and normalized by the spatial RMS amplitude of $u$ on the gray scale $\pm 1.355$ of Fig.~\ref{Fig_open_cells_formation}. Scattered phase defects of $u$ ($\arg u$ winding by $\pm 2\pi$) organize the surrounding cloud-water contours into spiral arms. Cyan boxes mark the periodic regions enlarged in (b,c)---with box (b) wrapping across the domain boundary---centered on the defect cores (green circles) exhibiting the strongest spiral winding under the index $S$ of Eq.~\eqref{Eq_cellular_spiral_index}. (b,c) Enlargements showing spiral arms of opposite handedness, $S=+0.52$ and $-0.48$, radiating from the cores, reminiscent of marine actinoform clouds~\cite{garay2004actinoform}; cyan curves trace the local contour orientation of $\operatorname{Re}u$ (bright inside the dashed scoring annulus, faint outside).}
\label{Fig_cellular_radial}
\vspace{-.4cm}
\end{figure*}

Notably, this self-organization occurs at $\beta=0$, where neither the homogeneous-cycle modulational instability (Eq.~\eqref{Eq_uniform_modulation_condition}) nor two-carrier sideband amplification (Eq.~\eqref{Eq_gain_damping_ratio}, $g_{pq}^2/(d_pd_q)=4/9<1$) is active, and where the linear growth rate $\lambda/(1+\epsilon|\kvec|^2)$ attains its maximum at $\kvec=0$ without any preferred finite wavenumber. Unlike the stochastic Swift--Hohenberg cloud model of Ref.~\cite{monroy2021description}, which generates cellular and roll patterns by postulating a fourth-order spatial operator $s_0-(k_c^2-|\kvec|^2)^2$ with a built-in characteristic wavenumber $k_c$, the PGLE selects a finite correlation scale and a spiral-arm geometry purely through the nonlinear saturation and spectral redistribution of screened cloud--rain oscillations growing from the reference state $u=0$.

\section{Discussion}
\label{Sec_concluding_perspective}

This study establishes an analytical route to warm-cloud organization by linking droplet-scale thermodynamic coupling to a stochastic field theory of cloud--rain oscillations. Eliminating the rapidly adjusting supersaturation bath from diffusional droplet growth yields the screened Helmholtz operator $(I-\epsilon\Delta)\partial_t$ and its associated dynamic inertia (Sec.~\ref{Sec_PGLE_physics}), while Hopf reduction of delayed cloud--rain feedback supplies the local Stuart--Landau tendency $F(u)$ (Sec.~\ref{Sec_PGLE_variables}). Embedding these two ingredients into a marked Poisson jump process for unresolved episodes of condensation, mixing, conversion, and fallout---and closing the third-order fluctuation moments in Eq.~\eqref{Eq_third_moment_expansion}---produces the deterministic PGLE as the leading ensemble-mean evolution law for the normalized cloud-water and rain-water anomalies $u=\rho+iq$ (Sec.~\ref{Sec_jump_construction}). In contrast to lattice cloud models whose deterministic mean-field limit collapses spatial structure into bulk area-fraction equations~\cite{khouider2010multicloud,Frenkel2013,khouider2019new}, this continuum closure preserves nonlocal spatial coupling and directly resolves cloud morphology, while remaining an open-system closure whose physical water inventory $M_l(t)$ (Sec.~\ref{Sec_liquid_water_budget}) and coordinate map must be calibrated against parent cloud-resolving dynamics.

Positioning the PGLE against classical pattern-forming equations clarifies why dynamic inertia is particularly suited to cloud fields (Sec.~\ref{Sec_dynamic_inertia_relaxation}). Whereas diffusive coupling in the standard CGLE or linear stochastic-diffusion cloud models~\cite{staliunas2001spectra,stechmann2016cloud} erases fine structure at rates proportional to $k^2$, dynamic inertia attenuates modal relaxation at wavelengths shorter than the screening length $L_{\rm scrn}=\sqrt{\epsilon}$, allowing sharp cloud boundaries, gradient filaments, and phase defects to persist while the broader oscillation evolves. Even when a linear diffusive-threshold model matches the static $k^{-2}$ Lorentzian spectrum of the parcel-driven PGLE, the two descriptions separate dynamically through temporal persistence and nonlinear spatial organization~\cite{stechmann2016cloud,chekroun_koren_liu_variance2026}. Mathematically, the PGLE splits into a Benjamin--Bona--Mahony-regularized nonlinear Schr\"odinger Hamiltonian flow and a pseudoparabolic Allen--Cahn gradient flow governed by the same metric operator $A_\epsilon=I-\epsilon\Delta$ (Sec.~\ref{Sec_Noether_balance}). Because the dispersion $\omega$ and twist $\beta$ reside solely in the Hamiltonian sector, they conserve the screening-weighted Noether charge $\mathcal{N}_\epsilon=\mathcal E_\epsilon$ and make zero direct contribution to net charge production, whereas linear growth and cubic saturation regulate the global amplitude balance.

This Hamiltonian--gradient structure unifies the diverse cloud-relevant transitions demonstrated in Secs.~\ref{Sec_Instab_Neimark}--\ref{Sec_modeling_results}. In one dimension, dynamic inertia and phase twist together destabilize traveling carriers through non-harmonic sidebands (Fig.~\ref{Fig_stab_chart}) and drive a transition from wave breakup to a two-frequency modulated state whose saturated spectrum organizes into a sharp combination-tone comb distinct from the initiating linear frequencies (Figs.~\ref{Fig_Space_Time}--\ref{Fig_PSD}), while resonant triads excite domain-mean liquid-water oscillations. In two dimensions, the cross-carrier gain--detuning criterion (Eq.~\eqref{Eq_gain_detuning_threshold}) identifies the twist threshold $|\beta/\alpha|>\sqrt{5}/2$ separating ordered \emph{undulatus}-like bands ($\beta/\alpha=0.25$, punctuated by isolated $\nu=\pm 1$ dislocations) and cellular-to-roll consolidation ($\beta/\alpha=1$, where mode competition eliminates spatial amplitude variance to reach the ceiling $R_*$) from the defect-rich \emph{asperitas}-like wave networks ($\beta/\alpha=2$) driven by conservative Hamiltonian phase-shearing into fine-scale gradients (Figs.~\ref{Fig_undulatus_vs_analytic}--\ref{Fig_biphasic}). Remarkably, even at $\beta=0$ from unstructured random anomalies, arresting global Noether-charge production ($\dot{\mathcal E}_\epsilon\simeq 0$) triggers nonlinear spectral redistribution that halts coarsening, establishes a stationary negative autocorrelation lobe indicative of sustained cellular alternation~\cite{koren2013adaptive,dagan2018organization}, and winds cloud-water contours into spiral arms around phase singularities reminiscent of marine actinoform clouds~\cite{garay2004actinoform} (Figs.~\ref{Fig_open_cells_formation}--\ref{Fig_cellular_radial})---all without prescribing a built-in preferred wavenumber as in Swift--Hohenberg descriptions~\cite{monroy2021description}. Together, these results invite joint observational and numerical tests comparing cloud-field morphology~\cite{harrison2017asperitas,ravichandran2022asperitas,stevens2020patterns,bony2020patterns}, cycle timing, and resolved water budgets.

Finally, the underlying jump formulation provides a concrete roadmap for restoring unresolved cloud variability beyond the deterministic mean field. Rather than treating stochasticity as ad hoc additive noise~\cite{LinNeelin2000,LinNeelin2003,stechmann2011stochastic}, Eqs.~\eqref{Eq_SIDE} and \eqref{Eq_third_moment_expansion} trace how finite parcel event rates, element areas, and mark correlations generate fluctuation variance and feed back onto the mean oscillation. The companion analysis~\cite{chekroun_koren_liu_variance2026} establishes how finite parcel footprints regularize short-wavelength variance and yields explicit intermittency conditions for the validity of the mean-field closure. Looking ahead, calibrating the marked counting measure $N(\d t,\d\y,\d\zeta)$ against large-eddy simulation diagnostics of entrainment, detrainment, precipitation, and column water budgets~\cite{dawe2013entrainment,yamaguchi2015precipitation,hoffmann2023roles}---while enforcing event-by-event liquid-water conservation $\mathscr L(\eta)=0$ for conversion-only episodes (Sec.~\ref{Sec_liquid_water_budget})---will allow the stochastic PGLE and jump-diffusion response tools~\cite{chekroun2025kolmogorov} to quantify how intermittent parcel physics shapes the persistence and transitions of real cloud fields.

Beyond LES calibration, disentangling how nonlinear mode interactions and jump noise jointly drive cloud-regime transitions presents a compelling mathematical challenge. Classical center-manifold and stochastic invariant-manifold reductions~\cite{DLS03,CLW15_vol1,DuanJ2015} rely on sectorial operators with strong spectral gaps separating slow active modes from rapidly damped fine scales. By contrast, dynamic inertia makes the pseudoparabolic operator $(I-\epsilon\Delta)^{-1}$ bounded, so modal rates accumulate at zero as $|\kvec|\to\infty$ and high-wavenumber structure adjusts slowly, eroding timescale separation. Moreover, because every Fourier mode around $u=0$ grows when $\lambda>0$, a meaningful Backward--Forward (BF) auxiliary system cannot be anchored at the trivial state; instead, the sideband instability chart (Fig.~\ref{Fig_stab_chart}) and the resonant gain--detuning criteria of Secs.~\ref{Sec_Instab_Neimark} and \ref{Sec_Nonlin_Wave_Interact} become indispensable, providing the spectral blueprint in the co-rotating frame of an oscillating state for splitting the sideband blocks $M_q$ (or $M_{pq}$) into the unstable backward sector $H_{\mathfrak c}$ and the genuinely damped forward sector $H_{\mathfrak f}$. Built on this splitting, the variational framework of optimal parameterizing manifolds and non-Markovian reduced models~\cite{chekroun2023optimal,chekroun2025unravel}---developed in continuity with~\cite{CLW15_vol1,CLW15_vol2,chekroun2023transitions} for weak timescale separation and jump-driven systems---provides a natural path for deriving amplitude closures in such challenging settings. Indeed, integrating finite-horizon BF systems across $H_{\mathfrak c}\oplus H_{\mathfrak f}$ makes it possible to regularize the weak pseudoparabolic spectral gap at high wavenumbers while encoding jump-noise and triad interactions through path-dependent coefficients. Adapting this formalism to the sideband and resonant-triad subspaces identified here thus holds promise for low-dimensional non-Markovian amplitude closures capable of predicting jump-induced transitions between banded, cellular, and irregular cloud-field states. This direction will be pursued elsewhere.

\medskip
\paragraph*{\bf Acknowledgments.} 
This work has been supported by the European Research Council (ERC) under the European Union's Horizon 2020 research and innovation program (grant agreement No.~810370), by the Knell Family Institute for Artificial Intelligence, Weizmann Institute of Science, and by the Institute for Environmental Sustainability (IES) at the Weizmann Institute of Science. This work has also been partially supported by the Office of Naval Research (ONR) Multidisciplinary University Research Initiative (MURI) Grant N00014-20-1-2023, and by the National Science Foundation Grant  DMS-2407483 and DMS-2407484.

\medskip
\paragraph*{\bf Data availability.} The codes that produced the figures in this study are available from the authors upon reasonable request.

\medskip
\appendix
\renewcommand{\thesubsection}{Appendix \thesection.\arabic{subsection}}
\makeatletter
\renewcommand{\p@subsection}{}
\makeatother
{\small
\section{Fourier Couplings and the Two-Sideband Criterion}
\label{App_wave_interactions}

\subsection{Residual and full linearized coupling}

Let $\mathcal Q=I-\epsilon\Delta$ and
$F(u)=(\lambda+i\omega)u-\kappa|u|^2u$.
For a prescribed field $u_0$, define
$\mathcal R=\mathcal Q\partial_tu_0-F(u_0)$.
Writing $u=u_0+v$ gives, to first order in $v$,
\be\label{Eq_wave_residual_linearization}
\mathcal Q\partial_tv
=(\lambda+i\omega)v
-\kappa(2|u_0|^2v+u_0^2\overline v)
-\mathcal R+O(|v|^2).
\ee
The residual enters with a minus sign under this definition. If each carrier in Eq.~\eqref{Eq_wave_carriers} is an isolated exact solution, direct expansion gives
\be\label{Eq_wave_residual}
\mathcal R=
\kappa\left(2|u_2|^2u_1+2|u_1|^2u_2
+u_1^2\overline u_2+u_2^2\overline u_1\right).
\ee
The first two terms change the original carriers, and the last two force mixed harmonics. Setting the residual aside holds the carrier background fixed and isolates perturbation amplification. Evolving the original superposition requires retaining the residual forcing.

To include any finite number $J$ of prescribed carriers, write $u_0=\sum_{j=1}^{J} A_j(t)e^{i\kvec_j\cdot\x}$, where $A_j(t)$ are physical Fourier amplitudes, and retain the orthonormal coefficients $\widehat v_{\kvec}$ for the perturbation. The two-carrier background corresponds to $J=2$. With both indices $j,l$ ranging from $1$ to $J$, the complete linearized Fourier system is
\begin{widetext}
\be\label{Eq_wave_full_linear_modes}
Q_{\kvec}\,\partial_t\widehat v_{\kvec}
=(\lambda+i\omega)\widehat v_{\kvec}
-\kappa\sum_{j,l}
\left[2A_j\overline A_l\,
\widehat v_{\kvec-\kvec_j+\kvec_l}
+A_jA_l\,
\overline{\widehat v_{\kvec_j+\kvec_l-\kvec}}\right]
-\widehat{\mathcal R}_{\kvec}.
\ee
\end{widetext}
The terms with $j\ne l$ in $2|u_0|^2v$ scatter perturbations between wavevectors separated by $\kvec_j-\kvec_l$. For the two-carrier background, these are precisely $2(u_1\overline u_2+u_2\overline u_1)v$, giving shifts $\pm(\kvec_1-\kvec_2)$. The terms $u_j^2\overline v$ from $u_0^2\overline v$ couple a perturbation at $\kvec$ to the conjugate mode at $2\kvec_j-\kvec$. The reduced system in Eq.~\eqref{Eq_wave_pair_matrix} sets $J=2$, retains the uniform contribution $2(|u_1|^2+|u_2|^2)v$ and the mixed product $2u_1u_2\overline v$, and omits the scattering and unmixed-square channels just identified. Its generic distinct sidebands are chosen so that the omitted channels do not coincide with the retained modes.

\paragraph*{Carrier frames and eigenvalues.} For the retained pair, write
$v=a_p(t)e^{i(\kvec_p\cdot\x+\nu_pt)}
+a_q(t)e^{i(\kvec_q\cdot\x+\nu_qt)}$.
The condition $\nu_p+\nu_q=\Omega_1+\Omega_2$ removes the explicit phase of the mixed-carrier coupling. This gives Eq.~\eqref{Eq_wave_pair_matrix}. For an eigenmode, $(a_p,\overline a_q)=(a,b)e^{\sigma t}$, and hence $a_q=\overline b e^{\overline\sigma t}$.  The two entries $(a_p,\overline a_q)$ therefore share the eigenvalue $\sigma$, while conjugation gives the evolution of $a_q$. This pairing follows from linearizing the complex cloud--rain amplitude.

Set $\mathcal C=4|\kappa|^2A^4$ and denote the diagonal entries of $M_{pq}$ by $m_{11},m_{22}$. The characteristic equation and its two roots are
\bea\label{Eq_wave_characteristic}
\left[(\sigma+i\nu_p)Q_p-\Lambda\right]
\left[(\sigma-i\nu_q)Q_q-\overline\Lambda\right]
-\mathcal C&=0,\\
\sigma_\pm=\frac{m_{11}+m_{22}}{2}
\ \pm\sqrt{\left(\frac{m_{11}-m_{22}}{2}\right)^2
+\frac{\mathcal C}{Q_pQ_q}}.&
\eea
These formulas require no division by $Q_p-Q_q$. All entries of $M_{pq}$ and $\sigma$ have units of inverse time; $\mathcal C$, $X_0^2$, and the squared detuning have units of inverse time squared.

\subsection{Neutral crossing and instability}\label{App_wave_neutral_crossing}

For $X_0<0$, substitute $\sigma=i\delta$ in Eq.~\eqref{Eq_wave_characteristic}. Its imaginary part gives
\be\label{Eq_wave_neutral_frequency}
\delta=\frac{Q_q\nu_q-Q_p\nu_p}{Q_p+Q_q}.
\ee
At this value, the two imaginary factors in the characteristic product are opposite. Its real part therefore yields
\be\label{Eq_wave_neutral}
\mathcal C_{\rm crit}
=X_0^2+\left[Y_0-\frac{Q_pQ_q}{Q_p+Q_q}
(\nu_p+\nu_q)\right]^2.
\ee
The physical sideband frequencies at neutrality are $\nu_p+\delta$ and $\nu_q-\delta$, whose sum remains the carrier-frequency sum. The shift is determined once, through the eigenvalue; inserting it again into the prescribed carrier frequencies would count it twice.

To see that this neutral boundary separates decay from growth, put
$d_p=-X_0/Q_p>0$, $d_q=-X_0/Q_q>0$,
$b_p=Y_0/Q_p-\nu_p$, and $b_q=-Y_0/Q_q+\nu_q$.
Then the characteristic equation is
$(\sigma+d_p-ib_p)(\sigma+d_q-ib_q)=\mathcal C/(Q_pQ_q)$.
At zero coupling both eigenvalues decay. In Eq.~\eqref{Eq_wave_characteristic}, increasing the nonnegative coupling adds to the real part of the square-root argument while leaving its imaginary part fixed; the larger eigenvalue real part is nondecreasing. The unique neutral crossing is
\be\label{Eq_wave_damping_detuning}
\frac{\mathcal C_{\rm crit}}{Q_pQ_q}
=d_pd_q\left[1+\frac{(b_p-b_q)^2}{(d_p+d_q)^2}\right],
\ee
which proves Eq.~\eqref{Eq_wave_pair_threshold}. At equality the other eigenvalue has negative real part, since
$\operatorname{Re}\operatorname{tr}M_{pq}=X_0(Q_p^{-1}+Q_q^{-1})<0$.
For $X_0>0$ this trace is positive and at least one eigenvalue grows regardless of the coupling. The degenerate case $X_0=0$ is decided directly from Eq.~\eqref{Eq_wave_characteristic}; the damping-based criterion above is stated only for $X_0<0$.

For isolated-wave amplitudes, introduce
$\eta_\kappa=\beta/\alpha$ and
$\mathcal H_{pq}=\frac{Q_pQ_q}{Q_p+Q_q}(Q_1^{-1}+Q_2^{-1})$.
Since $A^2=\lambda/\alpha$, $X_0=-3\lambda$, and
$\Omega_j=(\omega-\lambda\eta_\kappa)/Q_j$,
Eq.~\eqref{Eq_wave_pair_threshold} becomes
\be\label{Eq_wave_threshold_parameters}
\boxed{4\lambda^2(1+\eta_\kappa^2)
>9\lambda^2+
\left[(1-\mathcal H_{pq})\omega
+(\mathcal H_{pq}-4)\lambda\eta_\kappa\right]^2.}
\ee
In particular, $\eta_\kappa=0$ cannot destabilize this retained pair, and $|\eta_\kappa|>\sqrt5/2$ is necessary. In the unscreened limit $\epsilon=0$, all $Q_j=1$ and $\mathcal H_{pq}=1$; the inequality reduces to
$4\lambda^2(1+\eta_\kappa^2)>9\lambda^2(1+\eta_\kappa^2)$ and cannot hold. The retained pair is therefore stable in this limit, while the residual continues to generate mixed components in the full nonlinear evolution.

The omitted-channel error is assessed by comparing the retained-pair evolution with Eq.~\eqref{Eq_wave_full_linear_modes} or the full PGLE, including residual forcing and changes in the carriers. The algebraic instability criterion applies to the reduced matrix, and such comparisons determine its accuracy for a particular field.

\subsection{Exact stability of a single traveling carrier}\label{App_single_carrier_stability}

The paired system in Eq.~\eqref{Eq_system_perturb} follows by removing the carrier before resolving the modulation. With $\delta u=u_0V$ and $L=\lambda+i\omega-2\kappa A^2$, Eq.~\eqref{Eq_linearized_PGLE} becomes
\be\label{Eq_single_rotating_operator}
[1-\epsilon(\partial_x+ik_0)^2](\partial_t+i\Omega_0)V
=LV-\kappa A^2\overline V.
\ee
For $V=a(t)e^{iqx}+c(t)e^{-iqx}$, matching the two Fourier coefficients gives
\bea\label{Eq_single_scalar_pair}
Q_+(\dot a+i\Omega_0a)&=La-\kappa A^2\overline c,\\
Q_-(\dot c+i\Omega_0c)&=Lc-\kappa A^2\overline a.
\eea
Conjugating the entire second equation closes the system for $(a,\overline c)$ and yields $M_q$. In particular, conjugation changes the sign of its rotating-frame frequency but preserves its screening denominator $Q_-$. The characteristic polynomial is
\be\label{Eq_chiM}
\sigma^2-\operatorname{tr}(M_q)\sigma+\det(M_q)=0.
\ee
Writing $B_\pm=Y-\Omega_0Q_\pm$, its coefficients satisfy
\be\label{Eq_realTrM}
\operatorname{Re}\operatorname{tr}(M_q)=-\lambda(Q_+^{-1}+Q_-^{-1})<0,
\ee
\be\label{Eq_detM}
\det(M_q)=\frac{\lambda^2+B_+B_--|\kappa|^2A^4+i\lambda(B_--B_+)}{Q_+Q_-}.
\ee
To determine the sign of the largest growth rate, put $p=Q_+^{-1}$, $m=Q_-^{-1}$ and introduce
\bea\label{Eq_single_spectral_parameters}
\gamma&=\frac{\lambda}{2}(p+m),\qquad \nu=\frac{Y}{2}(p-m),\\
d&=-\frac{\lambda}{2}(p-m),\qquad H=\frac{Y}{2}(p+m)-\Omega_0,\\
C&=|\kappa|^2A^4pm.
\eea
Here $H$ measures the frequency mismatch and is distinct from the dimensionless screening ratio $h_q$. The diagonal entries are $-\gamma+i\nu\pm(d+iH)$, so
\be\label{Eq_single_exact_exponents}
\sigma_\pm=-\gamma+i\nu\pm\sqrt{C+(d+iH)^2}.
\ee
With $X=C+d^2-H^2$, the largest real part is
\be\label{Eq_single_exact_growth}
s_q=-\gamma+\sqrt{\frac{\sqrt{X^2+4d^2H^2}+X}{2}}.
\ee
This formula also treats a purely real square-root argument. To establish the instability criterion without ambiguity from squaring inequalities, let $t$ be the square of the nonnegative square-root term in Eq.~\eqref{Eq_single_exact_growth}. It is the larger, nonnegative root of
\be\label{Eq_single_sign_polynomial}
P(z)=z^2-Xz-d^2H^2.
\ee
The other root is nonpositive. Since $\gamma>0$, $s_q>0$ is equivalent to $t>\gamma^2$, hence to $P(\gamma^2)<0$. Using $\gamma^2-d^2=\lambda^2pm$ and $C=(\lambda^2+\beta^2A^4)pm$ gives
\be\label{Eq_single_polynomial_evaluation}
P(\gamma^2)=pm(\lambda^2H^2-\beta^2A^4\gamma^2).
\ee
Consequently,
\be\label{Eq_single_complete_sign}
\begin{aligned}
s_q>0&\ \Longleftrightarrow\ |H|<\tfrac12|\beta|A^2(p+m),\\
s_q=0&\ \Longleftrightarrow\ |H|=\tfrac12|\beta|A^2(p+m),\\
s_q<0&\ \Longleftrightarrow\ |H|>\tfrac12|\beta|A^2(p+m).
\end{aligned}
\ee
Dividing the strict inequality by $(p+m)/2$ gives Eq.~\eqref{Eq_stab_cond1}; inserting $A^2=\lambda/\alpha$ and $\Omega_0=(\omega-\lambda\eta)/Q_0$ yields Eq.~\eqref{Eq_stab_cond2}. For $\eta>0$, taking its square root and using
\[
(1-h_q)\omega+(h_q-2)\lambda\eta
=(1-h_q)(\omega-\lambda\eta)-\lambda\eta
\]
gives the intervals in Eq.~\eqref{Eq_single_frequency_intervals}. Finally, $Q_++Q_-=2(Q_0+\epsilon q^2)$ and $Q_+Q_-=(Q_0+\epsilon q^2)^2-4\epsilon^2k_0^2q^2$ give Eq.~\eqref{Eq_single_screening_sign}. These steps connect the exact eigenvalues to the frequency boundaries and their dependence on carrier and modulation scales.

\section{Rotating frame and characteristic exponents}
\label{App_PGLE_modulation}
Let $\mathcal Q=I-\epsilon\Delta$. Substitution of $u=e^{i\Omega_*t}(A+v)$ into Eq.~\eqref{Eq_PGLE} gives, to first order,
\be\label{Eq_uniform_rotating_linear}
\mathcal Q\partial_tv=
(\lambda+i\omega-2\kappa A^2)v-\kappa A^2\overline v
-i\Omega_*\mathcal Qv.
\ee
The rotating-frame term contains $\mathcal Q$ because this operator acts on the time derivative. Resolving $v$ into its real and imaginary fields yields Eq.~\eqref{Eq_uniform_modulation_matrix}. The Fourier coefficients of these real fields can be complex; coefficients at opposite wavenumbers satisfy the usual conjugacy relation. The characteristic polynomial and exponents are
\bea\label{Eq_uniform_modulation_exponents}
&\sigma^2+\frac{2\lambda}{Q_{\kvec}}\sigma+
\frac{\Omega_*r_{\kvec}(2\beta A^2+\Omega_*r_{\kvec})}{Q_{\kvec}^2}=0,\\
&\sigma_\pm(\kvec)=
\frac{-\lambda\pm\sqrt{\lambda^2-\Omega_*r_{\kvec}(2\beta A^2+\Omega_*r_{\kvec})}}{Q_{\kvec}}.
\eea
The negative trace precludes growth through a complex-conjugate pair crossing the imaginary axis. Growth instead requires a negative determinant, giving Eq.~\eqref{Eq_uniform_modulation_condition}. The eigenvalues here are Floquet exponents in the frame of the homogeneous cycle; the neutral $k=0$ exponent reflects time-translation symmetry.

\paragraph*{Band edges and the role of screening.} When $\beta\Omega_*<0$, the growing exponent starts at zero for $r_{\kvec}=0$ and returns to zero at $r_{\kvec}=-2\beta A^2/\Omega_*$. Between these endpoints it is positive and therefore has a positive interior maximum. At small wavenumber,
\be\label{Eq_uniform_longwave}
\sigma_+(\kvec)=-\frac{\beta\Omega_*}{\alpha}\epsilon|\kvec|^2
+O(\epsilon^2|\kvec|^4).
\ee
For $\beta=0$ the leading phase decay is instead $-\Omega_*^2\epsilon^2|\kvec|^4/(2\lambda)$. At large wavenumber the exponents approach $\pm i|\Omega_*|$ with real part $-\lambda/Q_{\kvec}$, so fine modulations become weakly damped rather than rapidly diffused. This is another expression of dynamic inertia. The continuum band identifies possible growth, but a finite periodic domain samples only its discrete wavenumbers.  The unstable modes seed spatial modulation of the cycle, whose saturated geometry and temporal evolution are determined by the nonlinear couplings.

\section{Amplitude-dependent timing, isochrons, and defects}
\label{App_phase_geometry}

\subsection{Reaction frequency and the screened phase response}
\label{App_screened_phase_response}
Define $F(u)=(\lambda+i\omega)u-(\alpha+i\beta)|u|^2u$ and $\mathcal Q=I-\epsilon\Delta$. Where $R=|u|>0$, write $u=R e^{i\vartheta}$. Multiplying $\partial_tu=\mathcal Q^{-1}F(u)$ by $e^{-i\vartheta}$ gives exact amplitude and phase equations in operator form:
\bea\label{Eq_exact_screened_polar}
\partial_tR&=\lambda R-\alpha R^3+\operatorname{Re}\!\left[e^{-i\vartheta}(\mathcal Q^{-1}-I)F(u)\right],\\
\partial_t\vartheta&=\omega-\beta R^2+
\frac{1}{R}\operatorname{Im}\!\left[e^{-i\vartheta}(\mathcal Q^{-1}-I)F(u)\right].
\eea
The first two terms in each equation give the local reaction tendency. The final term in each equation describes the screened response of the heterogeneous field. Taking a spatial gradient of the reaction frequency gives Eq.~\eqref{Eq_reaction_frequency_gradient}. Larger and smaller oscillations therefore have different intrinsic timings when $\beta\ne0$. Over time these differences can bend equal-phase bands, while the nonlocal term changes how the differences propagate. Following that process through amplitude minima requires the coupled amplitude and phase equations above. For an isolated plane wave, $F(u)$ has the same spatial wavenumber as $u$, so $\mathcal Q^{-1}$ multiplies it by $Q_{\kvec}^{-1}$ and recovers Eq.~\eqref{Eq_instantaneous_phase_rate}. For a superposition, the cubic response generates additional Fourier components, each acted on by its own screening factor. To understand the timing retained after an amplitude excursion, one must also account for relaxation toward the local cycle. \ref{App_twist_isochrons} does this through its asymptotic phase and relates it directly to the two screened tendencies in Eq.~\eqref{Eq_exact_screened_polar}.

\subsection{Twist and isochrons}\label{App_twist_isochrons}
In a spatially uniform field, $u(t)=R(t)e^{i\vartheta(t)}$, the screened corrections vanish and Eq.~\eqref{Eq_exact_screened_polar} reduces to $\dot R=\lambda R-\alpha R^3$ and $\dot\vartheta=\omega-\beta R^2$. For $\lambda>0$ and $\alpha>0$, the attracting radius is $A=\sqrt{\lambda/\alpha}$. On that circle the angular velocity is $\Omega_*=\omega-\beta A^2=\omega-\beta\lambda/\alpha$, so $u_*(t)=Ae^{i(\Omega_*t+\vartheta_0)}$. Thus $\Omega_*$ is the frequency determined by the reaction coefficients, rather than an additional parameter in the polar representation. When $\Omega_*\ne0$, this solution is a periodic orbit. Its asymptotic phase is
\be\label{Eq_PGLE_isochron}
\Theta=\vartheta-\frac{\beta}{\alpha}\log\frac{R}{A},
\qquad \dot\Theta=\Omega_*.
\ee
Differentiating gives $\dot\Theta=(\omega-\beta R^2)-(\beta/\alpha)(\lambda-\alpha R^2)=\Omega_*$, so the phase coordinate advances at the cycle frequency even during radial relaxation. States with equal $\Theta$ approach the same phase on the cycle: they lie on an isochron. These curves are radial rays for $\beta=0$ and logarithmic spirals for $\beta\ne0${~\cite{guckenheimer1975isochrons,Tantet_al_Hopf}}. At fixed instantaneous angle, an amplitude displacement from $R_1$ to $R_2$ changes the asymptotic phase by $-(\beta/\alpha)\log(R_2/R_1)$. The ratio $\beta/\alpha$ thus measures the conversion of amplitude disturbances into persistent timing offsets.

This phase geometry explains why a perturbation's direction matters. A displacement along an isochron relaxes without an asymptotic timing shift, whereas one across isochrons changes the recovered phase. Repeated disturbances followed by differential angular motion and relaxation can produce stretching and folding, the basis of shear-induced chaos under suitable forcing~\cite{wang2001strange,wang2003strange,lin2008shear,Young2016,Chekroun_al22SciAdv}. In the KTF cloud model, even weak twist could be exploited by suitably designed perturbations: the resulting stochastic dynamics reproduced important temporal features of simulated open-cell oscillations~\cite{Chekroun_al22SciAdv}.  This result motivates tracking the same amplitude--phase sensitivity as spatial coupling perturbs the oscillations of the PGLE.

The connection to \ref{App_screened_phase_response} can be made exactly. Define
\be\label{Eq_screened_complex_response}
\mathcal H(\x,t)=e^{-i\vartheta}(\mathcal Q^{-1}-I)F(u).
\ee
Using the same local phase coordinate $\Theta=\vartheta-(\beta/\alpha)\log(R/A)$ in the heterogeneous field and differentiating with Eq.~\eqref{Eq_exact_screened_polar} gives, wherever $R>0$,
\be\label{Eq_screened_isochron_response}
\partial_t\Theta=\Omega_*+
\frac{\operatorname{Im}\mathcal H-(\beta/\alpha)\operatorname{Re}\mathcal H}{R}.
\ee
The local reaction terms cancel, leaving a direct phase contribution from the screened response and a contribution through its effect on amplitude. The latter is weighted by the twist. Equation~\eqref{Eq_reaction_frequency_gradient} describes the instantaneous difference in reaction frequencies; Eq.~\eqref{Eq_screened_isochron_response} describes how environmental coupling changes the phase coordinate that already accounts for local amplitude relaxation. Together they connect spatial amplitude contrasts with persistent changes in the relative timing of cloud--rain oscillations.

Here $\Theta$ is a pointwise phase coordinate inherited from the uniform oscillator. Equation~\eqref{Eq_screened_isochron_response} determines how spatial coupling changes it, and therefore how the local isochron geometry enters the heterogeneous field. In the deterministic PGLE simulations, that coupling supplies the evolving amplitude and timing disturbances. Restoring parcel-event fluctuations would add intermittent disturbances to this same response, connecting the spatial theory to the stochastic mechanism examined in the KTF model \cite{Chekroun_al22SciAdv}.

\subsection{Cores and moving amplitude depressions}
\label{App_moving_defects}
 A nonzero winding in Eq.~\eqref{Eq_winding_nu} cannot be continued to a single-valued phase throughout the enclosed region without an amplitude zero.  At that zero the polar coordinates fail, whereas the complex field can remain smooth. Core evolution is therefore calculated directly in $u$, which tracks both water-path anomalies through their reference values as the surrounding phase winding reorganizes.

A useful check on any traveling-core approximation is supplied by a one-dimensional relative profile $u(x,t)=e^{i\Omega t}U(z)$, $z=x-ct$, with real $c$ and $\Omega$. Direct substitution gives
\begin{widetext}
\be\label{Eq_correct_traveling_profile}
(i\Omega-c\partial_z)(I-\epsilon\partial_z^2)U
=\epsilon c U^{\prime\prime\prime}-i\epsilon\Omega U''-cU'+i\Omega U
=(\lambda+i\omega)U-\kappa|U|^2U.
\ee
\end{widetext}
The mixed derivative generates a third spatial derivative when $c\ne0$. A second-order traveling-profile equation would therefore discard part of the dynamic inertia. Equation~\eqref{Eq_correct_traveling_profile} constrains moving amplitude depressions or phase slips in a banded field; admissible speeds depend on the full profile and its far-field conditions.  Constructing such a profile requires solving this boundary-value problem with the chosen far-field waves. For a two-dimensional defect, the corresponding evolution must also retain its phase winding. In both settings, motion of a band interruption involves the adjustment of the surrounding oscillation together with the core.

\section{Numerical evolution and diagnostics of the random-field experiment}
\label{App_cellular_numerics}

The calculations in Figs.~\ref{Fig_open_cells_formation}--\ref{Fig_cellular_radial} use the square torus $[0,L)^2$, with $L=20$, sampled on a $256^2$ grid. The initial values $10^{-4}(\xi_{jk}+i\zeta_{jk})$, built from independent standard normal samples drawn with the MATLAB default random seed, are projected onto the square Fourier band $|n_x|,|n_y|\le127$; this removes about $0.78\%$ of the initial power, and no randomness is supplied afterward.

For $\kvec=(2\pi/L)(n_x,n_y)$ and $Q_{\kvec}=1+\epsilon|\kvec|^2$, the retained coefficients satisfy
\be\label{Eq_cellular_Galerkin}
\frac{\d\widehat u_{\kvec}}{\d t}
=\frac{\lambda+i\omega}{Q_{\kvec}}\widehat u_{\kvec}
-\frac{\kappa}{Q_{\kvec}}\widehat{P_K(|u|^2u)}_{\kvec},
\ee
where $P_K$ is the square-band projection. The Laplacian acts through its squared-wavenumber symbol, the Nyquist row and column are excluded from the retained state, and the cubic term is formed on a $512^2$ zero-padded grid before projection, so that no wrapped cubic frequency can enter the retained band. Time integration uses the fourth-order exponential time-differencing Runge--Kutta scheme (ETDRK4)~\cite{cox2002exponential,kassam2005fourth} with step $\Delta t=0.004$ up to $T=60$: the diagonal linear part is integrated exactly, and the nonlinear stage coefficients are evaluated by full-circle contour quadrature, keeping their complex values because $(\lambda+i\omega)/Q_{\kvec}$ is complex. These coefficients agree with independently evaluated Taylor series to near machine precision, spatially uniform solutions converge at fourth order to their exact counterparts for both $\kappa=10$ and $\kappa=10+10i$, and reducing the step eightfold over a finite interval near saturation changes the field by less than $10^{-11}$. All analysis uses the actual integration times, and snapshot labels are rounded as stated in the captions.

The linear control in the figures is the exact solution
\be\label{Eq_cellular_linear_control}
\widehat u_{\mathrm{lin},\kvec}(t)
=\widehat u_{0,\kvec}\exp\!\left[
\frac{(\lambda+i\omega)t}{Q_{\kvec}}\right],
\ee
and the characteristic length in Fig.~\ref{Fig_open_cells_formation} is
\be\label{Eq_cellular_length}
\ell_u(t)=\frac{2\pi}{\langle|\kvec|\rangle_P},\qquad
\langle|\kvec|\rangle_P=
\frac{\sum_{\kvec\ne0}|\kvec|\,|\widehat u_{\kvec}|^2}
{\sum_{\kvec\ne0}|\widehat u_{\kvec}|^2}.
\ee
The spectral-shape comparison normalizes radially binned power to unit sum and takes half the summed absolute difference between nonlinear and linear bin fractions, so that it measures redistribution of power independently of overall amplification.

For the correlation diagnostic, the spatial mean of $|u|$ is subtracted at each time and a linear temporal trend is removed at each grid point. The periodic covariance $C_t(\boldsymbol r)=\langle a_t(x)a_t(x+\boldsymbol r)\rangle_x$ of the resulting anomaly $a_t$ is computed from its power spectrum, and $C_t(\boldsymbol r)/C_t(0)$ is averaged over radial bins of width $2/15$ model units. Figure~\ref{Fig_cellular_correlation} shows the median and interquartile range over the frames of each window; these bands describe temporal variability among correlated neighboring frames, and the medians agree closely with a pooled, variance-weighted estimate.

The spiral geometry of Fig.~\ref{Fig_cellular_radial} is diagnosed on $\operatorname{Re}u$ normalized by the spatial RMS amplitude of $u$. The field is Gaussian-smoothed at $1.5$ pixels before forming its gradient structure tensor, whose entries are smoothed at four pixels; rotating the principal gradient direction by $\pi/2$ defines the local contour orientation $\theta$, determined modulo $\pi$. Phase defects are identified as grid plaquettes around which $\arg u$ winds by $\pm 2\pi$. Around a defect core $c$, the spiral index is defined as
\be\label{Eq_cellular_spiral_index}
S(c)=\frac{\sum_{x\in\mathcal A_c} w(x)
\sin\!\left[2\bigl(\theta(x)-\varphi_c(x)\bigr)\right]}
{\sum_{x\in\mathcal A_c} w(x)},
\ee
where $\varphi_c$ is the polar angle about $c$, $\mathcal A_c$ is the periodic annulus between radii $6$ and $30$ pixels, and the weight $w$ is the difference between the two structure-tensor eigenvalues, emphasizing strong and coherently oriented contours. Doubling the angle difference $\theta-\varphi_c$ makes the index invariant to the $\pi$-ambiguity of unsigned contour orientation: $S(c)$ vanishes for purely radial ($\theta=\varphi_c$) or purely circumferential ($\theta=\varphi_c\pm\pi/2$) contours and reaches $\pm 1$ for logarithmic spirals crossing rays at $\pm 45^\circ$, with its sign recording the spiral handedness. For display only, the cyan curves in Fig.~\ref{Fig_cellular_radial} are traced bidirectionally through the same orientation field and terminated where directional coherence becomes weak.

\bibliographystyle{ieeetr}
\bibliography{PGLE_v16_references}
\end{document}